\documentclass[11pt]{article}

\usepackage[pdftex]{graphicx}
\usepackage[small,bf,sf,textfont={small,sf}]{caption}
\usepackage{vmargin}
\usepackage{subfig}
\usepackage{amsmath}
\usepackage{amssymb}
\usepackage{fancyhdr}
\usepackage{color}
\usepackage[colorlinks=true,
            linkcolor=blue,
            urlcolor=blue,
            citecolor=blue]{hyperref}
\usepackage[round]{natbib}
\usepackage{doi}

\usepackage{booktabs}
\usepackage{float}
\usepackage{lineno}
\usepackage{epstopdf}
\usepackage{bm}
\usepackage{dsfont}
\usepackage{multirow}
\usepackage{pdflscape}

\floatstyle{ruled}
\newfloat{floatbox}{thp}{lob}
\floatname{floatbox}{Algorithm}

\usepackage{titlesec}
\titleformat{\section}
   {\normalfont\sffamily\Large\bfseries}
   {\thesection}{1em}{}
\titleformat{\subsection}
   {\normalfont\sffamily\large\bfseries}
   {\thesubsection}{1em}{}
\titleformat{\subsubsection}
   {\normalfont\sffamily\bfseries}
   {\thesubsubsection}{1em}{}

\usepackage{float}
\usepackage{placeins}

\newcommand{\inv}{^{-1}}

\newcommand{\m}{\mathbf{m}}

\newcommand{\e}{\mathbf{e}}
\newcommand{\0}{\mathbf{0}}

\newcommand{\g}{\mathbf{g}}

\newcommand{\dsim}{\mathbf{d}}
\newcommand{\dobs}{\mathbf{d}_{\textrm{obs}}}

\newcommand{\C}{\mathbf{C}}

\newcommand{\Ce}{\mathbf{C}_{\mathbf{e}}}

\newcommand{\K}{\mathbf{K}}
\newcommand{\R}{\mathbf{R}}

\newcommand{\Ex}{\mathbb{E}}

\begin{document}

\title{\sffamily \textbf{Statistical Tapers for Correlation-Based Localization in Ensemble Data Assimilation}}
\author{Alexandre A. Emerick\footnote{Petrobras Research, Development, and Innovation Center, Brazil, \href{mailto://emerick@petrobras.com.br}{\texttt{emerick@petrobras.com.br}}} ~and Vinicius Luiz Santos Silva\footnote{Petrobras, Brazil, \href{mailto://viluiz@petrobras.com.br}{\texttt{viluiz@petrobras.com.br}}}}

\date{}

\maketitle

\section*{Abstract}
\label{sec:abstract}

Localization is essential in ensemble-based data assimilation because finite ensembles produce noisy covariance estimates, leading to spurious updates and excessive loss of ensemble variance. In subsurface applications, localization is usually based on spatial distance, but this criterion can be difficult to justify when parameter-data relationships are controlled by reservoir flow dynamics, nonlinear observation operators, non-local parameters, or prior conditioning effects.

This work investigates correlation-based localization as an alternative strategy in which tapering coefficients are computed directly from the statistical reliability of estimated model-data correlations. We interpret localization as a shrinkage problem in correlation space and propose three tapers: a generalized power-law taper motivated by mean-square-error correction, a logistic taper derived from a Bayesian spike-and-slab formulation, and a discrepancy-based taper inspired by Morozov's discrepancy principle. These methods are compared with existing correlation-based formulations from the literature.

The tapers are evaluated using synthetic reservoir data assimilation problems involving scalar parameters, grid-based parameters, localized flow responses, non-trivial spatial correlation patterns, and increasing model dimension. The results show that correlation-based localization can suppress spurious correlations while preserving meaningful parameter-data relationships. In several cases, the proposed power-law and logistic tapers retained more posterior ensemble variance than distance-based localization while maintaining acceptable data-match quality. The logistic taper generally provided the strongest variance preservation, whereas smoother tapers tended to favor slightly better data matches.

Overall, the results indicate that correlation-based localization is a statistically motivated alternative to distance-based localization, especially when spatial distance is unavailable or misleading. Remaining challenges include taper-parameter selection, computational cost, and robustness in large-scale applications.

\section{Introduction}
\label{sec:intro}

Ensemble-based data assimilation methods have become standard tools for reservoir history matching and uncertainty quantification \citep{emerick:25bk,evensen:25bk}. These methods update uncertain reservoir models by combining prior geological information with dynamic data collected during field operations.

In ensemble methods, the covariance between model parameters and data is estimated from a finite ensemble of reservoir realizations. Because the ensemble size is typically small compared to the dimension of the model and the number of data points, the resulting covariance estimates are affected by significant sampling errors. These errors manifest as spurious correlations between unrelated parameters and observations. In addition, small ensemble sizes limit the effective degrees of freedom available to assimilate data \citep{lorenc:03,aanonsen:09}. As a consequence, these limitations often lead to unrealistic parameter updates and to an underestimation of posterior uncertainty \citep{aanonsen:09}, ultimately degrading the performance of the data assimilation process.

Localization is commonly used to mitigate the adverse effects of small ensembles. In its classical form, localization is implemented by multiplying the ensemble-estimated covariance matrices (or alternatively the Kalman gain matrix) element-wise by a distance-based tapering function \citep{houtekamer:01}. Although very effective in many applications, distance-based localization presents some limitations:

\begin{itemize}
    \item localization radii typically require empirical tuning;
    \item spatial distance may not adequately represent the correlation structure induced by reservoir flow dynamics;
    \item distance is not well-defined for certain types of parameters and observations.
\end{itemize}

A natural alternative to distance-based localization is to design schemes that do not rely on spatial distance, but instead operate directly on the statistical reliability of the estimated correlations. These strategies are often refereed to as correlation-based localization \citep{luo:18a,vossepoel:25a}. In this perspective, each sample correlation is evaluated according to its signal-to-noise ratio, accounting for the uncertainty induced by the finite ensemble size. Correlations that are likely to arise from sampling noise are then attenuated, while statistically meaningful correlations are preserved. Correlation-based localization provides a principled mechanism to control spurious long-range interactions without imposing arbitrary spatial assumptions, making it particularly attractive for complex reservoir settings where the correlation structure is driven by flow dynamics, observation operators, and model nonlinearities rather than geometric proximity.

Despite their attractive features, the effectiveness of correlation-based localization remains limited when compared to distance-based approaches, which continue to be the standard practice in reservoir data assimilation \citep{emerick:25bk}. In this context, the present work aims to:

\begin{enumerate}
    \item Propose statistically motivated correlation-based taper functions and compare them with existing formulations.
    \item Assess the performance of correlation-based localization in the context of non-local model parameters.
    \item Investigate whether correlation-based localization can replace distance-based localization in subsurface applications without compromising the quality of the results.
\end{enumerate}

The remainder of the manuscript is organized as follows. Section~\ref{sec:works} reviews related work on correlation-based localization. Section~\ref{sec:ens_corr} introduces the problem and discusses the role of localization in ensemble smoothers. Section~\ref{sec:cbt} presents the proposed correlation-based localization formulations, and Section~\ref{sec:test_cases} describes the test cases used to evaluate them. Section~\ref{sec:discussion} discusses the main results, while Section~\ref{sec:remarks} highlights important remarks regarding the applicability and implementation of the proposed methods. Finally, Section~\ref{sec:conc} summarizes the main conclusions. Additional derivations omitted from the main text are provided in the appendix.

\section{Related Works}
\label{sec:works}

Early developments in distance-free localization include the hierarchical filter of \citet{anderson:07}, in which the ensemble is partitioned into smaller subsets that are used to estimate localization coefficients. This idea was later extended by \citet{zhang:10}, who employed bootstrap resampling to generate such subsets. \citet{furrer:07} derived a taper function for general covariance structures by minimizing, element-wise, the discrepancy between the true covariance matrix and its localized counterpart, without enforcing positive definiteness. Although the resulting expression depends on the unknown true covariance, the authors suggest replacing it with the ensemble-based estimate to obtain a practical, distance-independent localization. \citet{bishop:07a} proposed constructing the localization matrix by applying a nonlinear transformation to the sample correlation matrix, specifically by raising each element to a prescribed power.

\citet{anderson:12a} proposed deriving localization weights from the expected sampling error of correlations, interpreting localization as a statistically motivated, flow-dependent shrinkage of noisy correlations. This work motivated further investigations aiming to develop a theoretical basis for designing adaptive, non-distance-based localization schemes \citep{flowerdew:15a,menetrier:15a}.

These works laid the foundation for what is now referred to as correlation-based localization, a term introduced in \citep{luo:18a}, where the localization coefficients are defined through thresholding: coefficients are set to one when the estimated correlation exceeds a prescribed threshold, and to zero otherwise. Subsequently, \citet{luo:20a} proposed replacing this hard-thresholding scheme with a continuous tapering function. In their formulation, the Gaspari-Cohn correlation function \citep{gaspari:99} is adopted using a pseudo-distance defined in terms of the estimated correlation and a scaling parameter.

Building on these ideas, \citet{ranazzi:22a} incorporated correlation thresholding into the framework of \citet{furrer:07} by introducing an additional penalty term in the taper function. More recently, \citet{ranazzi:26a} proposed determining a scaling factor that minimizes the expected Frobenius norm of the difference between the true and estimated covariance matrices. In contrast to localization methods that assign different tapering coefficients to each entry of the covariance matrix, this covariance-scaling approach applies a single coefficient to the entire matrix. In the same work, the authors also introduced an improved version of the localization taper originally proposed by \citet{furrer:07}.

\citet{lee:21a} proposed computing localization coefficients as powers of the absolute values of the estimated correlations, providing a simple nonlinear transformation to control the strength of the tapering. More recently, \citet{vishny:24a} introduced NICE (noise-informed covariance estimation), in which the Morozov's discrepancy principle is used to adaptively select the exponent applied to the correlations when constructing localization coefficients. \citet{vossepoel:25a} proposed an approach based on correlation truncation in the Fisher space, combined with an empirical function that tapers model updates by inflating the observation errors associated with distant or weakly correlated observations.

\section{Ensemble Correlations}
\label{sec:ens_corr}

Let $\m \in \mathbb{R}^{N_m}$ denote the vector of model parameters, representing the uncertain quantities to be estimated through data assimilation. In reservoir applications, $\m$ typically includes rock properties such as porosity and permeability, but it may also contain other scalar variables, such as parameters describing relative permeability curves. The observed data are denoted by $\dobs \in \mathbb{R}^{N_d}$ and are related to the model through a nonlinear forward operator $\dsim = \g(\m)$. In this context, $\dobs$ usually consists of measurements acquired at well locations, such as fluid rates and pressure data, as well as seismic observations.

Consider an ensemble of $N_e$ reservoir models, $\m_k$, and the corresponding predicted data, $\dsim_k = \g(\m_k)$,  for $k=1,\dots,N_e$. The goal of data assimilation is to update the ensemble of $\m_k$ by incorporating the information available in $\dobs$. In the context of reservoir applications, ensemble smoothers are typically the preferred choice \citep{emerick:25bk,evensen:25bk}. These methods require estimating the covariance $\widetilde{c}_{ij}$ between each model parameter $m_i$ and each predicted datum $d_j$ from the ensemble:

\begin{equation}
  \widetilde{c}_{ij} = \frac{1}{N_e-1} \sum_{k=1}^{N_e} 
  (m_{i,k} - \overline{m}_i)(d_{j,k} - \overline{d}_j),
\end{equation}

\noindent where $\overline{m}_i$ and $\overline{d}_j$ denote the corresponding ensemble means. Because the ensemble size is limited, the estimated covariance $\widetilde{c}_{ij}$ may differ significantly from the actual covariance $c_{ij}$.

\subsection{Localization}
\label{sec:ens_corr.local}

Localization is an \emph{ad hoc} strategy used to improve estimated covariance values. Two main approaches are commonly employed in the literature: covariance (or Kalman gain) localization and domain localization (also known as local analysis) \citep{sakov:11}. Covariance or Kalman gain localization applies a Schur (element-wise) product between a tapering matrix and the estimated covariance (or Kalman gain). In contrast, domain localization divides the analysis into a set of independent local problems, each assimilating a subset of the observations.

Although these approaches are not mathematically equivalent, they tend to produce similar results when the analysis is largely dominated by the prior \citep{sakov:11}. Domain localization is particularly attractive in implementations where the covariance or Kalman gain are not explicitly formed, such as square-root filters \citep{tippett:03,sakov:08a} and the subspace iterative ensemble smoother \citep{raanes:19b,evensen:19b}.

Here, we adopt the ensemble smoother with multiple data assimilation (ES-MDA) \citep{emerick:13b}, in which case Kalman gain localization provides a convenient and efficient implementation \citep[Chap.~8]{emerick:25bk}. The ES-MDA update equation with Kalman gain localization can be written as

\begin{equation}
    \m_k^{\ell + 1} = \m_k^\ell + \left(\R \circ \widetilde{\K}^\ell \right) \left(\dobs + \sqrt{\alpha_\ell}\e^\ell_k - \g\left(\m_k^\ell\right) \right),\quad \text{for}~k=1,\dots,N_e,
\end{equation}

\noindent with

\begin{equation}
    \widetilde{\K}^\ell = \widetilde{\C}^\ell_{\m\dsim} \left(\widetilde{\C}^\ell_{\dsim\dsim} + \alpha_\ell \Ce \right)\inv
\end{equation}

\noindent denoting the Kalman gain. $\widetilde{\C}^\ell_{\m\dsim}$ and $\widetilde{\C}^\ell_{\dsim\dsim}$ are the covariance matrices between $\m$ and $\dsim$ at the $\ell$-th assimilation step. The vector $\e^\ell_k$ is a random sample drawn from $\mathcal{N}(\0, \Ce)$, where $\Ce$ is the data-error covariance matrix. The parameter $\alpha_\ell$ is the MDA inflation factor. $\R$ is the localization matrix, whose entries are computed using a predefined tapering function, typically defined in terms of the distance between the spatial locations of model parameters and data points.

In practical implementations, it is undesirable to store the full matrices $\R$ and $\widetilde{\K}$, as this would require prohibitive memory for realistic reservoir problems. Instead, the computation is performed in blocks: only small groups of rows of $\widetilde{\K}$ are allocated, and the corresponding localization coefficients are computed on demand. A consequence of this strategy is that the localization operator is not formed explicitly as a matrix. Moreover, the operator is inherently rectangular, as only the $N_m \times N_d$ terms are required for the update, rather than a full square matrix. Therefore, in the following, we derive taper expressions for localization based on correlations for individual model-data pairs, without enforcing the localization matrix to be positive definite.

\section{Correlation-Based Tapers}
\label{sec:cbt}

In correlation-based localization the entries of the $\R$ are computed based on the estimated correlation coefficient between model parameter $m_i$ and data component $d_j$ given by

\begin{equation}
  \widetilde{\rho}_{ij} = \frac{\widetilde{c}_{ij}}{\sqrt{\widetilde{c}_{ii}\widetilde{c}_{jj}}}.
\end{equation}

In this setting, it is convenient to formulate the problem as finding a tapering function $r(\widetilde{\rho})$ such that the localized version of the correlation coefficient 

\begin{equation}
  \widetilde{\rho}_{ij}^{\,\textrm{loc}} = r(\widetilde{\rho}_{ij})\,\widetilde{\rho}_{ij}
\end{equation}

\noindent is a better approximation of the true correlation $\rho_{ij}$. In the remainder of this paper, we drop the subscripts $i$ and $j$ to simplify the notation, noting that we always refer to the correlation between the $i$-th model parameter and the $j$-th data point.

\subsection{Mean Squared Error Taper}
\label{sec:cbt.mse}

Consider the optimal taper $r^\star$ obtained by minimizing the mean-squared error

\begin{equation}
  \mathcal{J}(r) = \Ex\left[(r\widetilde{\rho} - \rho)^2\right].
\end{equation}

Minimizing this expression yields (see Appendix~\ref{sec:app.mse}) to

\begin{equation}\label{eq:mse.rstar}
  r^\star = \frac{\rho^2}{\rho^2 + \sigma^2},
\end{equation}

\noindent where $\sigma^2 = \textrm{var}[\widetilde{\rho}]$ represents the sampling variance of the estimated correlation. There are several expressions proposed in the literature \citep{gnambs:23a} as practical estimators of $\sigma$. Overall these expressions indicate that $\sigma$ is a inverse function of the square root of the ensemble size

\begin{equation}
  \sigma \approx \frac{1}{\sqrt{N_e}}.
\end{equation}

Here, we use the estimator proposed in \citep{soper:13}, with $\widetilde{\rho}$ used as a plug-in estimator:

\begin{equation}\label{eq:bayes_taper.params.sigma}
  \sigma = \frac{1-\widetilde{\rho}^2}{\sqrt{N_e-1}},
\end{equation}

\noindent which indicates that $\sigma$ depends on the ensemble size and in the value of the estimated correlations. This expression is consistent with the fact that small correlations are inherently difficult to estimate from finite ensembles, as their signal-to-noise ratio is low and sampling variability is large \citep{fisher:21a}.

Eq.~\ref{eq:mse.rstar} is a oracle type of taper as it requires knowing the true value of $\rho$. Since $\rho$ is unknown, a natural approximation is to replace it by the ensemble estimate $\widetilde{\rho}$:

\begin{equation}\label{eq:mse.r}
  r(\widetilde{\rho}) = \frac{\widetilde{\rho}^2}{\widetilde{\rho}^2 + \sigma^2}.
\end{equation}

Variations of this MSE the taper have been previously proposed \citep{furrer:07,anderson:12a,flowerdew:15a}. Unfortunately, this taper is often too permissive and fails to effectively suppress spurious correlations in realistic data assimilation problems \citep{lacerda:19a,lacerda:21a,silva:25a}.

\subsubsection{Power-law taper}
\label{sec:cbt.power}

Defining the standardized correlation coefficient

\begin{equation}
  t = \frac{|\widetilde{\rho}|}{\sigma},
\end{equation}

\noindent Eq.~\ref{eq:mse.r} can be written as

\begin{equation}\label{eq:mse.rt2}
  r(t) = \frac{t^2}{t^2 + 1}.
\end{equation}

\noindent Note that $r(t) \approx 0$ for $t \ll 1$ and $r(t) \approx 1$ for $t \gg 1$. Eq.~\ref{eq:mse.rt2} suggests a generalized form 

\begin{equation}\label{eq:power.rtbeta}
  r(t) = \frac{t^\beta}{t^\beta + t_0^\beta},
\end{equation}

\noindent where the exponent $\beta \geq 2$ controls the shape of the taper and the coefficient $t_0$ represents the standardized correlation at which the taper equals $1/2$. Note that once we replace the unknown true correlation by the noisy ensemble estimate in Eq.~\ref{eq:mse.rstar}, we introduced a second layer of estimation error. In this setting, Eq.~\ref{eq:power.rtbeta} can be justified as an extra shrinkage step that compensates for plug-in bias and low signal-to-noise ratios. In Appendix~\ref{app:generalized_tapers}, we provide a motivation for the use of this generalized form based on spike-and-slab distributions, showing that the power-law taper can be interpreted as a smooth posterior probability that an estimated correlation belongs to the signal component. Note that using $\beta > 2$ is analogous to raising the estimated correlation coefficient to an exponent larger than one, a strategy that has also been adopted in other localization approaches \citep{bishop:07a,lee:21a,vishny:24a}.

\begin{figure}
\centering
  \subfloat[\scriptsize{Effect of $t_0$}]{\includegraphics[width=0.4\linewidth]{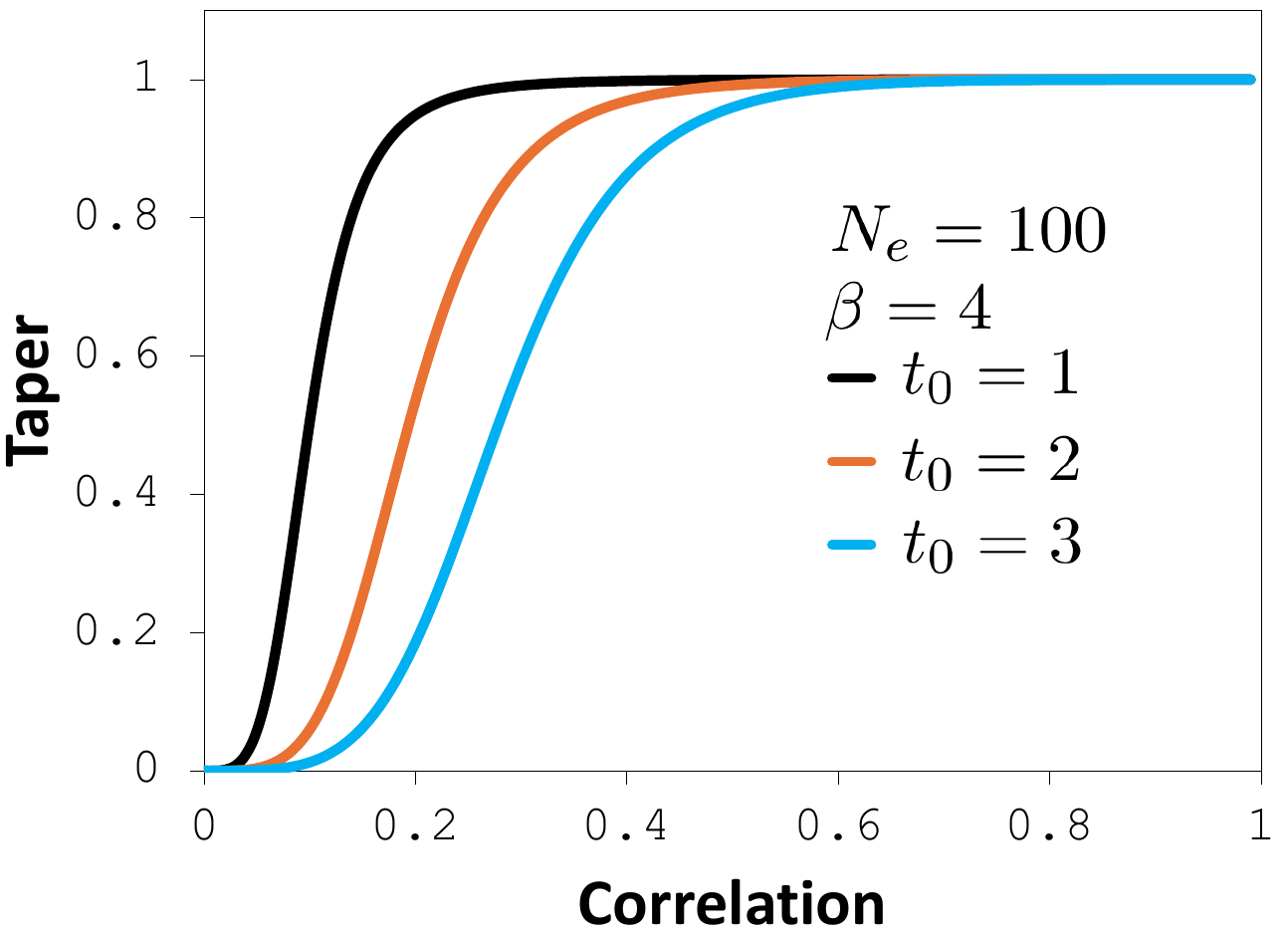}}
  \subfloat[\scriptsize{Effect of $\beta$}]{\includegraphics[width=0.4\linewidth]{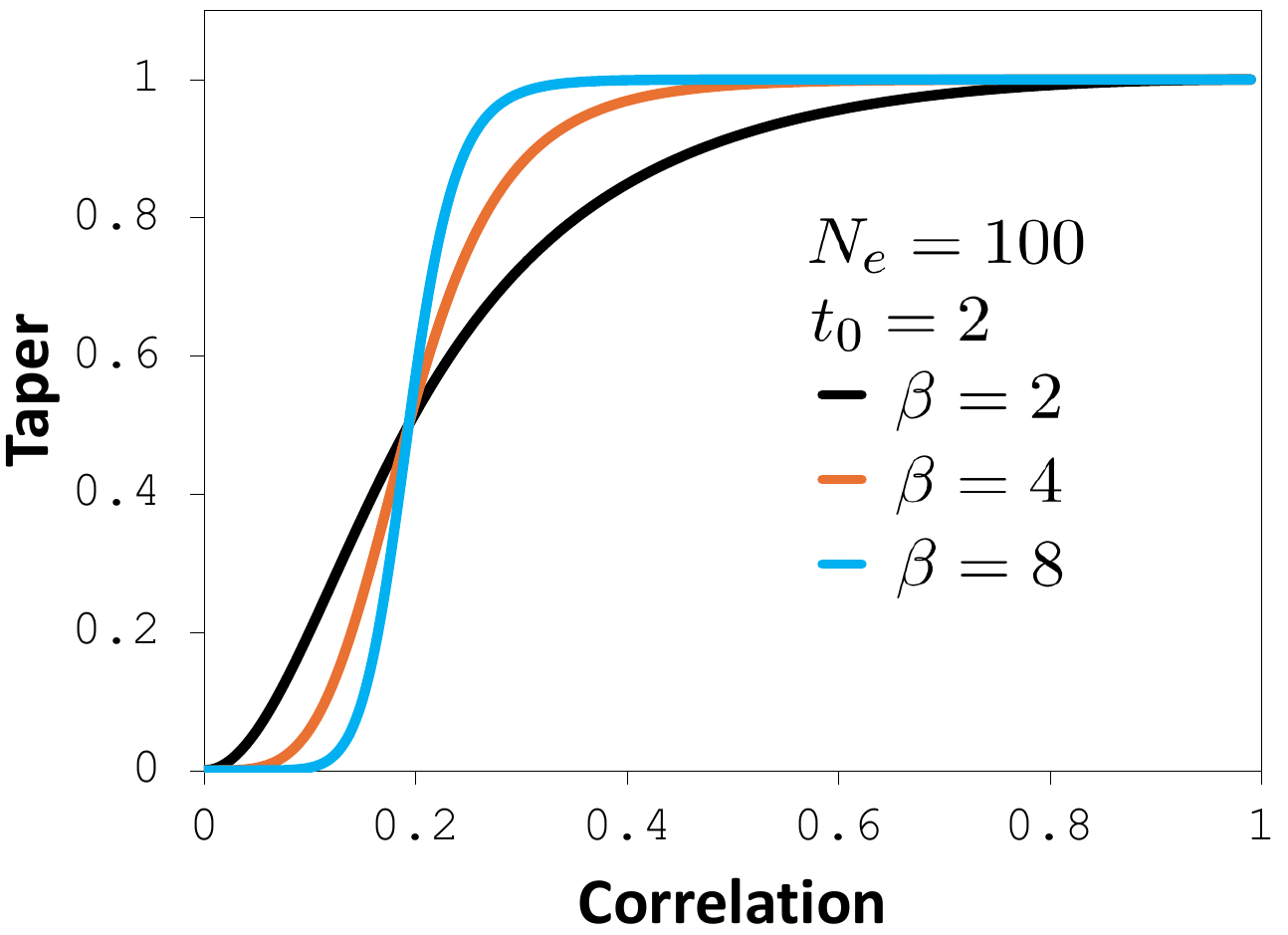}}\\
   \subfloat[\scriptsize{Effect of $N_e$}]{\includegraphics[width=0.4\linewidth]{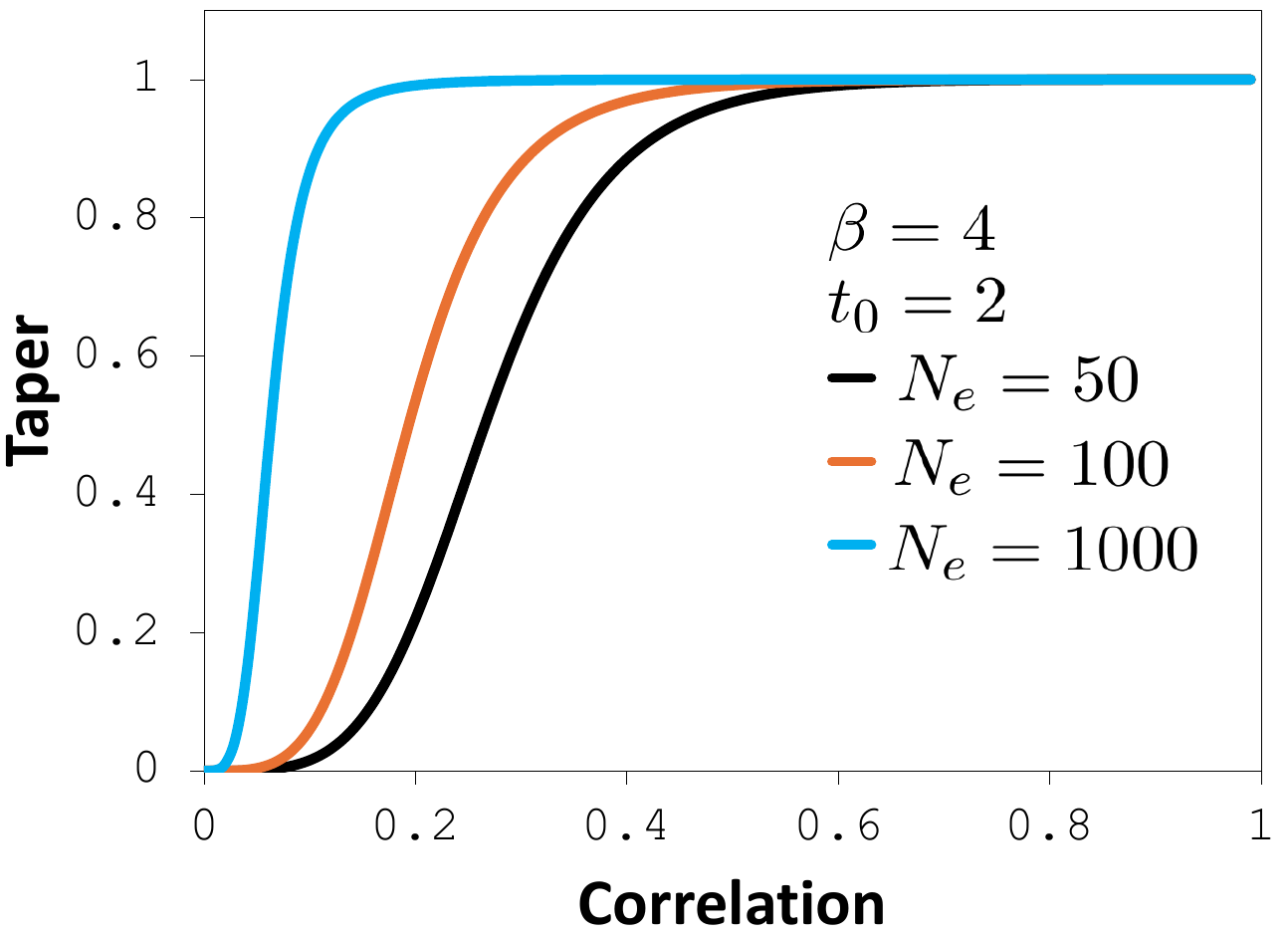}}
\caption{Power taper as function of the correlation coefficient for different combinations of the parameters $t_0$ and $\beta$, and ensemble size.}
\label{fig:cbt.power.taper}
\end{figure}

The resulting taper function has two free parameters, $t_0$ and $\beta$. The parameter $t_0$ defines the transition level at which the correlation is equally likely to be spurious or genuine, while $\beta$ controls the sharpness of this transition, with larger values producing a more abrupt, near hard-threshold behavior. Fig.~\ref{fig:cbt.power.taper} illustrates the resulting taper values for different choices of $t_0$, $\beta$, and ensemble size.

The ensemble size enters the taper computation through $\sigma$, which is used to define the standardized correlation. (An expression for computing $\sigma$ is presented later in Section~\ref{sec:cbt.bayes_taper.params}.) As a result, the taper automatically adapts to the ensemble size, with larger ensembles requiring less tapering (Fig.~\ref{fig:cbt.power.taper}c).

\subsection{Sparse Bayesian Tapering in the Correlation Space}
\label{sec:cbt.bayes_taper}

In the previous section, we used the MSE criterion to motivate the power-law taper function. In this section, we derive a taper function within a Bayesian framework. We begin with a Gaussian prior and then extend the formulation to a more realistic spike-and-slab distribution.

\subsubsection{Gaussian prior taper}
\label{sec:cbt.bayes_taper.gauss}

We begin by assuming the likelihood model

\begin{equation}
\widetilde{\rho} \mid \rho \sim \mathcal{N}(\rho,\sigma^2).
\end{equation}

\noindent As before, $\sigma^2$ represents the sampling variance of the estimated correlation.

Suppose further that the true correlation follows a Gaussian prior

\begin{equation}
  \rho \sim \mathcal{N}(0,\upsilon^2),
\end{equation}

\noindent where $\upsilon^2$ denotes the variance of genuine correlations, representing the typical magnitude of the true correlations. Note that this is different from $\sigma^2$ which is the sampling variance of the estimated correlation and therefore represents noise rather than signal magnitude. Under squared-error loss, the optimal estimator of $\rho$ is the posterior mean

\begin{equation}
  \widetilde{\rho}^{\,\textrm{loc}} = \Ex[\rho \mid \widetilde{\rho}].
\end{equation}

Since both prior and likelihood are Gaussian, the posterior distribution is also Gaussian (see derivation in Appendix~\ref{sec:app.gauss_taper}):

\begin{equation}
  \rho \mid \widetilde{\rho} \sim \mathcal{N} \left( \frac{\upsilon^2}{\upsilon^2+\sigma^2}\widetilde{\rho}, \frac{\upsilon^2\sigma^2}{\upsilon^2+\sigma^2} \right),
\end{equation}

\noindent which yields

\begin{equation}
  \widetilde{\rho}^{\,\textrm{loc}} =  \frac{\upsilon^2}{\upsilon^2+\sigma^2}\,\widetilde{\rho}.
\end{equation}

Therefore the corresponding taper is

\begin{equation}\label{eq:bayes_taper.gauss.r}
  r = \frac{\upsilon^2}{\upsilon^2+\sigma^2}.
\end{equation}

Introducing the ratio

\begin{equation}
  \tau = \frac{\upsilon}{\sigma},
\end{equation}

\noindent this taper can be written as

\begin{equation}\label{eq:bayes_taper.r}
  r = \frac{\tau^2}{\tau^2+1}.
\end{equation}

Eq.~\ref{eq:bayes_taper.r} has the same algebraic structure as the MSE taper (Eq.~\ref{eq:mse.r}), with the prior variance computed entrywise as $\widetilde{\rho}^2$. This expression has a natural interpretation: the taper depends on the signal-to-noise ratio between the typical magnitude of genuine correlations, $\upsilon$, and the sampling uncertainty, $\sigma$.

Although the Gaussian-prior taper is mathematically convenient, it does not fully capture the structure typically expected in practical localization problems. The Gaussian prior assumes that all correlations belong to a single continuous population centered at zero.

In ensemble data assimilation, however, correlations between model parameters and data points are expected to be sparse. Many correlations arise purely from sampling noise and should therefore be regarded as effectively zero, whereas only a relatively small subset corresponds to genuine physical or statistical dependence. Hence, the empirical distribution of correlations is expected to consist of a large concentration of zero or near-zero values, together with a smaller set of significant nonzero correlations.

The Gaussian prior does not represent this structure properly, since it assigns continuous probability mass around zero rather than distinguishing between null and nonzero correlations. As a result, the Gaussian taper may be too permissive: weak and likely spurious correlations are not suppressed aggressively enough, while stronger correlations are shrunk.

\subsubsection{Spike-and-slab prior taper}
\label{sec:cbt.bayes_taper.spike-slab}

To represent this sparse structure, we propose to use a spike-and-slab prior (see Appendix~\ref{sec:app.spike-and-slab_distribution}) for the true correlation,

\begin{equation}
  p(\rho)=(1-\lambda)\delta(\rho)+\lambda\,p_{\textrm{slab}}(\rho),
\end{equation}

\noindent where $\delta$ denotes a point mass at zero, $\lambda = P(\rho \neq 0)$ is the prior probability that a correlation is nonzero, and $p_{\textrm{slab}}$ describes the distribution of genuine nonzero correlations.

Under the likelihood model

\begin{equation}
  \widetilde{\rho} \mid \rho \sim \mathcal{N}(\rho,\sigma^2),
\end{equation}

\noindent the posterior distribution becomes a mixture

\begin{equation}
p(\rho \mid \widetilde{\rho}) = (1-f(\widetilde{\rho}))\delta(\rho) + f(\widetilde{\rho})p_{\textrm{slab}}(\rho \mid \widetilde{\rho}),
\end{equation}

\noindent where

\begin{equation}
f(\widetilde{\rho}) = P(\rho \neq 0 \mid \widetilde{\rho})
\end{equation}

\noindent is the posterior probability that the correlation belongs to the slab. The posterior mean therefore becomes

\begin{equation}
\Ex[\rho \mid \widetilde{\rho}] = f(\widetilde{\rho}) \Ex_{\textrm{slab}}[\rho \mid \widetilde{\rho}].
\end{equation}

Assuming a Gaussian slab

\begin{equation}
\rho \mid (\rho\neq0) \sim \mathcal{N}(0,\upsilon^2),
\end{equation}

\noindent yields the localization rule (see derivation in the Appendix~\ref{sec:app.spike-slab}):

\begin{equation}
  \widetilde{\rho}^{\,\textrm{loc}} = f(\widetilde{\rho}) \frac{\upsilon^2}{\upsilon^2+\sigma^2}\widetilde{\rho},
\end{equation}

\noindent where 

\begin{equation}\label{eq:bayes_taper.spike-slab.gamma}
  f(\widetilde{\rho}) = \left[1 + \frac{1-\lambda}{\lambda}\sqrt{\frac{\sigma^2+\upsilon^2}{\sigma^2}}\exp\left(-\frac{\upsilon^2\widetilde{\rho}^2}{2\sigma^2(\sigma^2+\upsilon^2)}\right)\right]\inv.
\end{equation}

Hence the taper becomes

\begin{equation}\label{eq:bayes_taper.spike-slab:r1}
  r(\widetilde{\rho}) = f(\widetilde{\rho}) \frac{\upsilon^2}{\upsilon^2+\sigma^2}.
\end{equation}

This expression reveals that the spike-and-slab taper combines two mechanisms:

\begin{enumerate}
\item a detection component $f(\widetilde{\rho})$ that determines whether the correlation is likely to be genuine,
\item a shrinkage component $\upsilon^2/(\upsilon^2+\sigma^2)$ that controls the damping of nonzero correlations.
\end{enumerate}

Thus the spike-and-slab prior naturally produces the behavior desired for localization: weak correlations are aggressively suppressed because they are likely to belong to the spike component, whereas stronger correlations are preserved with moderate shrinkage.

\subsubsection{Interpretation of the taper parameters}
\label{sec:cbt.bayes_taper.params}

Using standardized correlation $t$, Eq.~\ref{eq:bayes_taper.spike-slab:r1} assumes the following form (see Appendix~\ref{sec:app.spike-slab}) 

\begin{equation}\label{eq:bayes_taper.spike-slab:r2}
  r(t) = \frac{\tau^2}{\tau^2+1} \left[ 1 + \frac{1-\lambda}{\lambda}\sqrt{\tau^2+1}\exp\!\left( -\frac{\tau^2}{2(1+\tau^2)}t^2\right) \right]\inv,
\end{equation}

\noindent which shows that the localization coefficient depends only on the standardized correlation magnitude $t$, together with two parameters, $\lambda$ and $\tau$, that control the sparsity and magnitude of genuine correlations.

\paragraph{Parameter $\lambda$:} The parameter $\lambda$ represents the prior probability that a correlation is nonzero, $\lambda = P(\rho \neq 0)$. Equivalently, $1-\lambda$ is the probability that the true correlation belongs to the spike component and is therefore zero. Hence $\lambda$ directly controls the assumed sparsity of the correlation structure.

Small values of $\lambda$ correspond to strongly sparse settings, where most correlations are assumed to vanish. In this regime, the taper strongly suppresses small estimated correlations because they are likely to originate from sampling noise. Conversely, larger values of $\lambda$ imply a denser correlation structure and therefore lead to less aggressive suppression. In the limit of $\lambda = 1$, we return to the Gaussian prior taper.

\paragraph{Parameter $\tau$:} The parameter $\tau$ represents the ratio between $\upsilon$, which characterizes the typical magnitude of genuine correlations, and $\sigma$, which represents the sampling uncertainty of the estimated correlation. Thus, $\tau$ measures the signal-to-noise ratio of the correlations. When $\tau$ is small, genuine correlations are weak relative to sampling noise, and the taper remains small even for moderately large estimated correlations. When $\tau$ is large, genuine correlations dominate sampling variability, and the taper approaches one for sufficiently large values of $t$.

\paragraph{Standardized correlation $t$:} The standardized correlation $t$ represents the magnitude of the estimated correlation relative to its sampling uncertainty. Hence $t$ plays a role analogous to a statistical significance measure. Small values of $t$ correspond to correlations that are comparable to sampling noise and should therefore be suppressed. Larger values indicate correlations that are unlikely to have arisen purely from sampling variability.

In particular,

\begin{itemize}
  \item if $t \lesssim 1$, the correlation is comparable to noise and the taper is close to zero;
  \item if $t \approx 2$, the taper begins to increase significantly;
  \item if $t \gtrsim 3$, the taper approaches its asymptotic value $\tau^2/(\tau^2+1)$.
\end{itemize}

\subsubsection{Logistic parameterization}
\label{sec:cbt.bayes_taper.logistic}

The spike-and-slab formulation presented in the previous subsection provides a probabilistic interpretation of the localization taper. However, the resulting expression involves parameters $\lambda$ and $\tau$ that are not easy to tune in practice. 

For practical applications, it is convenient to rewrite the taper in the equivalent logistic form (see Appendix~\ref{sec:app.spike-slab.logistic}):

\begin{equation}\label{eq:cbt.bayes_taper.logistic.r1}
  r(t) = r_{\max} \frac{1}{1+\exp\!\left(-c(t^2-t_0^2)\right)},
\end{equation}

where

\begin{align}
  r_{\max} &= \frac{\tau^2}{\tau^2+1}, \\
  c &= \frac{\tau^2}{2(\tau^2+1)} = \frac{r_{\max}}{2} \\
  t_0^2 &=\frac{2(\tau^2+1)}{\tau^2}\ln\!\left(\frac{1-\lambda}{\lambda}\sqrt{\tau^2+1}\right).
\end{align}

In this representation, $t_0$ has the same meaning as before: it defines the standardized correlation at which the taper reaches half of its maximum value; $c$ controls the sharpness of the transition; and $r_{\max}$ determines the maximum retention of strong correlations.

Eq.~\ref{eq:cbt.bayes_taper.logistic.r1} produces a sharp transition between negligible and fully retained correlations, as illustrated in Fig.~\ref{fig:cbt.logist.taper}. This abrupt behavior may lead to an undesirable switching effect, in which moderate correlations are either excessively damped or almost fully retained. To attenuate this effect and increase the flexibility of the taper, we follow the same strategy used for the power-law taper and replace $t^2$ with $t^\gamma$:

\begin{equation}\label{eq:cbt.bayes_taper.logistic.ralpha}
    r(t) = r_{\max} \frac{1}{1+\exp\!\left[-c \left(t^\gamma-t_0^\gamma \right)\right]},
\end{equation}

\noindent where $\gamma > 0$, with typical values in the range $\gamma \in [0.5,2]$. Values of $\gamma < 2$ produce smoother transitions than the original formulation, whereas larger values lead to sharper transitions. A probabilistic justification for using $\gamma < 2$, based on a spike-and-slab interpretation of the taper, is provided in Appendix~\ref{app:generalized_tapers}.

\subsubsection*{Practical parameter choices}

In the spike-and-slab formulation, the taper satisfies $r_{\max}=\tau^2/(\tau^2+1)<1$, implying that even highly significant correlations are subject to residual shrinkage. In practice, however, it is often desirable to preserve strong correlations without attenuation. For this reason, we simplify the expression by enforcing $r_{\max}=1$, so that sufficiently significant correlations are fully retained. This choice decouples the asymptotic behavior of the taper from its transition region and leads to a more interpretable and flexible formulation.

Under the choice $r_{\max} = 1$, the logistic taper cannot satisfy $r(0)=0$ exactly. Instead, we enforce the practical condition $r(0) = \varepsilon$, where $\varepsilon$ is a small tolerance (for example, $\varepsilon = 10^{-2}$). Since

\begin{equation}
  r(0)=\frac{1}{1+\exp(ct_0^\gamma)},
\end{equation}

\noindent this requirement leads to

\begin{equation}
  c = \frac{\ln\!\left(\frac{1-\varepsilon}{\varepsilon}\right)}{t_0^\gamma}.
\end{equation}

\noindent Thus, once $t_0$ and tolerance $\varepsilon$ are specified, the steepness parameter $c$ is uniquely determined. This construction ensures that correlations comparable to sampling noise are effectively suppressed, while maintaining a smooth transition toward full retention for increasingly significant correlations.

The parameter $t_0$ controls the transition between shrinking and retaining correlations and can therefore be interpreted as a \emph{threshold} separating statistically insignificant from significant correlations. In practice, $t_0$ is the key parameter of the taper function, as it directly governs the balance between suppressing spurious correlations and preserving meaningful ones.

Although $t_0$ can be motivated using the Student-$t$ statistic (see Appendix~\ref{sec:app.t0}), this approach relies on assumptions, such as independence and Gaussianity of samples, that are not strictly satisfied in ensemble data assimilation, potentially leading to overly restrictive thresholds. In addition, the optimal value of $t_0$ may depend on the problem dimension: in high-dimensional settings, true correlations tend to be weaker and their estimates noisier, so large values of $t_0$ may cause relevant correlations to fall below nominal significance levels. 

\begin{figure}
\centering
  \subfloat[\scriptsize{Effect of $t_0$}]{\includegraphics[width=0.4\linewidth]{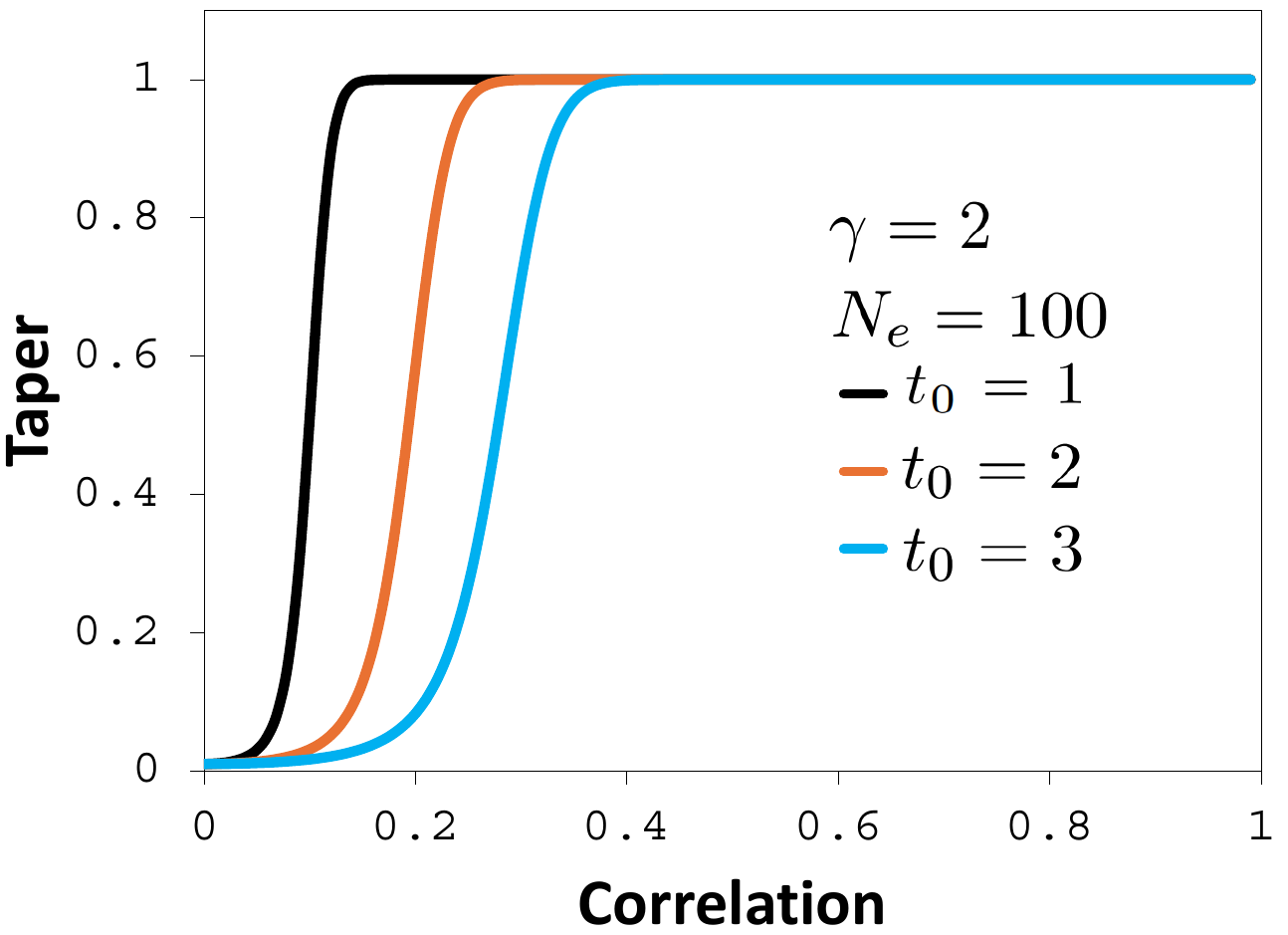}}
  \subfloat[\scriptsize{Effect of $\gamma$}]{\includegraphics[width=0.4\linewidth]{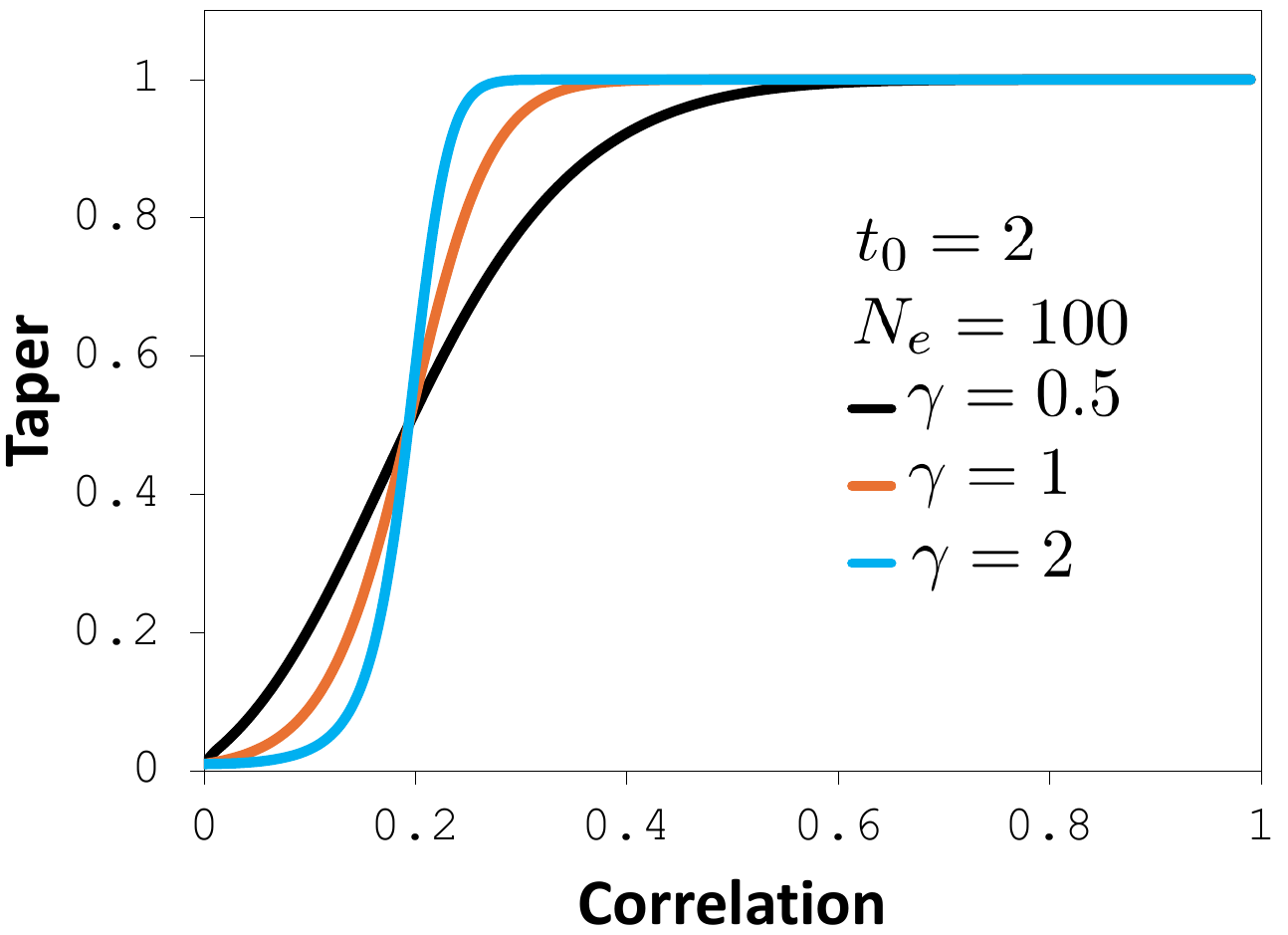}} \\
  \subfloat[\scriptsize{Effect of $N_e$}]{\includegraphics[width=0.4\linewidth]{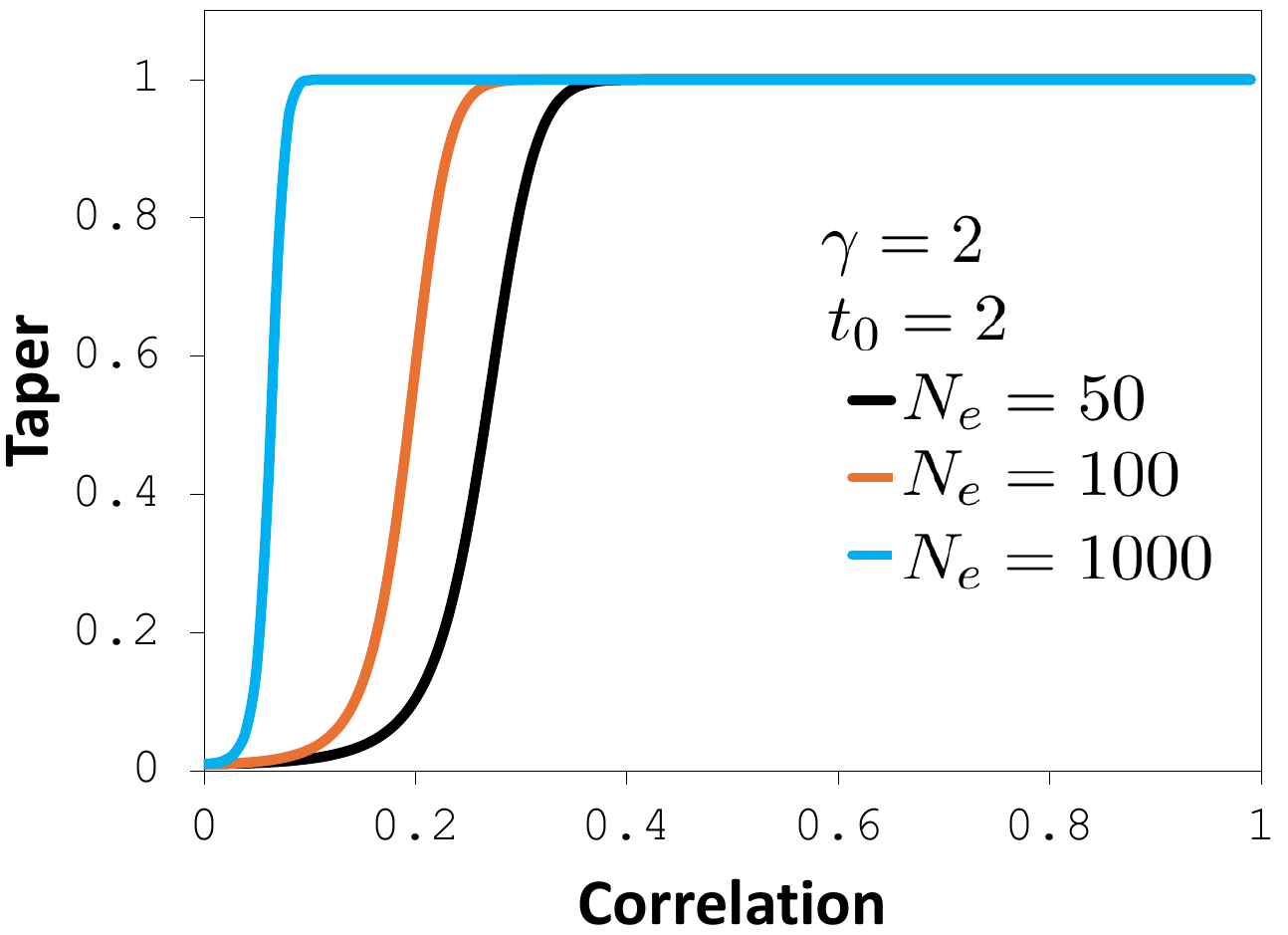}}
\caption{Logistic taper as function of the correlation coefficient for different combinations of the coefficients $t_0$, $\gamma$, and the ensemble size $N_e$.}
\label{fig:cbt.logist.taper}
\end{figure}

Fig.~\ref{fig:cbt.logist.taper} illustrates the resulting taper function for different values of $t_0$, $\gamma$, and ensemble size. Compared to the power-law taper, the logistic taper presents a sharper transition, particularly for $\gamma = 2$. Reducing $\gamma$ leads to a smoother taper.

\subsection{Discrepancy Taper}
\label{sec:cbt.discrepancy}

In inverse problems, the Morozov discrepancy principle \citep{morozov:84a} is often used to select the regularization strength so that the residual is consistent with the expected noise level, rather than driving the fit as low as possible. \citet{vishny:24a} used this principle to select parameters of the NICE taper function. Here, we use a similar idea but, instead of selecting parameters of the previous tapers, we show that the discrepancy principle induces a new taper formulae.

A natural Morozov-type discrepancy rule at the level of a single correlation is to require that the amount removed by localization be equal to the expected noise level, up to a factor, $\eta > 0$: 

\begin{equation}
  | \widetilde{\rho} - \widetilde{\rho}^{\,\textrm{loc}}| = \eta \sigma
\end{equation}
 
Equivalently,

\begin{equation}
  \left(1 - r(t) \right) |\widetilde{\rho}| = \eta \sigma
\end{equation}

\noindent Using $t = |\widetilde{\rho}|/\sigma$ leads to the taper 

\begin{equation}
  r(t) = 1 - \frac{\eta}{t}, \quad t > 0.
\end{equation}

Because a taper must lie in [0,1], the practical form is the clipped version

\begin{equation}\label{eq:cbt.discrepancy_taper}
  r(t) = \max \left( 0, 1 - \frac{\eta}{t} \right).
\end{equation}

\begin{figure}[ht!]
\centering
  \subfloat[\scriptsize{Effect of $\eta$}]{\includegraphics[width=0.4\linewidth]{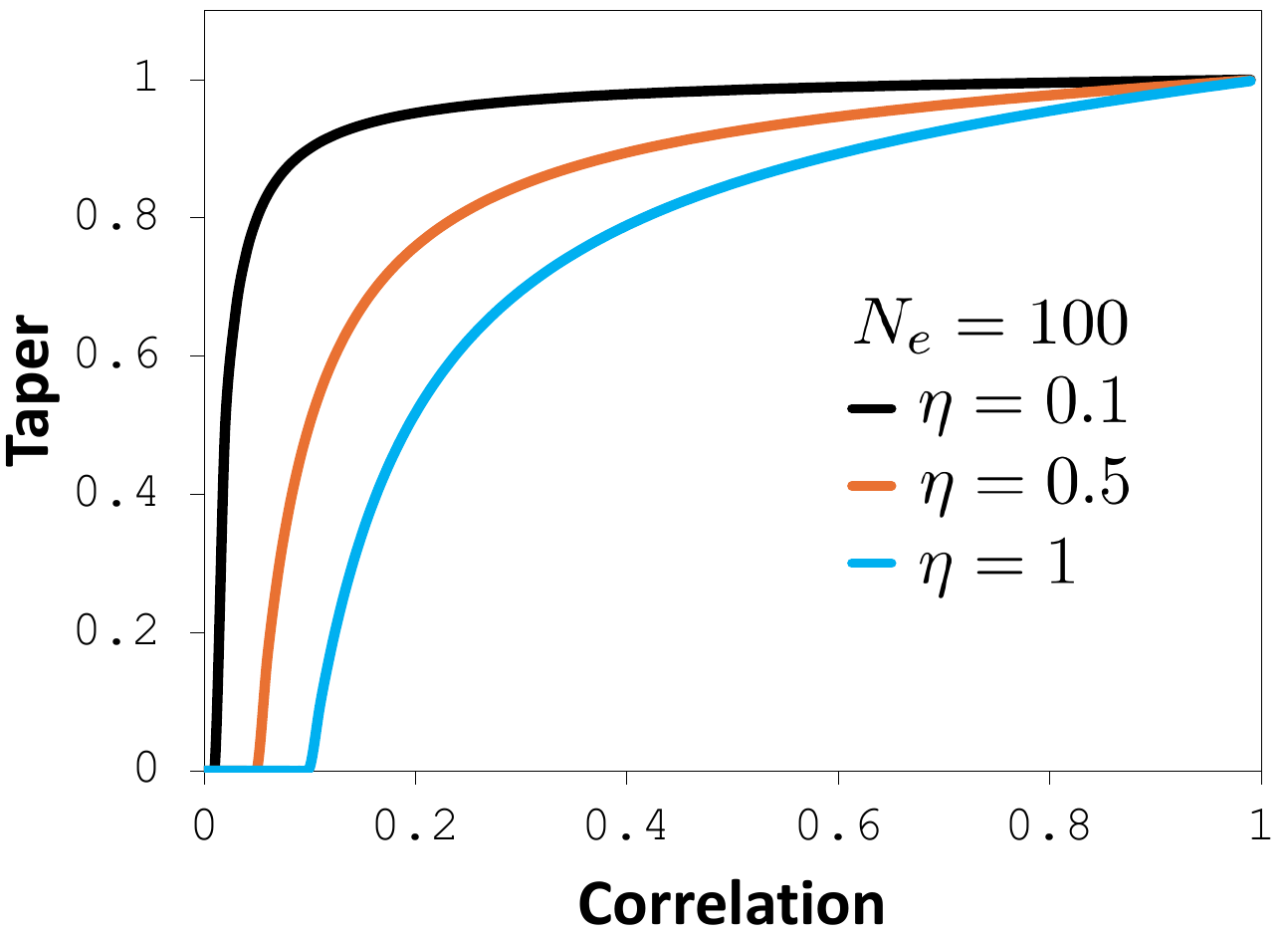}}
  \subfloat[\scriptsize{Effect of $N_e$}]{\includegraphics[width=0.4\linewidth]{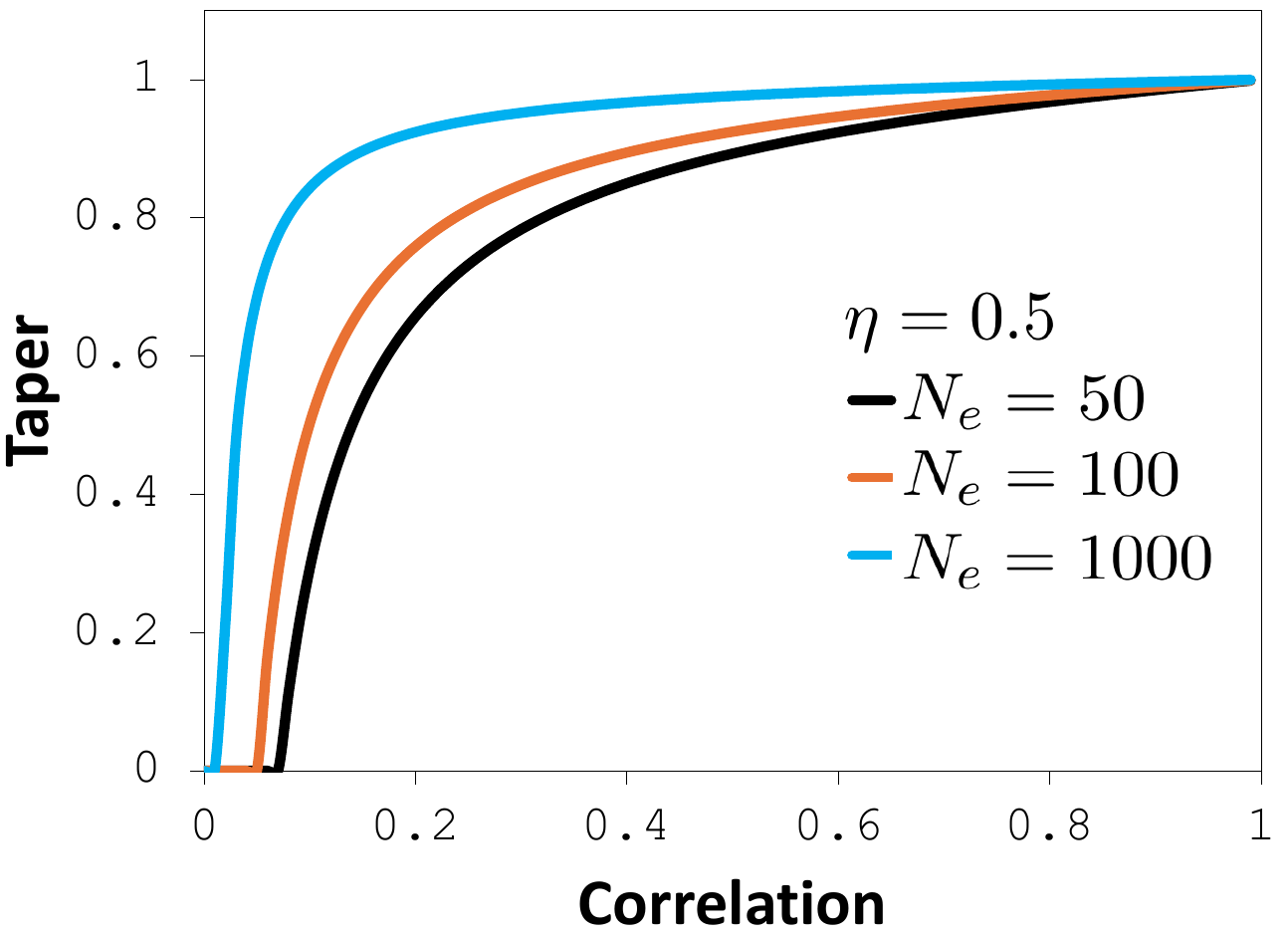}}
\caption{Discrepancy taper as function of the threshold coefficient $\eta$ and ensemble size.}
\label{fig:cbt.discrepancy.taper}
\end{figure}

Fig.~\ref{fig:cbt.discrepancy.taper} illustrates the resulting taper for different values of the parameter $\eta$ and for varying ensemble sizes. The taper has a single parameter, $\eta$, which acts as a hard-threshold in $t$: correlations with standardized magnitude below $\eta$ are fully suppressed, while larger values are progressively retained. As expected, the need for tapering decreases as the ensemble size increases, leading to less aggressive attenuation of the correlations.

\subsection{Existing Correlation-Based Tapers}
\label{sec:cbt.existing}

As discussed in Section~\ref{sec:works}, a variety of correlation-based taper functions have been proposed in the literature. Here, we present the equations of three taper functions that have been previously applied in the context of reservoir data assimilation. The objective is to use these tapers in the comparisons presented in the test case problems discussed in the following section. We selected tapers that do not require forming the full localization matrix, since this may become infeasible for large-scale problems.

\paragraph{Correlation taper from \citet{luo:20a}:} This taper function extends the hard-thresholding localization proposed by \citet{luo:18a} by introducing a continuous formulation. In this approach, a pseudo-distance defined in terms of the correlation coefficient and a scaling factor is used as the argument of the Gaspari-Cohn correlation function:

\begin{equation}\label{eq.cbt.existing.cgc}
  r(\widetilde{\rho}) = f_{\textrm{GC}}\left( \frac{1 - |\widetilde{\rho}|}{1 - \theta} \right),
\end{equation}

\noindent where $f_{\textrm{GC}}(\cdot)$ denotes the Gaspari-Cohn correlation function. \citet{luo:20a} discussed strategies to define the scale parameter $\theta$ as a function of the noise level in $\widetilde{\rho}$, which is expected to be on the order of $1/\sqrt{N_e}$. Although \citet{luo:20a} refer to $\theta$ as a threshold, it appears that they do not explicitly set the taper values to zero when $|\widetilde{\rho}| < \theta$ (see Fig.~2 in \citealp{luo:20a}). In the experiments presented in the next section, we adopt a similar procedure and use $\theta = \sigma$, with $\sigma$ estimated from Eq.~\ref{eq:bayes_taper.params.sigma}. Hereafter, we refer to this formulation as CGC (correlation-based Gaspari-Cohn) taper.

\paragraph{Correlation taper from \citet{furrer:07}:} This is one of the earliest correlation-based tapers proposed in the literature. It is closely related to the MSE taper discussed in Section~\ref{sec:cbt.mse}. The main difference is that Eq.~\ref{eq:mse.r} requires a plug-in estimate for $\sigma$, whereas \citet{furrer:07} derived the taper function from the eigen-decomposition of the prior covariance matrix. Although originally expressed in terms of covariances, the taper can be equivalently written in terms of correlation coefficients as

\begin{equation}\label{eq.cbt.existing.po}
    r(\widetilde{\rho}) = \frac{\widetilde{\rho}^{2}}{\widetilde{\rho}^{2} + \frac{1 + \widetilde{\rho}^{2}}{N_e}}.
\end{equation}

\noindent The resulting taper function has no extra parameters. Following \citet{furrer:07}, we refer to this expression as PO (pseudo-optimal) taper hereafter.

\paragraph{Correlation taper from \citet{ranazzi:26a}:} This taper was proposed as an improved version of the PO taper:

\begin{equation}\label{eq.cbt.existing.mpo}
    r(\widetilde{\rho}) = \max\left(0, \frac{N_e - \frac{1}{\widetilde{\rho}^{2}}}{N_e + 1} \right).
\end{equation}

\noindent Unlike the original PO taper, Eq.~\ref{eq.cbt.existing.mpo} implies in a hard-threshold of $1/\sqrt{N_e}$. In the following, we refer to Eq.~\ref{eq.cbt.existing.mpo} as MPO (modified pseudo-optimal) taper. 

\begin{figure}
\centering
    \includegraphics[width=0.5\linewidth]{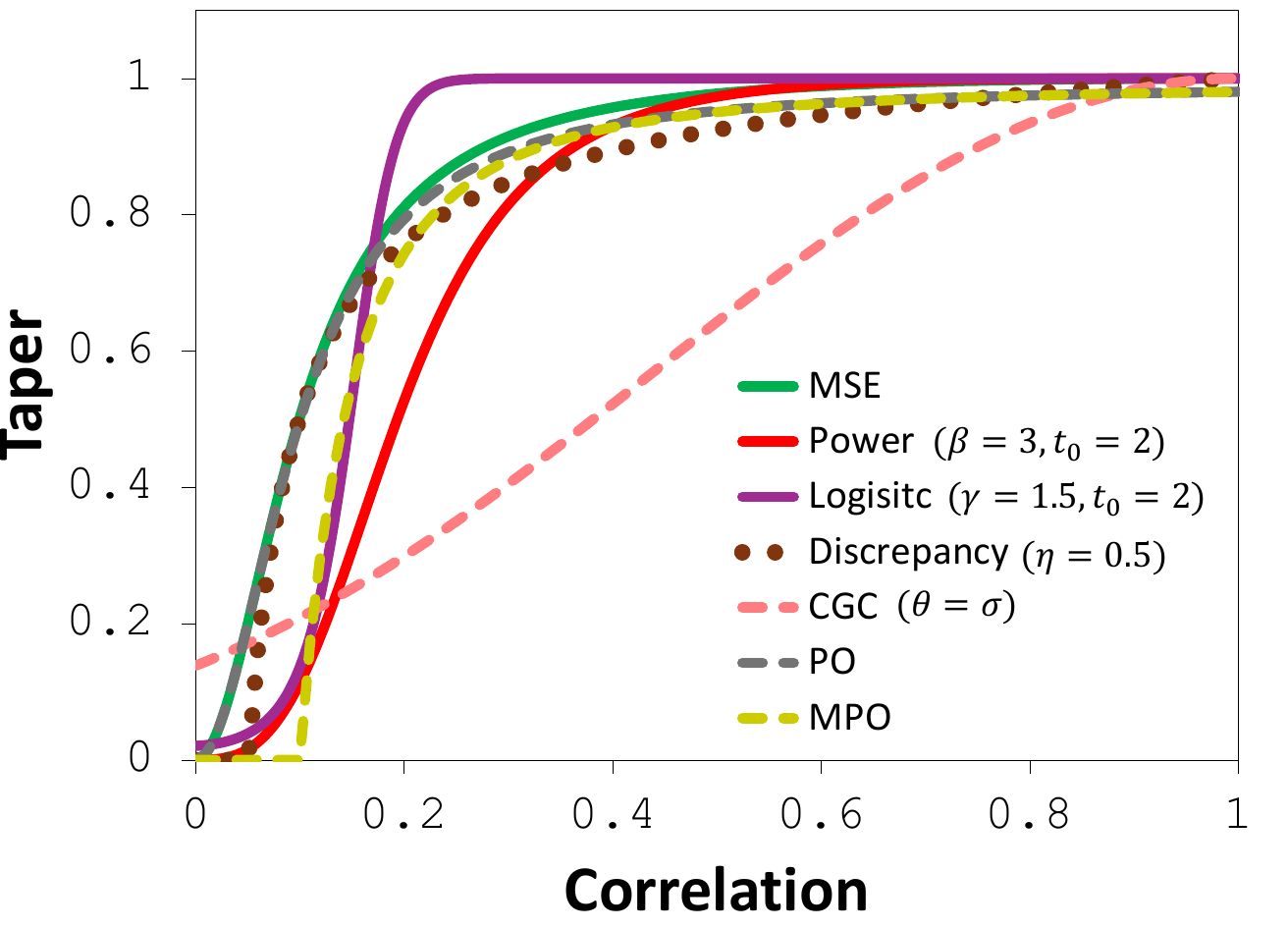}    
\caption{Correlation-based taper functions considered in the test cases ($N_e=100$).}
\label{fig:cbt.summary.tapers}
\end{figure}

Table~\ref{tab:cbt.summary} summarizes the taper functions discussed in this work, along with their corresponding typical parameter ranges. Fig.~\ref{fig:cbt.summary.tapers} shows the taper profiles with corresponding parameters adopted in the test cases discussed in Section~\ref{sec:test_cases}.

\begin{table}[ht!]
\centering
\renewcommand{\arraystretch}{2.1}
\setlength{\tabcolsep}{6pt}
\caption{Summary of correlation-based taper functions for localization.}
\label{tab:cbt.summary}
\begin{tabular}{lll}
\toprule
\textbf{Taper} & \textbf{Expression} & \textbf{Parameters} \\
\midrule

MSE
& $\displaystyle r = \frac{t^2}{t^2 + 1}$ 
& No parameters \\

Power-law
& $\displaystyle r = \frac{t^\beta}{t^\beta + t_0^\beta}$ 
& $\beta \geq 2, \quad t_0 \in [1, 3]$ \\

Logistic 
& $\displaystyle r = \frac{1}{1 + \exp\!\left[-c\left(t^\gamma - t_0^\gamma\right)\right]}$ 
& $\gamma \in [0.5,2], \quad t_0 \in [1, 3]$ \\

Discrepancy 
& $\displaystyle r = \max\!\left(0,\; 1 - \frac{\eta}{t}\right)$ 
& $\eta \in [0.1, 1], \quad t > 0$ \\

CGC 
&  $r = f_{\textrm{GC}}\left( \frac{1 - |\widetilde{\rho}|} {1 - \theta} \right)$
& $\theta \in (0, 1), \quad \theta \approx 1/\sqrt{N_e}$ \\

PO 
& $r = \frac{\widetilde{\rho}^{2}}{\widetilde{\rho}^{2} + \left(1 + \widetilde{\rho}^{2} \right)/N_e}$
& No parameters \\

MPO 
& $r = \max\left(0, \frac{N_e - \frac{1}{\widetilde{\rho}^2}}{N_e + 1} \right)$
& No parameters \\

\bottomrule
\end{tabular}
\end{table}

\section{Test Cases}
\label{sec:test_cases}

We selected a set of synthetic reservoir data assimilation problems designed to investigate different aspects of the performance of the correlation-based tapers. The first test case considers only scalar model parameters, representing a scenario in which distance-based localization is not applicable. Test case~2 considers a problem in which distance-based localization can be applied. However, an analysis of the spatial structure of the correlations shows that relevant model parameters are not necessarily located in the vicinity of the wells. Therefore, a distance-based taper centered at the data location may not be optimal in this case.

Test case~3 represents a situation in which distance-based localization is expected to perform well, as the relevant updates are predominantly local. The objective is to investigate whether correlation-based localization can effectively replace distance-based localization in this setting. Finally, test case~4 extends the third test case to investigate the impact of increasing problem dimensionality on the performance of the localization methods.

For comparisons, we are particularly interested in two metrics:

\paragraph{The average data-mismatch objective function:}
    \begin{equation}
        \overline{\mathcal{O}_d(\m)} = \frac{1}{N_e} \sum_{k=1}^{N_e} \left[\frac{1}{2N_d} \sum_{j=1}^{N_d} \left( \frac{d_{\textrm{obs},j} - g_j(\m_k)}{\sigma_{e,j}}\right)^2 \right],
    \end{equation}
    \noindent where $\sigma_{e,j}$ denotes the standard deviation of the error associated with the $j$-th data point. We use $\overline{\mathcal{O}_d(\m)}$ as a measure of data-match quality. For linear-Gaussian problems, the expected value of this normalized objective function for a posterior ensemble should be approximately $1/2$ \citep{oliver:18b}. Although this result does not strictly apply to the nonlinear test cases considered here, it provides a useful reference for interpreting the results.

    In this work, we adopt the interval $1/2 \leq \overline{\mathcal{O}_d(\m)} \leq 1$ as a practical range for evaluating the performance of the localization methods. The lower bound is motivated by the linear-Gaussian reference value: values substantially below $1/2$ may indicate overfitting, with the ensemble predictions fitting the observed data more closely than warranted by the assumed data uncertainty. The upper bound provides a more conservative tolerance for nonlinear effects and sampling variability. Values above this range suggest that the posterior ensemble still presents a relatively poor data match, whereas values within the interval indicate that the method achieves a reasonable compromise between fitting the observations and avoiding excessive updates.    
    
\paragraph{The normalized variance:}
    \begin{equation}
        \textrm{NV} = \frac{1}{N_m} \sum_{i=1}^{N_m} 
        \frac{\textrm{var}\left[m_{\textrm{post},i}\right]}
        {\textrm{var}\left[m_{\textrm{prior},i}\right]} .
    \end{equation}

$\textrm{NV}$ provides a scalar measure of the average variance retained by the posterior ensemble relative to the prior ensemble. We use $\textrm{NV}$ as an approximate measure of uncertainty reduction after data assimilation \citep{oliver:08bk}. A value of $\textrm{NV}=1$ indicates that the posterior ensemble preserves the prior variance, corresponding to no uncertainty reduction. In contrast, $\textrm{NV}=0$ indicates a complete collapse of the ensemble variance.

In general, data assimilation is expected to reduce uncertainty, leading to values of $\textrm{NV}$ below one. However, excessively small values may indicate overcorrection or artificial loss of ensemble spread, especially when the corresponding data match is not substantially improved. Therefore, when comparing two methods with similar data-match quality, we prefer the method that retains a larger value of $\textrm{NV}$, since it provides a better preservation of posterior variability while achieving a comparable fit to the observations.

\subsection{Test Case 1: Scalar Parameters}
\label{sec:test_cases.scalar_par}

The first test case corresponds to a modified version of the PUNQ-S3 problem \citep{floris:01}, considering only scalar model parameters. This is the same problem used in \citet{lacerda:19a,silva:25a}, designed to evaluate localization strategies in a setting where distance-based localization is not applicable. 

The model uses a corner-point grid of $19 \times 28 \times 5$ cells (2,660 total gridblocks, 1,761 active). Horizontal cell dimensions are uniform at $180 \text{ m} \times 180 \text{ m}$, with variable thickness. The geological structure consists of sand channels embedded in low-porosity shale, with facies variations across layers. The model simulates compressible multiphase flow of gas, oil, and water. We use a commercial black-oil simulator to run the model. 

The data assimilation problem consists of 15 model parameters with varying levels of impact on the reservoir’s dynamic response. These include five porosity multipliers and five permeability multipliers, one for each reservoir layer. Additional parameters comprise rock compressibility, two parameters controlling the size of the lateral aquifer (which provides pressure support), and two parameters defining the positions of the original gas-oil and oil-water contacts.

In addition, five dummy parameters are included (total $N_m=20$), sampled from a standard Gaussian distribution. These parameters have no influence on the simulation responses and are uncorrelated \emph{a priori}. Therefore, any reduction in their posterior variances can be attributed solely to sampling errors. 

The production history comprises one year of extended well testing, a three-year shut-in period, and four years of subsequent production of six wells. Observed data include monthly measurements of well water cut, gas-oil ratio, and bottom-hole pressure, obtained by adding Gaussian noise to the reference model predictions. The noise has zero mean and standard deviation equal to $10\%$ of the data value for water cut and gas-oil ratio, and $1\%$ for bottom-hole pressure. The total number of data points is $N_d = 1{,}530$ .

For this test case, we performed data assimilation using ES-MDA with $\alpha_\ell = N_a = 4$. This test case was designed to assess the performance of the correlation-based tapers in a setting considering scalar model parameters. This represents a typical scenario in which distance-based localization is not applicable.

\begin{figure}
\centering
    \subfloat[\scriptsize{Objective function}]{\includegraphics[width=0.5\linewidth]{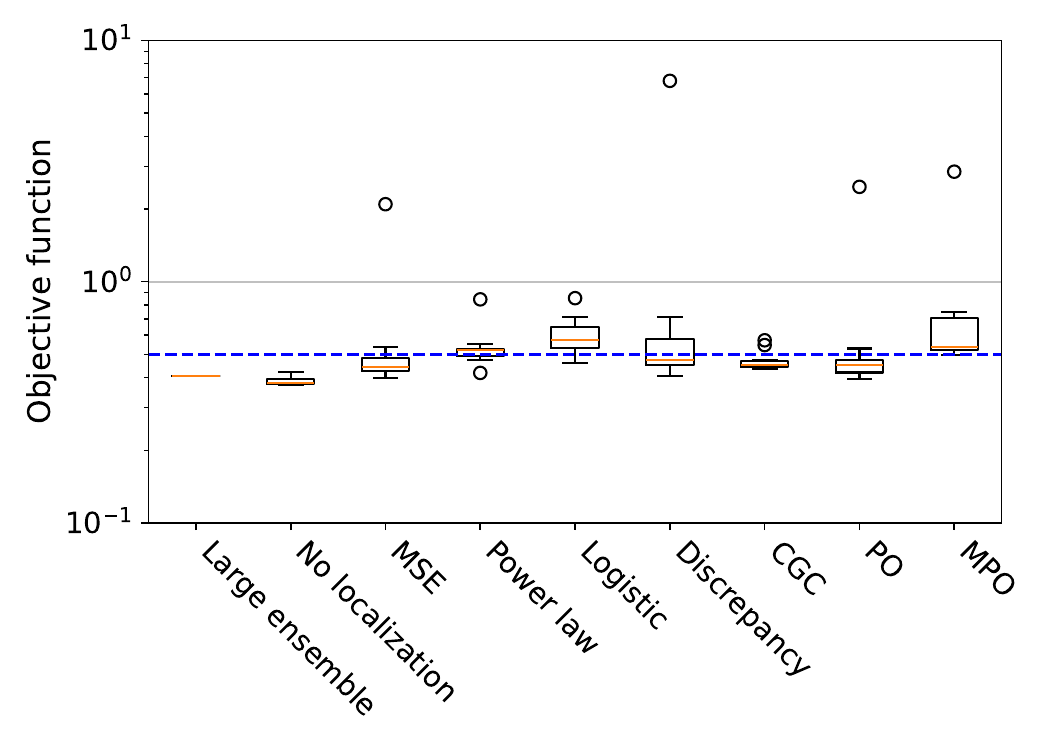}}
    \subfloat[\scriptsize{Normalized variances}]{\includegraphics[width=0.50\linewidth]{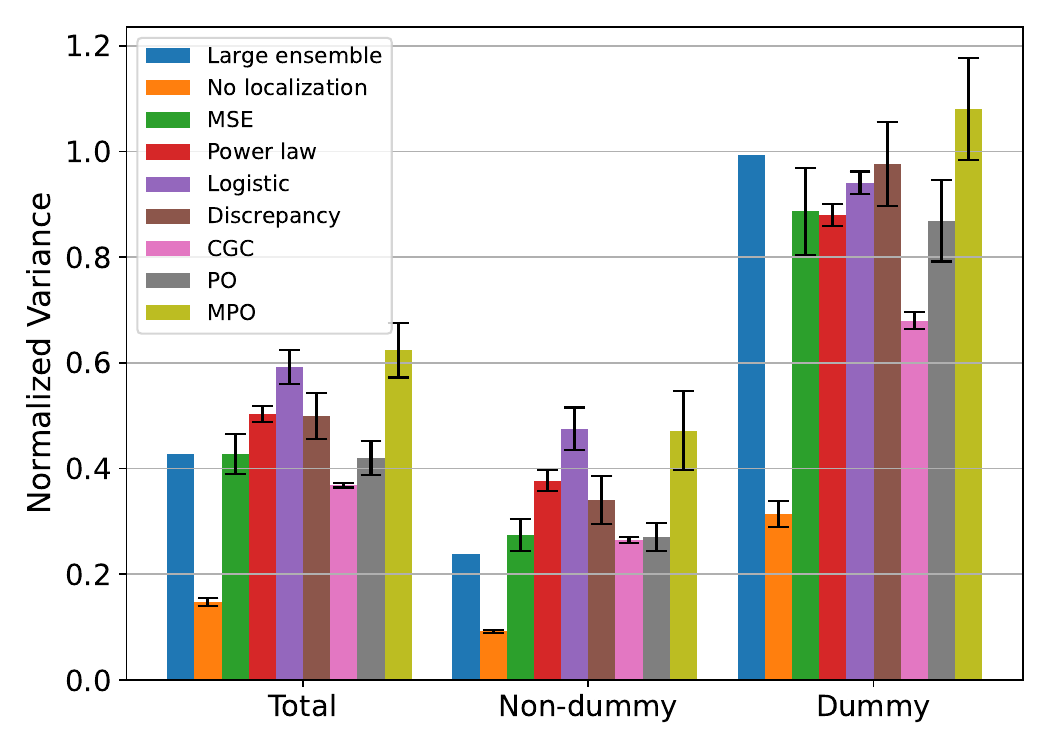}}
\caption{Average data-mismatch objective function, normalized variance and mean offset for different localization methods. Each case represents the average over 10 runs with different initial ensembles, except for the case labeled ``Large ensemble,'' which corresponds to data assimilation results obtained with an ensemble of size $N_e =20{,}000$ without localization. The dashed blue line in the boxplot highlights an objective function of 0.5. The error lines in the bars plots correspond to the 95\% confidence intervals. Test case 1.}
\label{fig:test_cases.scalar_par.obj.nv}
\end{figure}

Fig.~\ref{fig:test_cases.scalar_par.obj.nv} shows the results of running 10 independent data assimilations using different initial ensembles of size $N_e = 100$. As a reference, we also include the results obtained with a large ensemble ($N_e = 20{,}000$) without localization. Fig.~\ref{fig:test_cases.scalar_par.obj.nv}a shows boxplots of $\overline{\mathcal{O}_d(\m)}$, Fig.~\ref{fig:test_cases.scalar_par.obj.nv}b shows a bar plot of the corresponding average $\textrm{NV}$. 

In terms of data-match quality, most cases resulted in values of $\overline{\mathcal{O}_d(\m)} < 1$. However, the MSE, discrepancy, PO, and MPO tapers produced at least one data assimilation run with $\overline{\mathcal{O}_d(\m)} > 1$, indicating reduced robustness with respect to the choice of the initial ensemble. Regarding normalized variance, the case without localization resulted in a clear underestimation of $\textrm{NV}$, while all localization methods substantially increased the $\textrm{NV}$. Considering that the correct value of the $\textrm{NV}$ for the dummy parameters should be $\textrm{NV}=1$, we notice that the large-ensemble case generated $\textrm{NV} \approx 1$, the discrepancy and MPO overestimate it for some runs, and the CGC generated a value considerably below the other tapers. For the non-dummy parameters the logistic and MPO showed the highest $\textrm{NV}$ values. 

\begin{figure}
\centering
    \subfloat[\scriptsize{MSE}]{\includegraphics[width=0.25\linewidth]{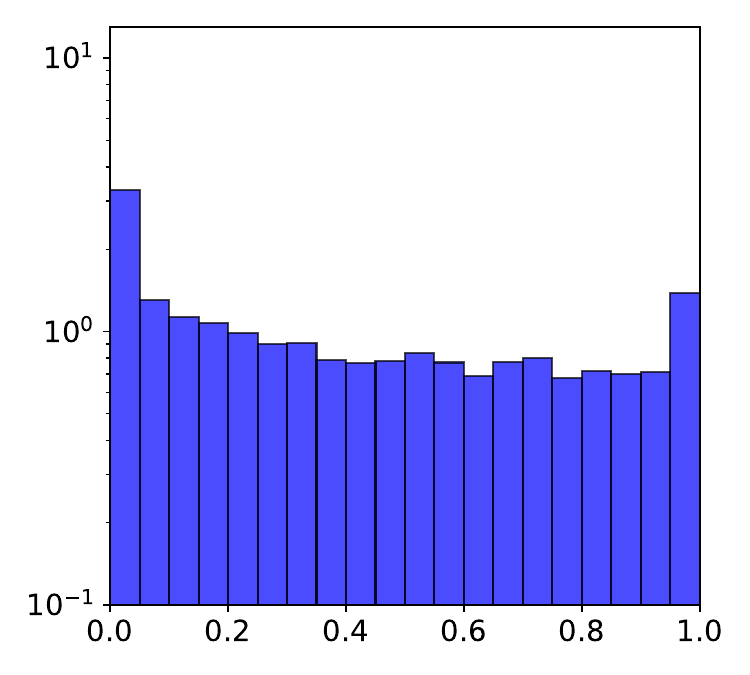}}
    \subfloat[\scriptsize{Power-law}]{\includegraphics[width=0.25\linewidth]{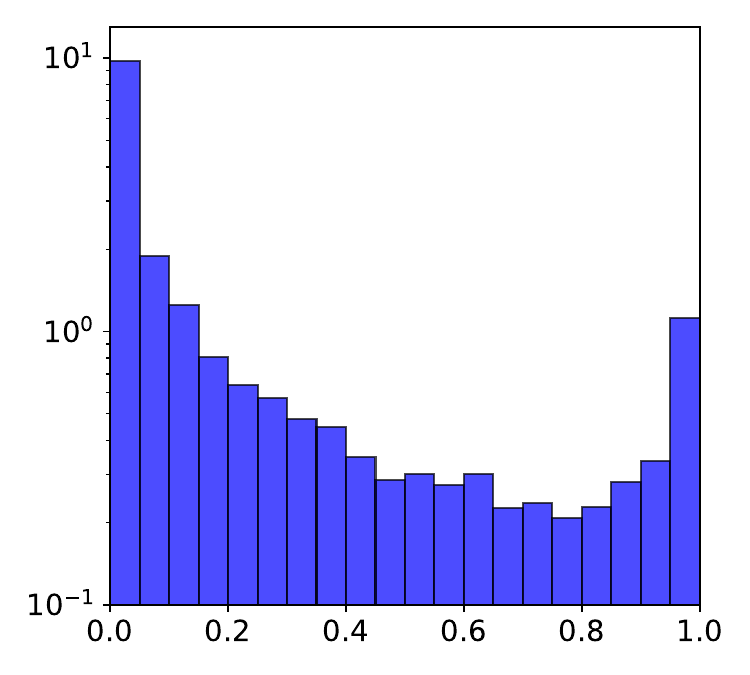}}
    \subfloat[\scriptsize{Logistic}]{\includegraphics[width=0.25\linewidth]{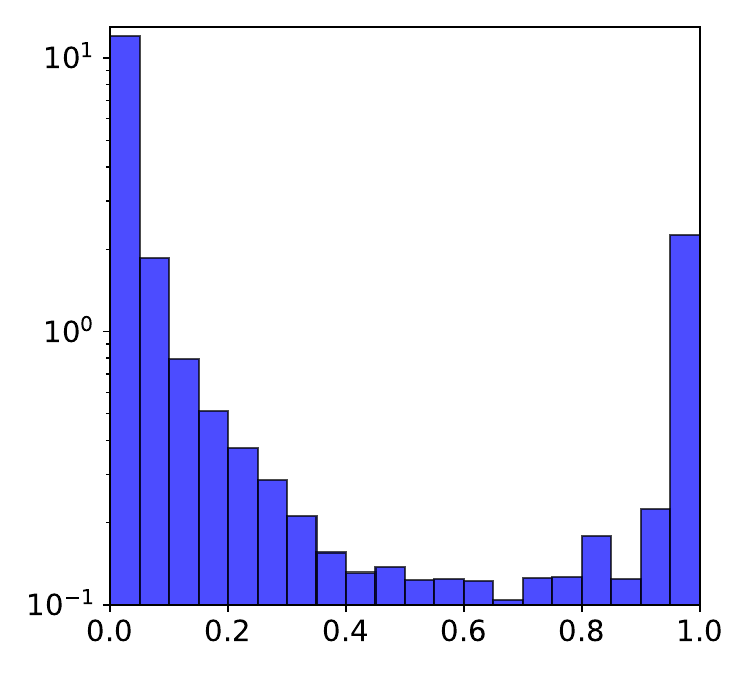}}
    \subfloat[\scriptsize{Discrepancy}]{\includegraphics[width=0.25\linewidth]{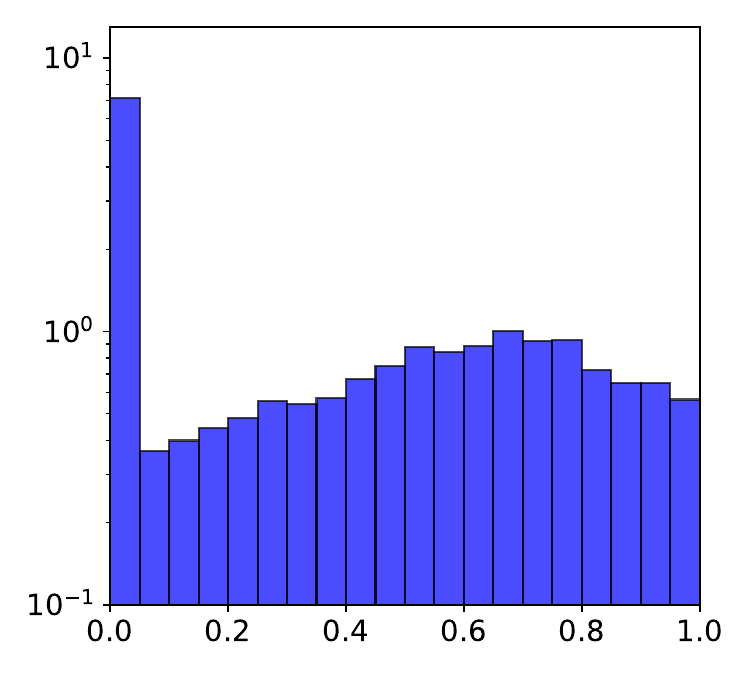}} \\
    \subfloat[\scriptsize{CGC}]{\includegraphics[width=0.25\linewidth]{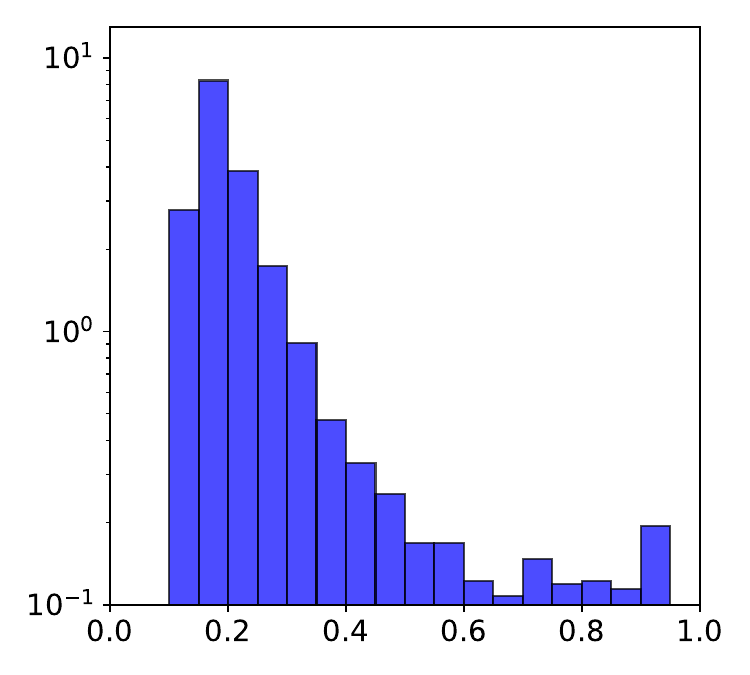}}
    \subfloat[\scriptsize{PO}]{\includegraphics[width=0.25\linewidth]{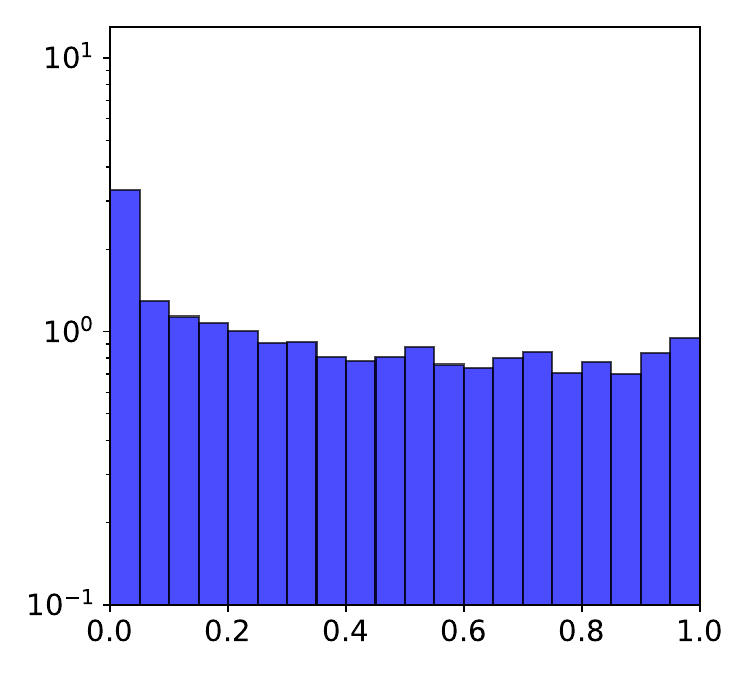}}
    \subfloat[\scriptsize{MPO}]{\includegraphics[width=0.25\linewidth]{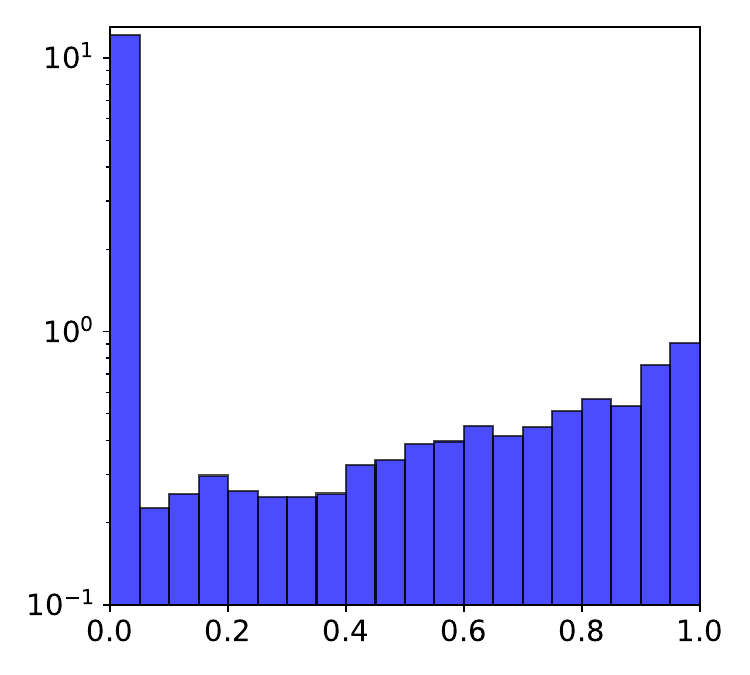}}    
\caption{Histogram of the taper values for different localization methods. Test case 1.}
\label{fig:test_cases.scalar_par.histtaper}
\end{figure}

Fig.~\ref{fig:test_cases.scalar_par.histtaper} shows the distribution of taper values for one data assimilation run. The MSE and PO tapers produced nearly uniform distributions, which supports the previous observation that these tapers are too permissive and do not sufficiently suppress spurious correlations. In contrast, the power-law, logistic, discrepancy-based, and MPO tapers exhibit a pronounced peak near zero, indicating a stronger suppression of weak and potentially spurious correlations. The power-law and logistic tapers also show a concentration of values near one, suggesting a clearer separation between correlations that are strongly suppressed and those that are largely retained. This behavior indicates that these tapers are more effective in discriminating between meaningful and spurious correlations. The CGC taper presents the most distinct distribution among the cases analyzed. In particular, it does not produce taper values equal to zero for the choice $\theta=\sigma$, as also shown in Fig.~\ref{fig:cbt.summary.tapers}.

\subsection{Test Case 2: Grid Parameters}
\label{sec:test_cases.grid_par}

The second test case uses the PUNQ-S3 model, but considers only grid parameters: porosity, horizontal log-permeability, and vertical log-permeability. This results in $N_m = 5{,}283$ parameters updated during data assimilation. The prior ensemble was generated using sequential Gaussian simulation \citep{deutsch:02bk}, based on the geostatistical description presented in \citep{floris:01}. The observed data are the same as those described in Section~\ref{sec:test_cases.scalar_par}, with $N_d = 1{,}530$. This is the same problem considered by \citet{silva:25b}.

For this test case, data assimilation was performed using ES-MDA with $\alpha_\ell = N_a = 4$. This test case was chosen to assess the performance of the correlation-based tapers in a setting where distance-based localization is applicable. However, analysis of the spatial correlation structure reveals that the relevant model parameters are not necessarily concentrated near the wells \citep{silva:25b}. Consequently, a distance-based taper centered at the data location is suboptimal in this case. Nevertheless, we included a distance-based localization case using the Gaspari-Cohn function with a critical radius equal to 10 gridblocks, corresponding to one third of the largest reservoir dimension.

\begin{figure}
\centering
    \subfloat[\scriptsize{Objective function}]{\includegraphics[width=0.50\linewidth]{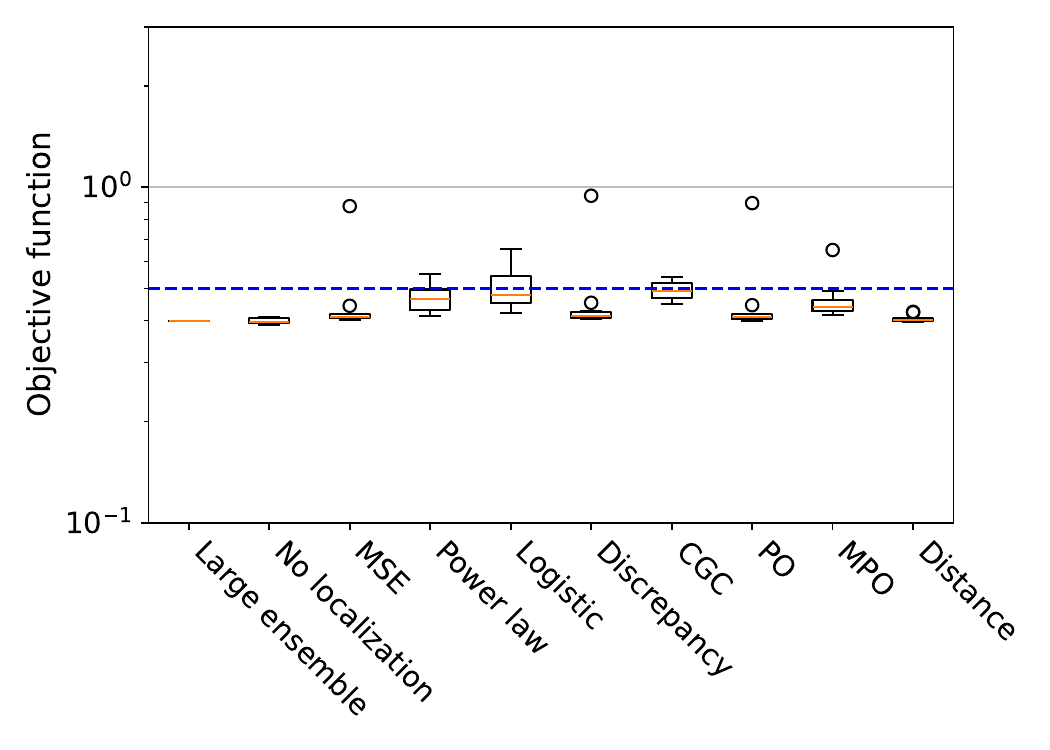}}
    \subfloat[\scriptsize{Normalized variance and mean offset}]{\includegraphics[width=0.50\linewidth]{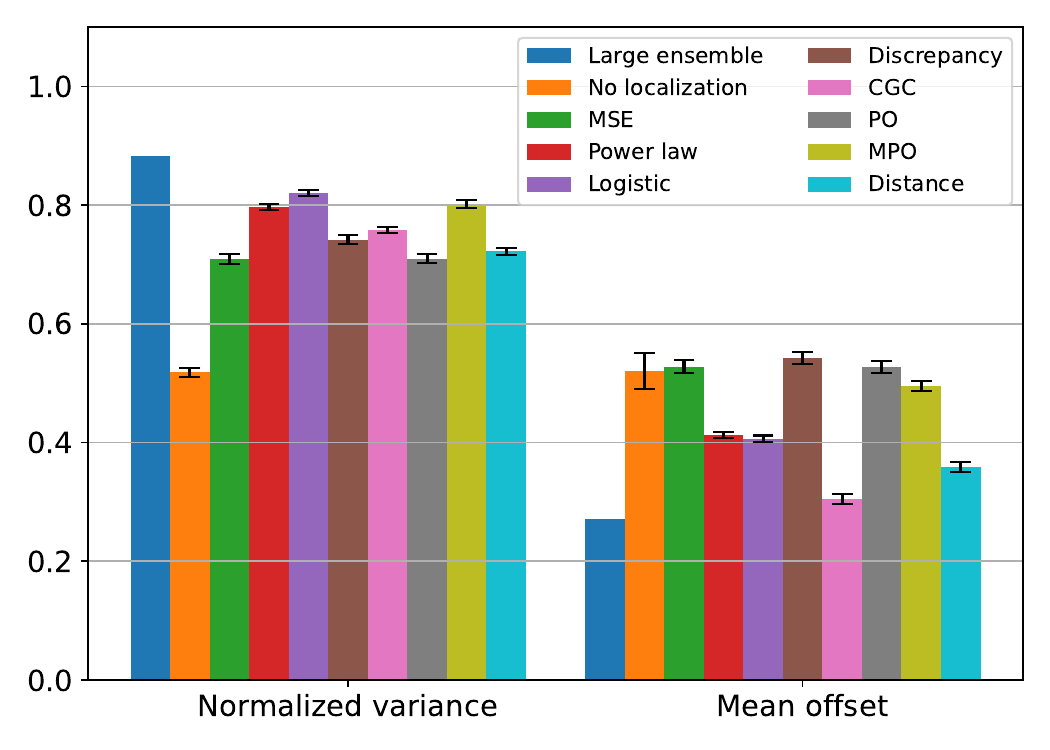}}
\caption{Average data-mismatch objective function, normalized variance and mean offset for different localization methods. Each case represents the average over 10 runs with different initial ensembles, except for the case labeled ``Large ensemble,'' which corresponds to data assimilation results obtained with an ensemble of size $N_e = 5{,}000$ without localization. The dashed blue line in the boxplot highlights an objective function of 0.5. The error lines in the bars plots correspond to the 95\% confidence intervals. Test case 2.}
\label{fig:test_cases.grid_par.obj.nv}
\end{figure}

Fig.~\ref{fig:test_cases.grid_par.obj.nv} shows the results of running 10 independent data assimilations using different initial ensembles of size $N_e = 100$. As a reference, we also include the results obtained with a large ensemble ($N_e = 5{,}000$) without localization. Fig.~\ref{fig:test_cases.grid_par.obj.nv}a shows boxplots of $\overline{\mathcal{O}_d(\m)}$, whereas Fig.~\ref{fig:test_cases.grid_par.obj.nv}b shows bar plots of the corresponding average $\textrm{NV}$ and the average offset of the mean comparing prior and posterior distributions of model parameters. 

In terms of data-match quality, all cases resulted in relatively low values of $\overline{\mathcal{O}_d(\m)}$, often below 0.5. The MSE, discrepancy-based, and PO tapers produced at least one data assimilation run with a larger objective-function value, although still with $\overline{\mathcal{O}_d(\m)} < 1$. In terms of normalized variance, the case without localization resulted in the lowest value of $\textrm{NV}$. All localization methods substantially increased $\textrm{NV}$. However, in most cases, the resulting values remained lower than those obtained with the large ensemble. The power-law, logistic, and MPO tapers produced results closer to the reference solution. Regarding the mean offset, the large ensemble produced the lowest values, indicating the smallest average change from the prior, followed by the CGC, distance-based, logistic, and power-law tapers.

\begin{figure}
\centering
    \subfloat[\scriptsize{MSE}]{\includegraphics[width=0.25\linewidth]{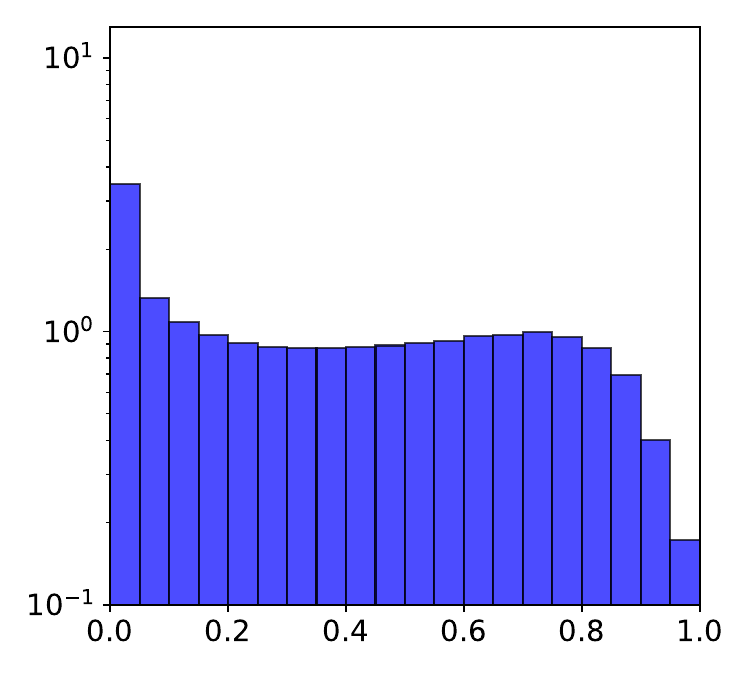}}
    \subfloat[\scriptsize{Power-law}]{\includegraphics[width=0.25\linewidth]{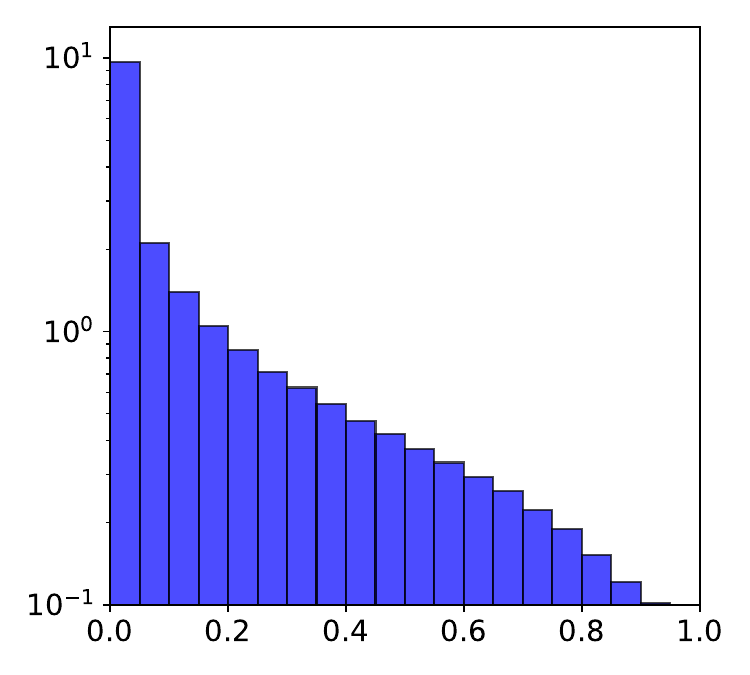}}
    \subfloat[\scriptsize{Logistic}]{\includegraphics[width=0.25\linewidth]{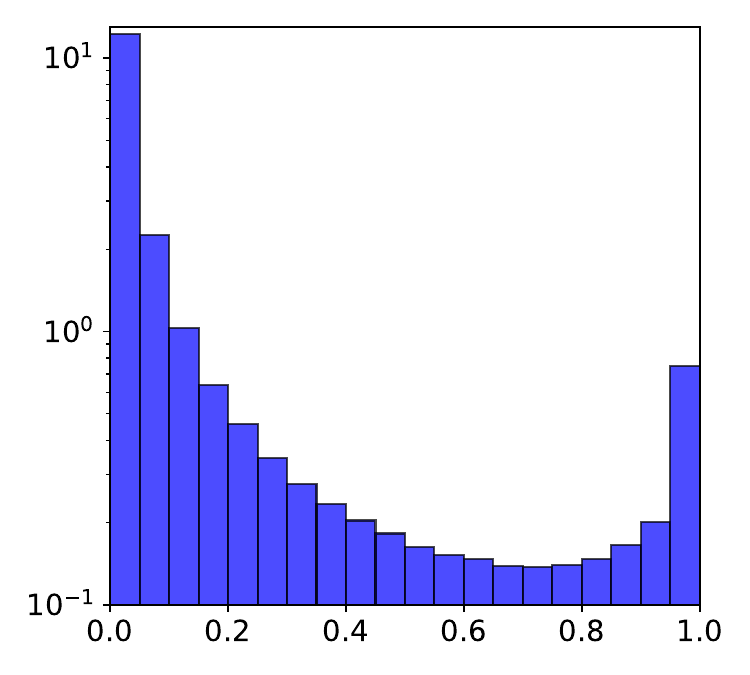}}
    \subfloat[\scriptsize{Discrepancy}]{\includegraphics[width=0.25\linewidth]{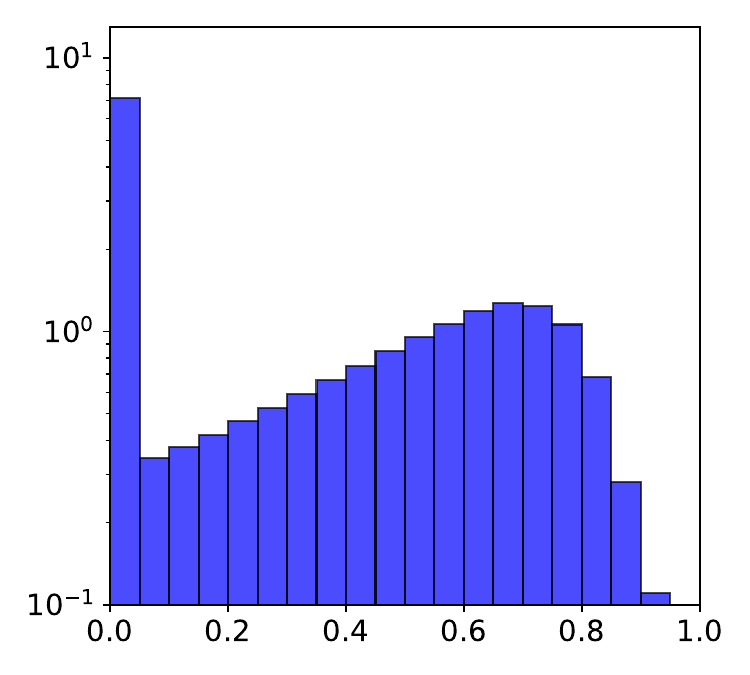}} \\
    \subfloat[\scriptsize{CGC}]{\includegraphics[width=0.25\linewidth]{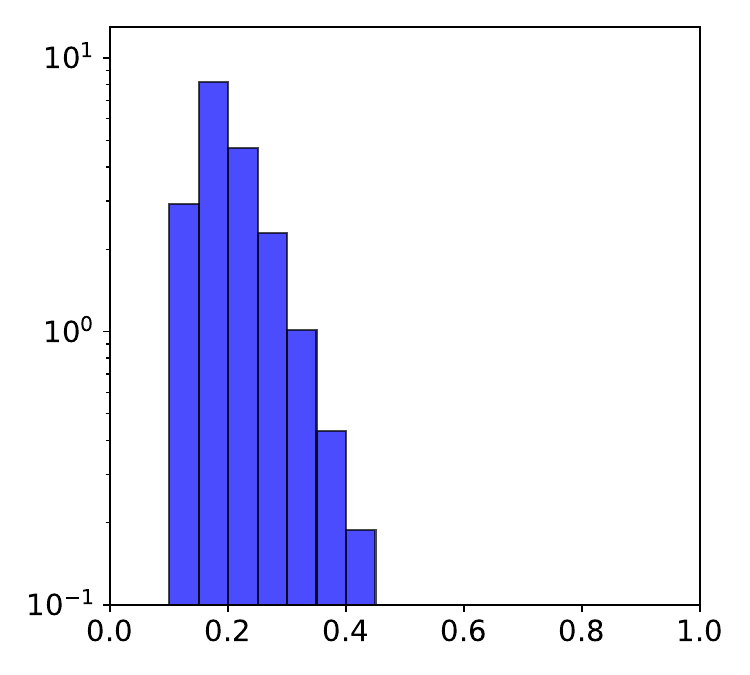}}
    \subfloat[\scriptsize{PO}]{\includegraphics[width=0.25\linewidth]{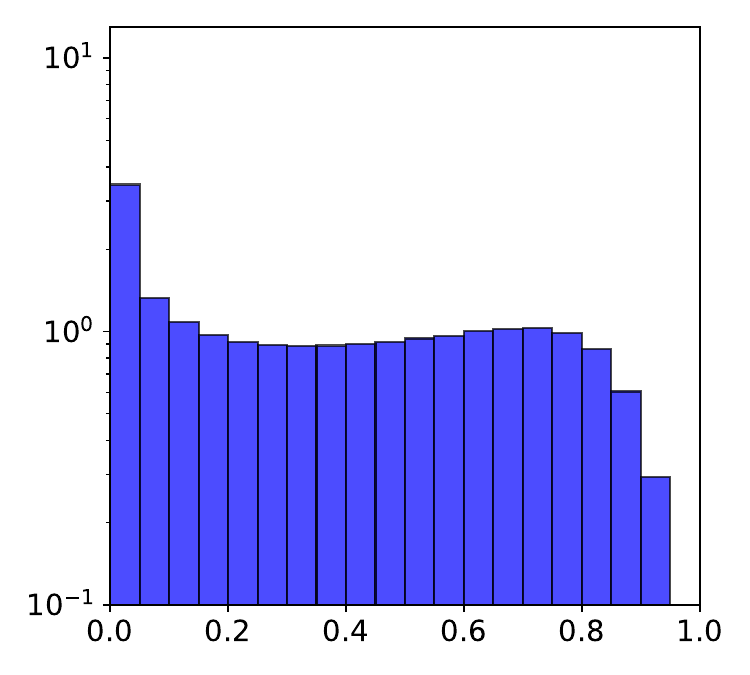}}
    \subfloat[\scriptsize{MPO}]{\includegraphics[width=0.25\linewidth]{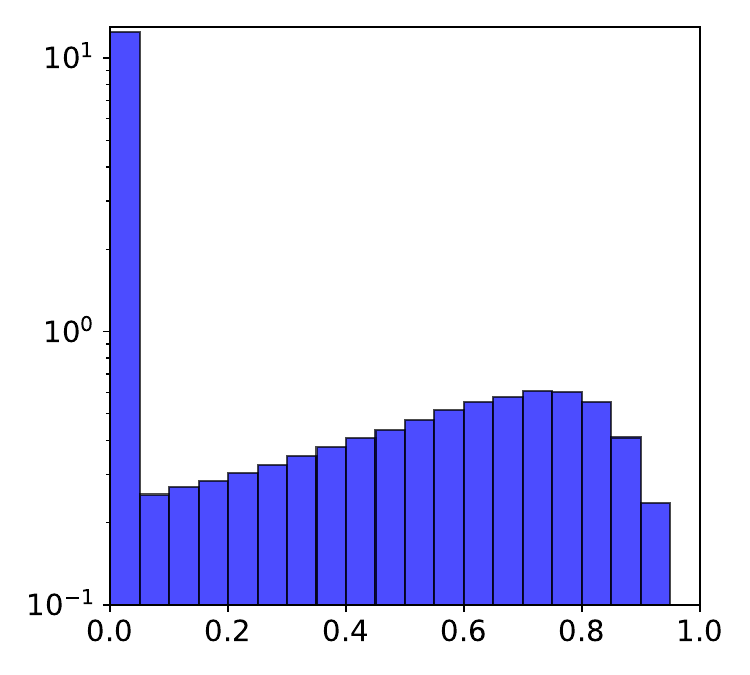}}    
    \subfloat[\scriptsize{Distance}]{\includegraphics[width=0.25\linewidth]{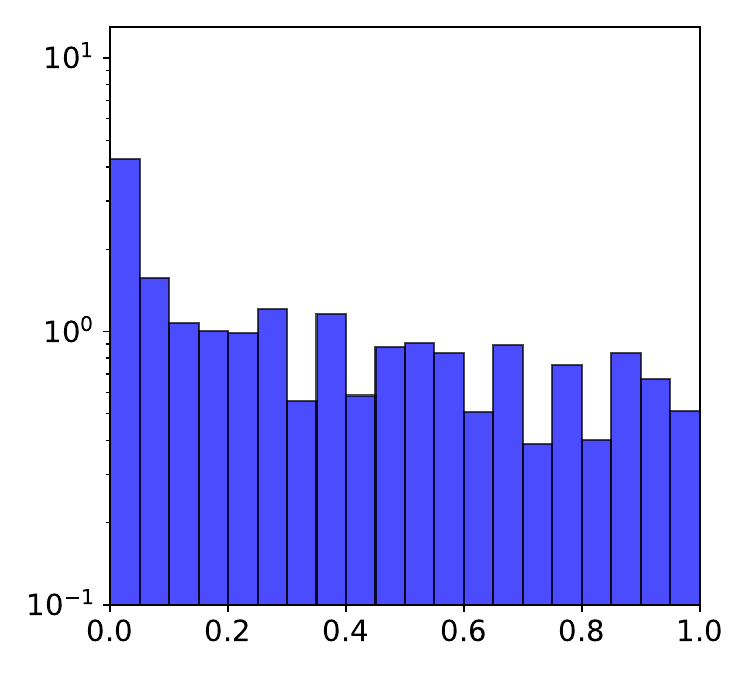}}    
\caption{Histogram of the taper values for different localization methods. Test case 2.}
\label{fig:test_cases.grid_par.histtaper}
\end{figure}

\begin{figure}
\centering
    \subfloat[\scriptsize{Bottom-hole pressure}]{\includegraphics[width=1.0\linewidth]{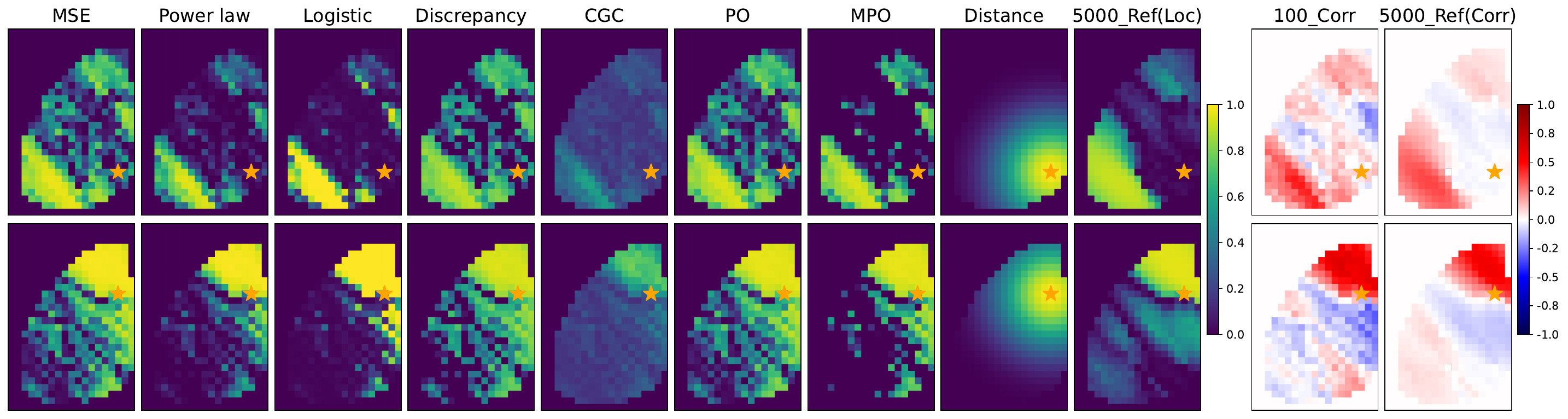}} \\
    \subfloat[\scriptsize{Gas-oil ratio}]{\includegraphics[width=1.0\linewidth]{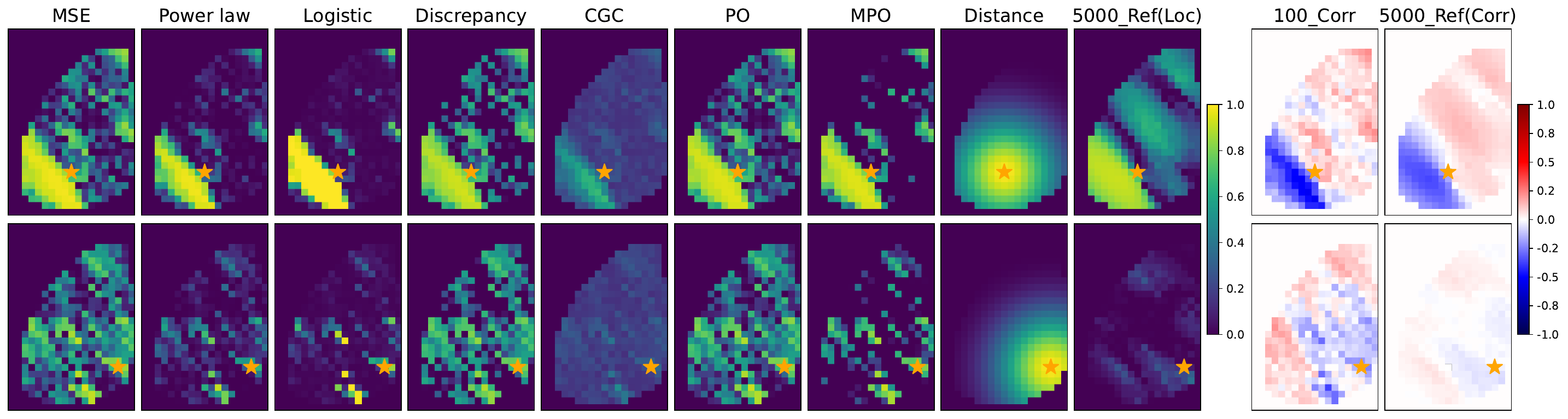}} \\
    \subfloat[\scriptsize{Water cut}]{\includegraphics[width=1.0\linewidth]{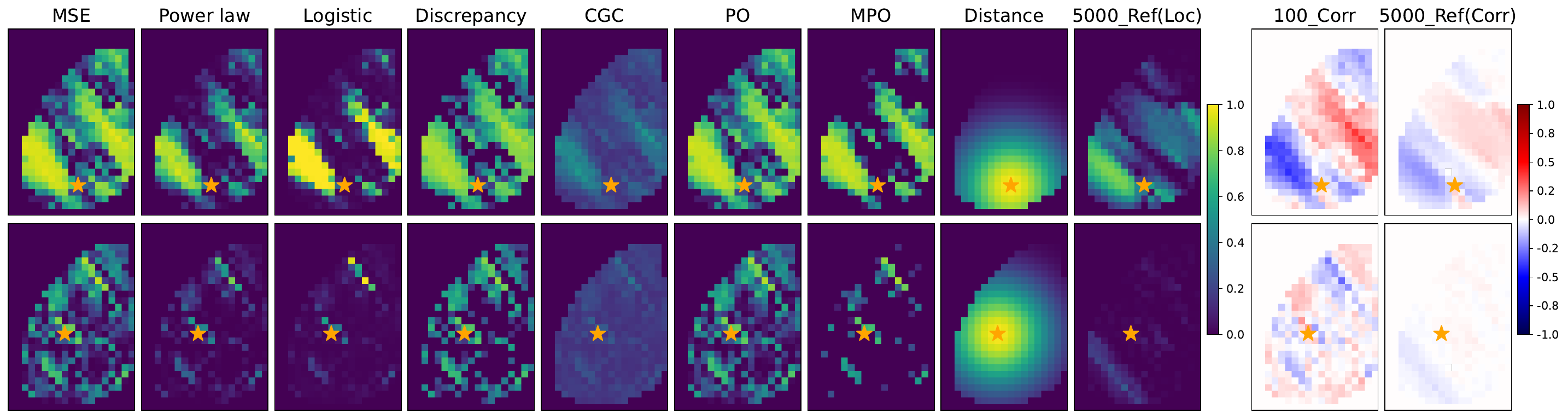}}
\caption{Localization values for horizontal log-permeability in a middle layer. Each row corresponds to an observed data from a different well (orange star). 5000\_Ref(loc) refers to localization computed using the reference ensemble with 5,000 realizations. 100\_Corr and 5000\_Ref(Corr) represent the correlation coefficients computed using ensembles of 100 and 5,000 members, respectively. Test case 2.}
\label{fig:test_cases.grid_par.tapers}
\end{figure}

Fig.~\ref{fig:test_cases.grid_par.histtaper} shows the histogram of the tapers for one data assimilation. The results are similar to Fig.~\ref{fig:test_cases.scalar_par.histtaper}; however, for the test case 2 only the logistic taper shows a concentration of values near one. Fig.~\ref{fig:test_cases.grid_par.tapers} shows localization maps for the third layer of the PUNQ-S3 model, focusing on horizontal permeability and considering data points from bottom-hole pressure (Fig.~\ref{fig:test_cases.grid_par.tapers}a), gas-oil ratio (Fig.~\ref{fig:test_cases.grid_par.tapers}b), and water cut (Fig.~\ref{fig:test_cases.grid_par.tapers}c). In all cases, the selected data correspond to the last time step of each time series. For bottom-hole pressure, this point represents a static pressure measurement during a shut-in period. For comparison, we include a reference localization map computed using the MSE taper function, where correlations were estimated from $N_e = 5{,}000$ model runs. This reference case is labeled as 5000\_Ref(loc). Each plot also includes correlation coefficient maps estimated with both 100 (labeled 100\_Corr) and 5,000 (labeled 5000\_Ref(Corr)) model runs.

The results in Fig.~\ref{fig:test_cases.grid_par.tapers} indicate that the resulting localization patterns are considerably more intricate than the circular or elliptical regions centered at the well locations that are typically assumed in distance-based localization. In several instances, the well associated with the observed data lies outside the region with the highest taper values. Although this may seem counterintuitive, it arises because the prior realizations are conditioned on porosity and permeability data at the well locations. This conditioning significantly reduces the prior variance near the wells and, consequently, influences the computed localization values. Moreover, reservoir pressure support is provided by a lateral aquifer, which makes the flow patterns more complex than those typically observed in injector-producer configurations. A key limitation of distance-based localization in this context is that it assigns small, or even zero localization coefficients to regions far from the well, even when meaningful correlations exist. This limitation is evidenced by the results labeled 5000\_Ref(Corr).

We also observe from Fig.~\ref{fig:test_cases.grid_par.tapers} that all correlation-based tapers were able to capture the regions with the highest correlations, differing mainly in the regions with lower correlation values. The CGC taper exhibits a narrower range of taper values compared with the other tapers. The power-law, logistic, and MPO tapers were able to eliminate most spurious correlations; however, in some cases, they also suppressed correlations that appear to be meaningful.

\subsection{Test Case 3: Localized Updates}
\label{sec:test_cases.local_updates}

The third test problem consists of a simple 2D reservoir model with water injection and oil production wells arranged in five-spot patterns. The objective is to estimate the porosity and log-permeability of all gridblocks. The model has $150 \times 150$ gridblocks, resulting in a total of $N_m = 45{,}000$ model parameters. The data consist of monthly measurements of water cut at 36 oil-producing wells and water rate at 25 injection wells. The total number of data points is $N_d = 6{,}226$. 

This test case was designed to assess the performance of the correlation-based tapers in a setting where model updates are expected to be spatially localized, primarily along the flow paths connecting injectors and producers. This represents a typical scenario in which distance-based localization is expected to perform well, providing a useful benchmark for comparison.

\begin{figure}
\centering
    \subfloat[\scriptsize{Objective function}]{\includegraphics[width=0.5\linewidth]{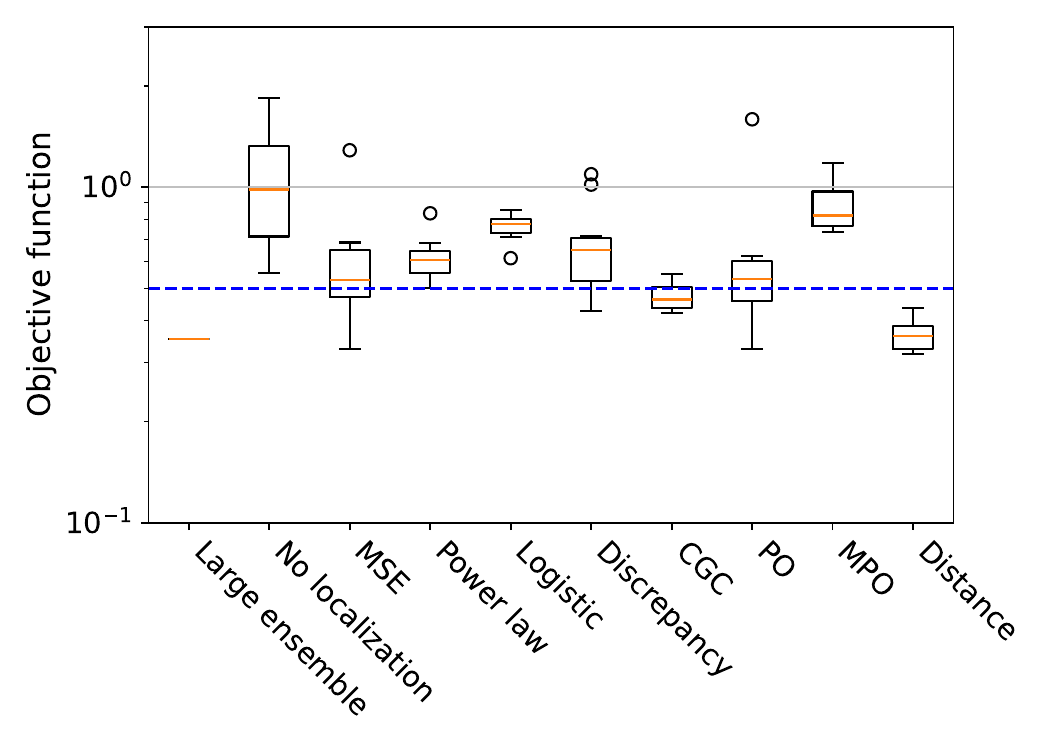}} 
    \subfloat[\scriptsize{Normalized variance}]{\includegraphics[width=0.5\linewidth]{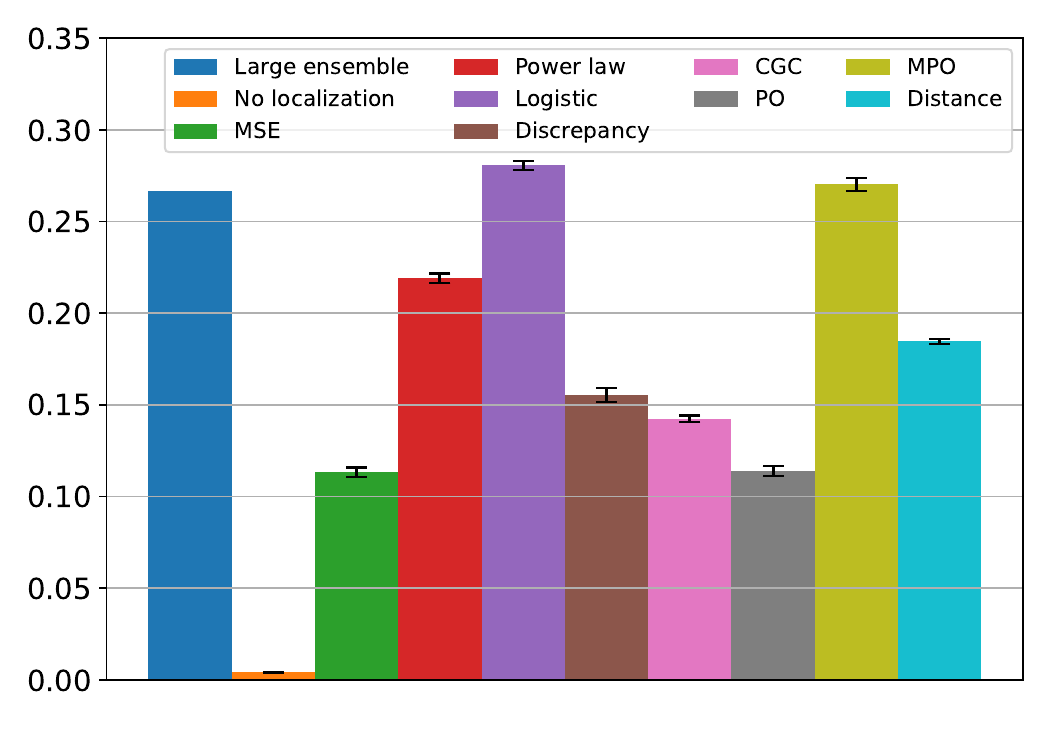}} 
\caption{Average data-mismatch objective function and normalized variance for different localization methods. Each case represents the average over 10 runs with different initial ensembles, except for the case labeled ``Large ensemble,'' which corresponds to data assimilation results obtained with an ensemble of size $N_e = 10{,}000$ without localization. The dashed blue line in the boxplot highlights an objective function of 0.5. The error lines in the bars plots correspond to the 95\% confidence intervals. Test case 3.}
\label{fig:test_cases.local_updates.of_nv}
\end{figure}

Fig.~\ref{fig:test_cases.local_updates.of_nv}a shows boxplots of $\overline{\mathcal{O}_d(\m)}$, whereas Fig.~\ref{fig:test_cases.local_updates.of_nv}b shows a bar plot of the corresponding average $\textrm{NV}$. The results were obtained from 10 independent runs using different initial ensembles of size $N_e = 200$. The data assimilation was performed with ES-MDA using $N_a = 4$ assimilation steps and $\alpha_\ell = N_a$. As a reference, we also include the results obtained with a large ensemble ($N_e = 10{,}000$) without localization.

In terms of data-match quality, most cases resulted in values of $\overline{\mathcal{O}_d(\m)}$ within the reference range $[0.5, 1]$. However, the MSE, discrepancy, PO, and MPO tapers produced at least one data assimilation run with $\overline{\mathcal{O}_d(\m)} > 1$, indicating reduced robustness with respect to the choice of the initial ensemble. As before, the large-ensemble case and the cases with distance-based localization resulted in $\overline{\mathcal{O}_d(\m)} < 0.5$.

In terms of normalized variance, the case without localization resulted in $\textrm{NV} \approx 0$, indicating near ensemble collapse. All localization methods substantially increased $\textrm{NV}$. However, in most cases, including distance-based localization, the resulting values remained lower than those obtained with the large ensemble. The exceptions were the logistic and MPO tapers, which produced values closer to the reference solution.

\begin{figure}
\centering
    \subfloat[\scriptsize{MSE with large ensemble}]{\includegraphics[width=0.3\linewidth]{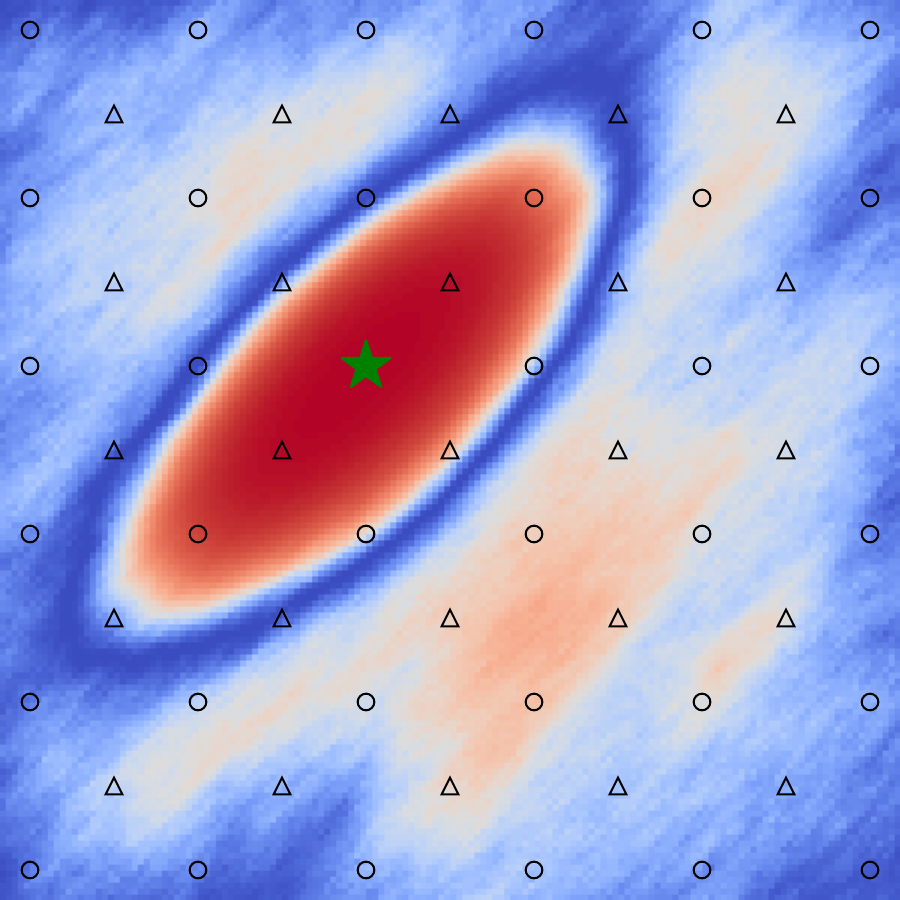}} \hspace{2mm}
    \subfloat[\scriptsize{Distance}]{\includegraphics[width=0.3\linewidth]{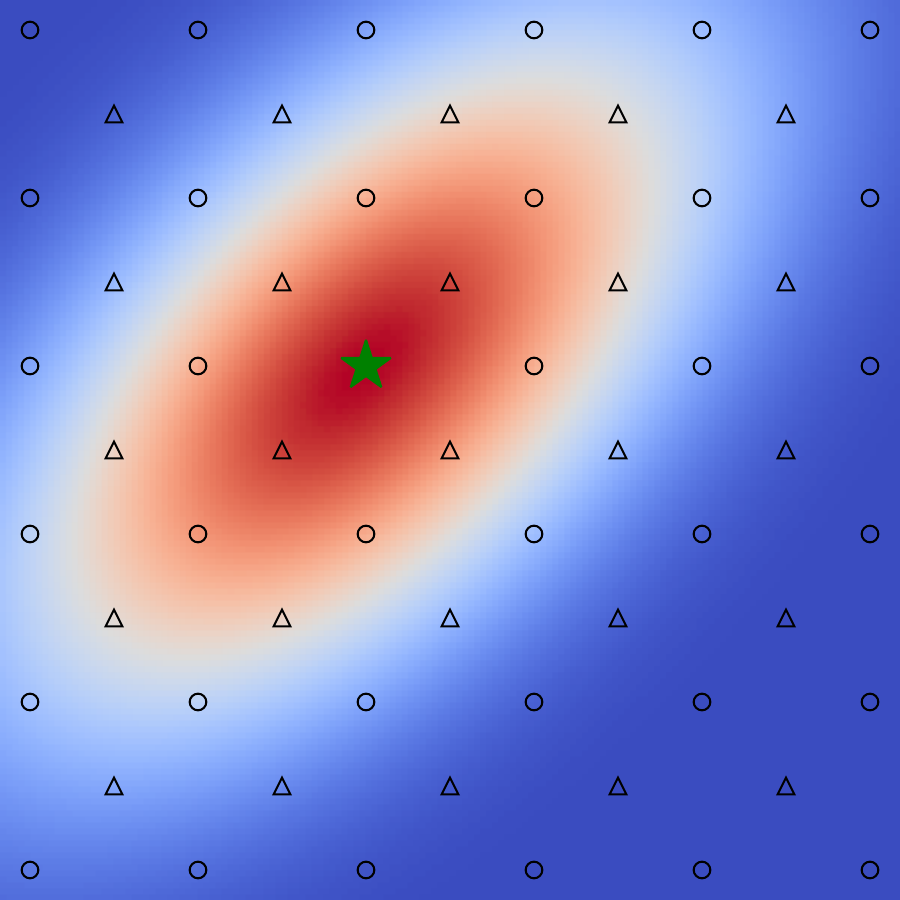}} \hspace{2mm}
    \subfloat[\scriptsize{MSE}]{\includegraphics[width=0.3\linewidth]{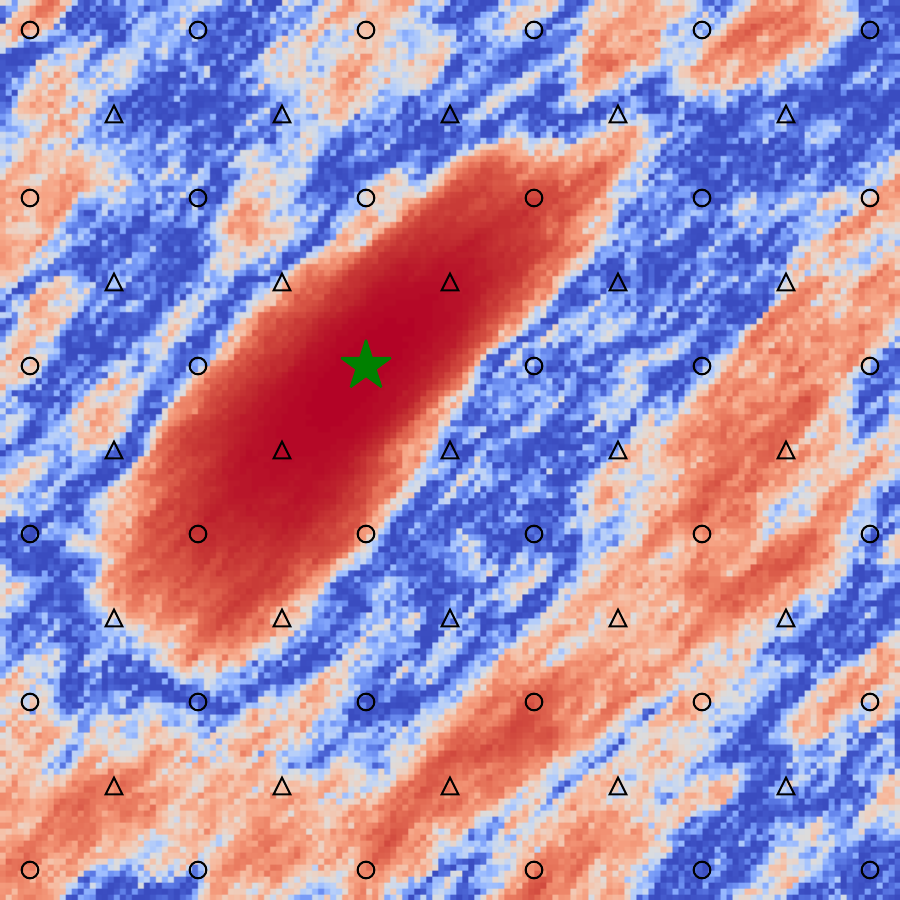}} \\
    \subfloat[\scriptsize{Power-law ($\beta = 3, t_0 = 2$)}]{\includegraphics[width=0.3\linewidth]{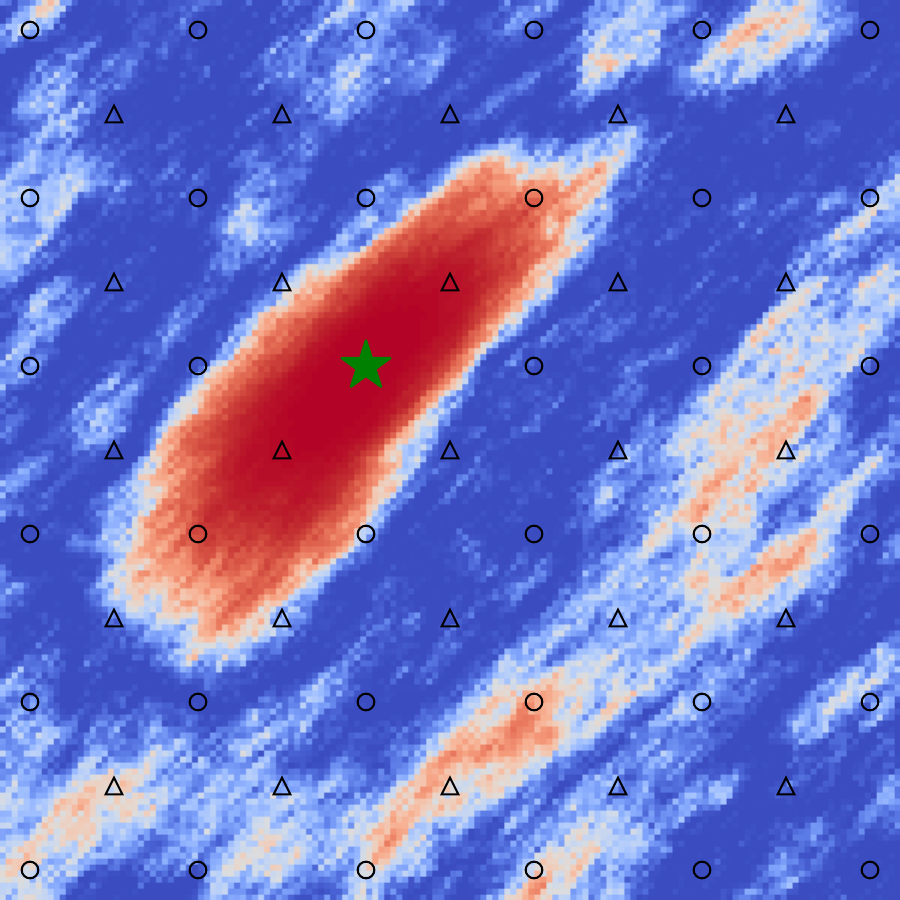}} \hspace{2mm}
    \subfloat[\scriptsize{Logistic ($\gamma = 1.5, t_0 = 2$)}]{\includegraphics[width=0.3\linewidth]{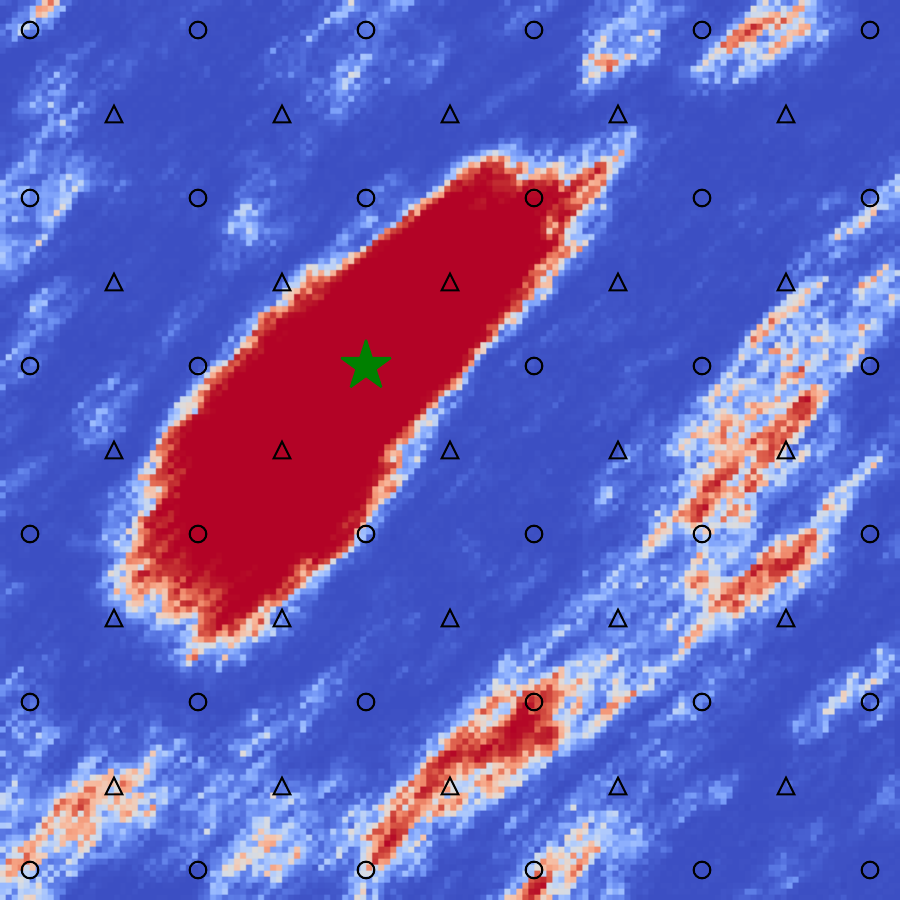}} \hspace{2mm}
    \subfloat[\scriptsize{Discrepancy ($\eta = 0.5$)}]{\includegraphics[width=0.3\linewidth]{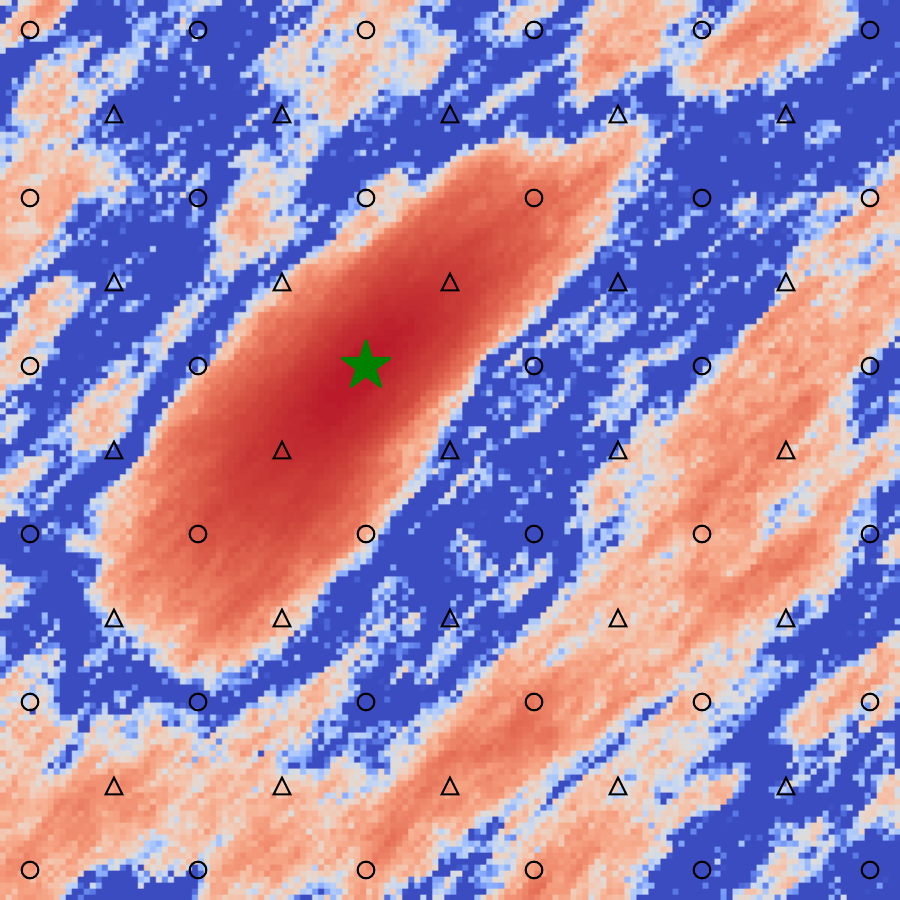}} \\
    \subfloat[\scriptsize{CGC ($\theta = \sigma$)}]{\includegraphics[width=0.3\linewidth]{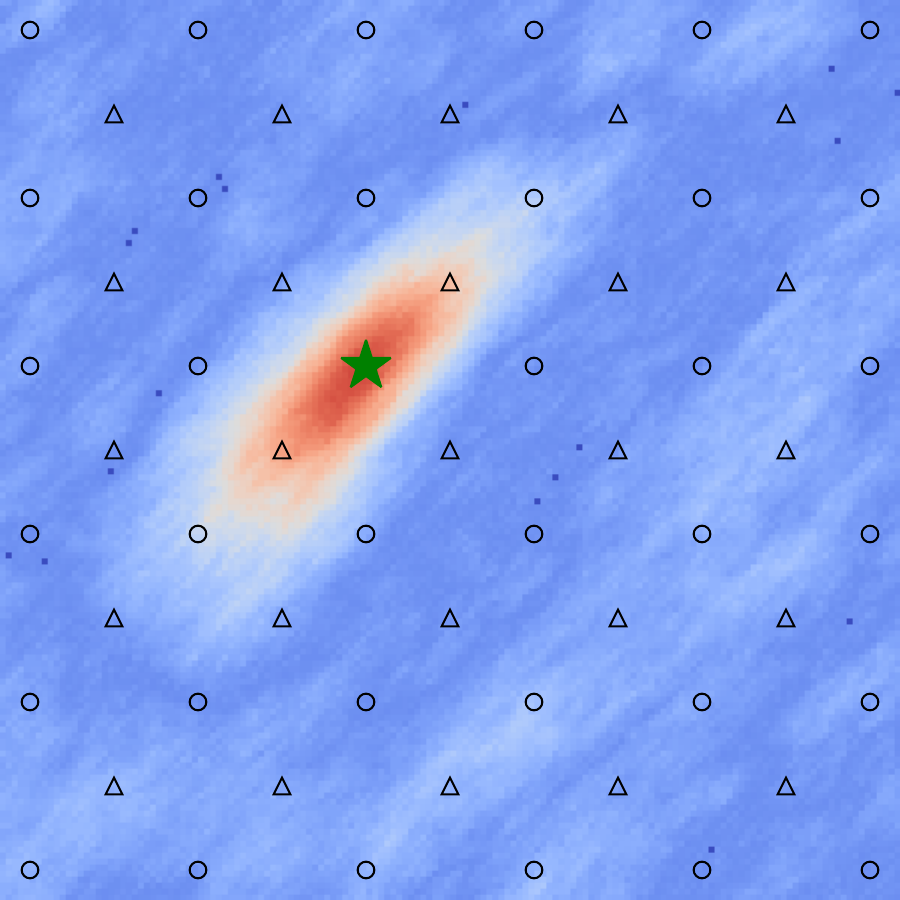}} \hspace{2mm}
    \subfloat[\scriptsize{PO}]{\includegraphics[width=0.3\linewidth]{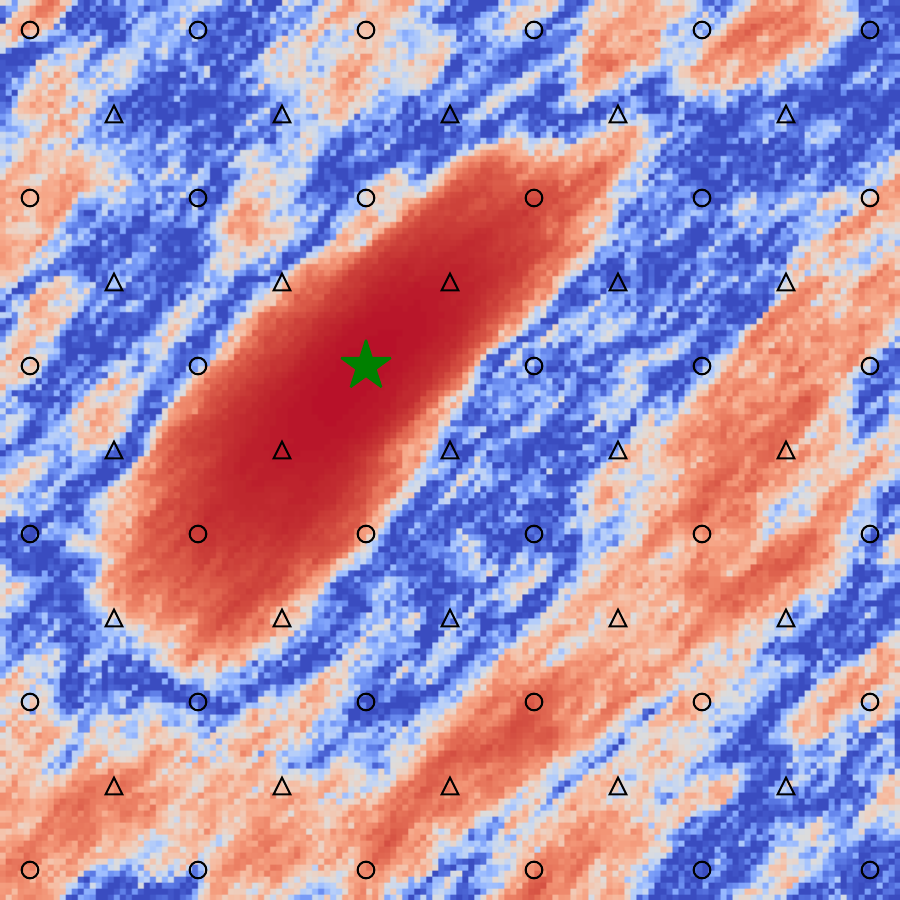}} \hspace{2mm}
    \subfloat[\scriptsize{MPO}]{\includegraphics[width=0.3\linewidth]{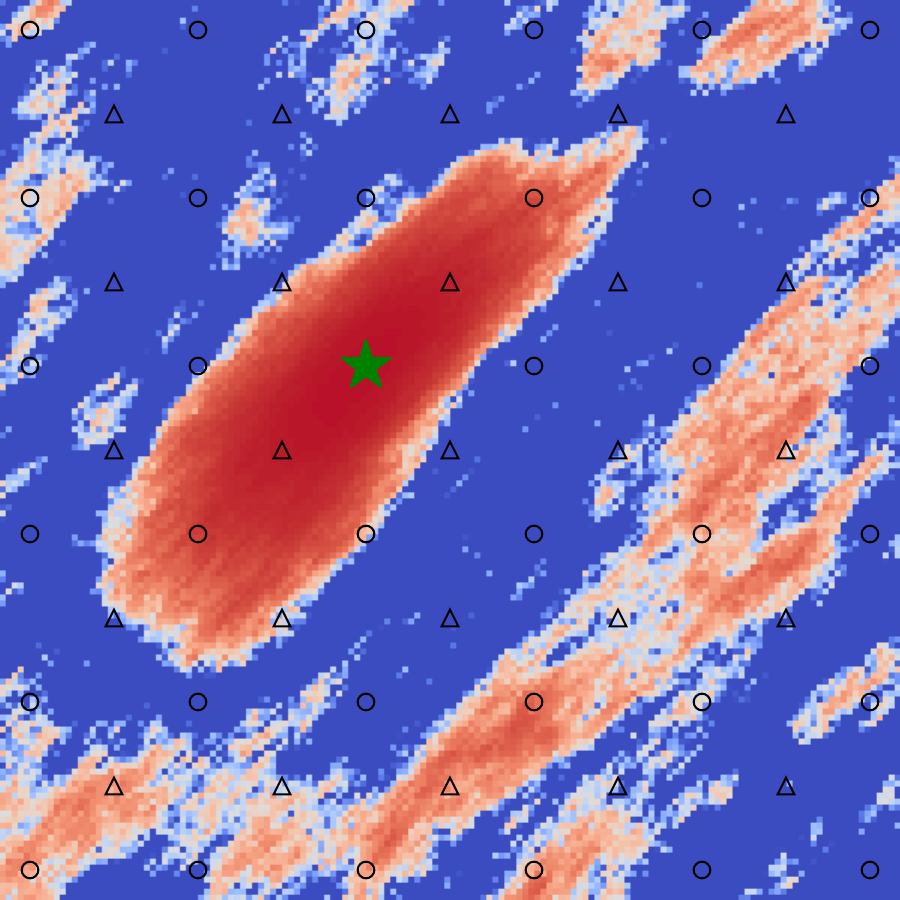}} \\
    {\includegraphics[width=0.7\linewidth]{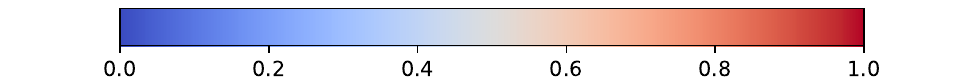}}
\caption{Taper values computed for the water-cut data of well 21 (green star) at the final time step. Test case 3.}
\label{fig:test_cases.local_updates.tapers}
\end{figure}

Fig.~\ref{fig:test_cases.local_updates.tapers} shows images of the taper generated for the water-cut data with respect to log-permeability using each localization method for the first of the 10 ensembles used to generate the results of Fig.~\ref{fig:test_cases.local_updates.of_nv}. To serve as reference, Fig.~\ref{fig:test_cases.local_updates.tapers} includes the taper values obtained with the MSE taper using correlations estimated from the ensemble with $N_e = 10{,}000$. For distance-based localization, we used an anisotropic version of the Gaspari-Cohn correlation function \citep[Chap.~7]{emerick:25bk}, with critical lengths corresponding to the size of 90 and 45 gridblocks along the principal directions oriented at 45$^\circ$, matching the anisotropy direction used to generate the permeability realizations. This configuration was selected based on the recommendations of \citet{chen:10b} and \citet{emerick:11e}, which suggest that the localization critical length should account for both the prior spatial correlation structure and the region of sensitivity associated with the data.

The results in Fig.~\ref{fig:test_cases.local_updates.tapers} show that all methods produce high taper values in the regions surrounding the data-generating well, consistent with the behavior observed for the MSE taper using the large ensemble. It is interesting to note that the large-ensemble MSE taper still presents nonzero values far from the well location. This likely indicates that even an ensemble with $N_e = 10{,}000$ is not sufficient to completely eliminate spurious correlations.

Overall, the tapers exhibit different levels of effectiveness in suppressing long-distance correlations. The MSE, PO, and discrepancy tapers produce several distant regions with relatively large taper values, suggesting that these approaches are overly permissive and fail to adequately filter out long-range spurious correlations. Compared to the PO taper, the modified version (MPO) is slightly more restrictive due to the embedded truncation level of $1/\sqrt{N_e}$. On the other hand, the CGC taper results in a relatively small region with large taper values around the well. Moreover, this taper does not approach zero far from the well location, since the argument of the Gaspari-Cohn function converges to $1/(1-\theta)$ when $\widetilde{\rho}=0$. This behavior is consistent with the nearly linear taper profile illustrated in Fig.~\ref{fig:cbt.summary.tapers}. The power-law and logistic tapers resulted in a stronger suppression of spurious correlations, with the logistic taper exhibiting a sharper transition between near-zero and near-one taper values.

\begin{figure}
\centering
    \subfloat[\scriptsize{Large ensemble (no local.)}]{\includegraphics[width=0.3\linewidth]{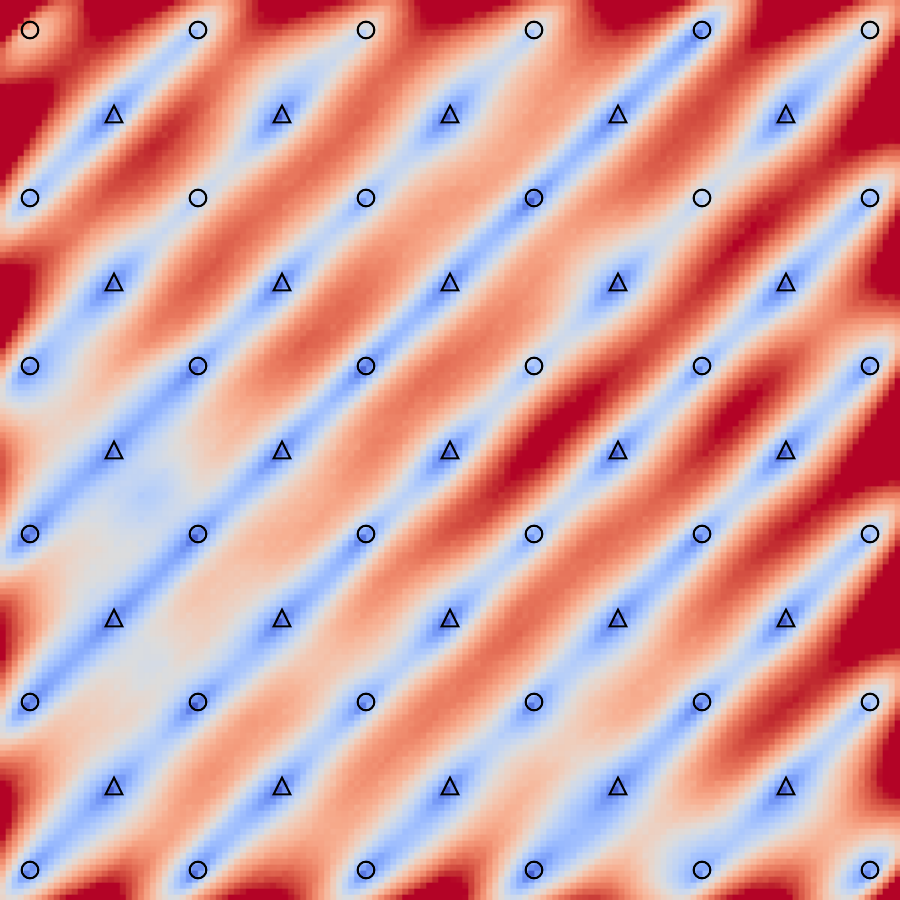}} \hspace{2mm}
    \subfloat[\scriptsize{Distance}]{\includegraphics[width=0.3\linewidth]{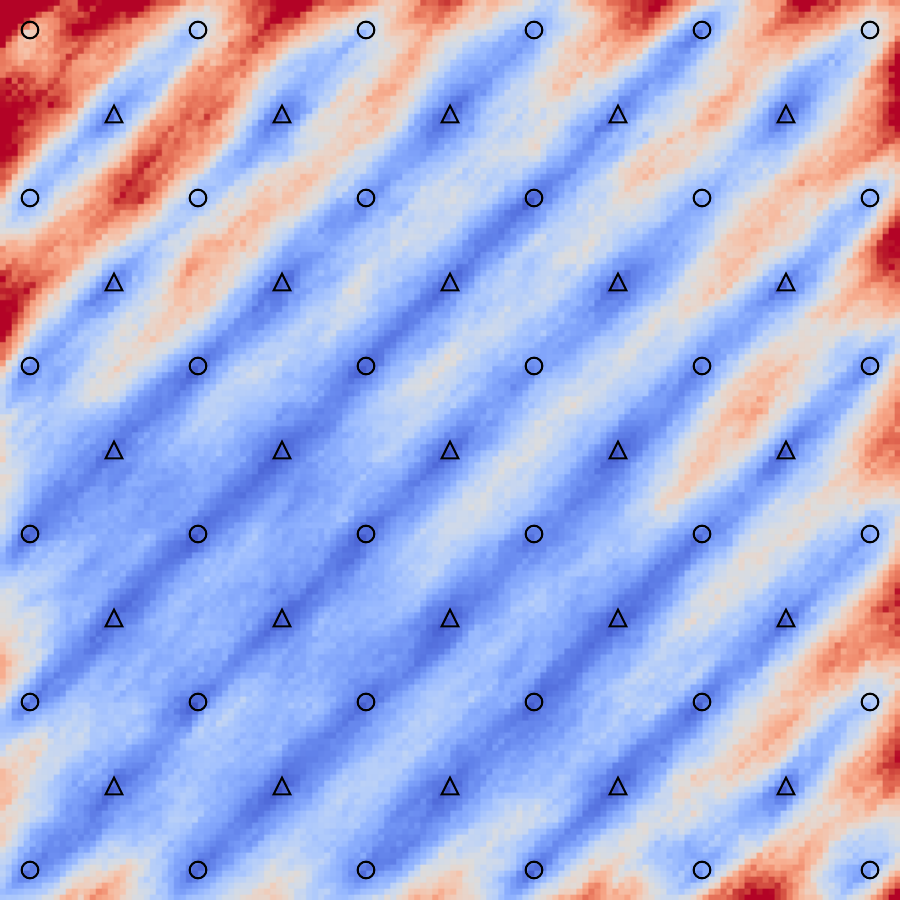}} \hspace{2mm}
    \subfloat[\scriptsize{MSE}]{\includegraphics[width=0.3\linewidth]{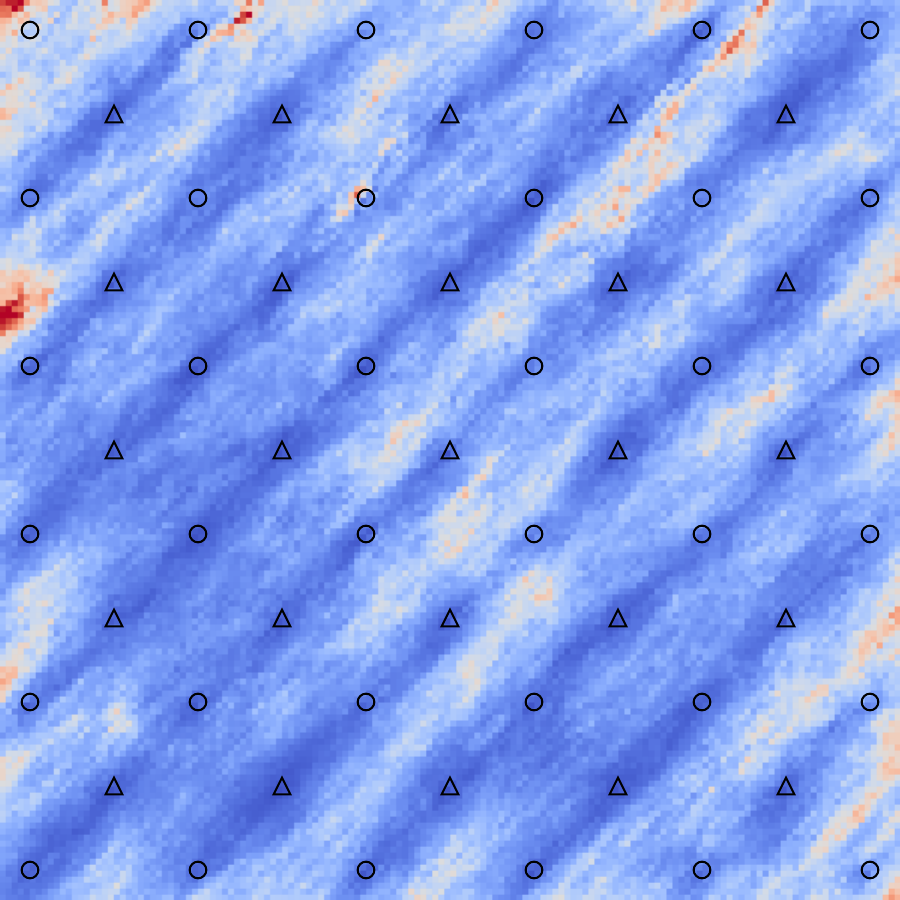}} \\
    \subfloat[\scriptsize{Power-law ($\beta = 3, t_0 = 2$)}]{\includegraphics[width=0.3\linewidth]{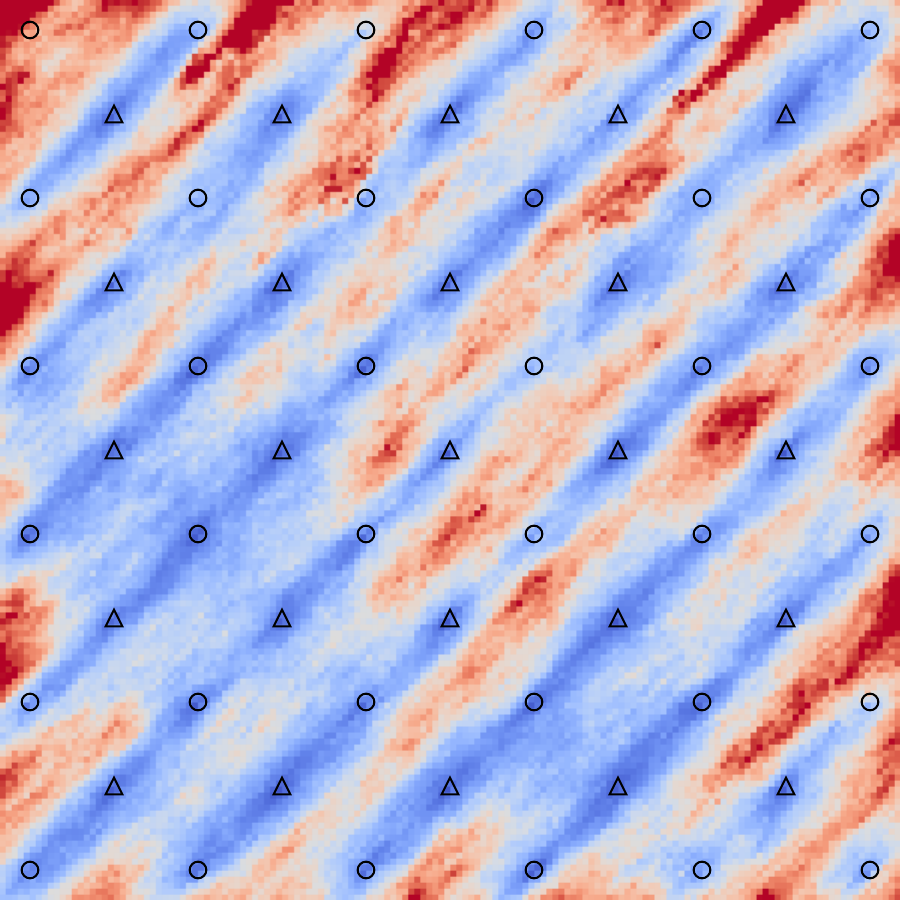}} \hspace{2mm}
    \subfloat[\scriptsize{Logistic ($\gamma = 1.5, t_0 = 2$)}]{\includegraphics[width=0.3\linewidth]{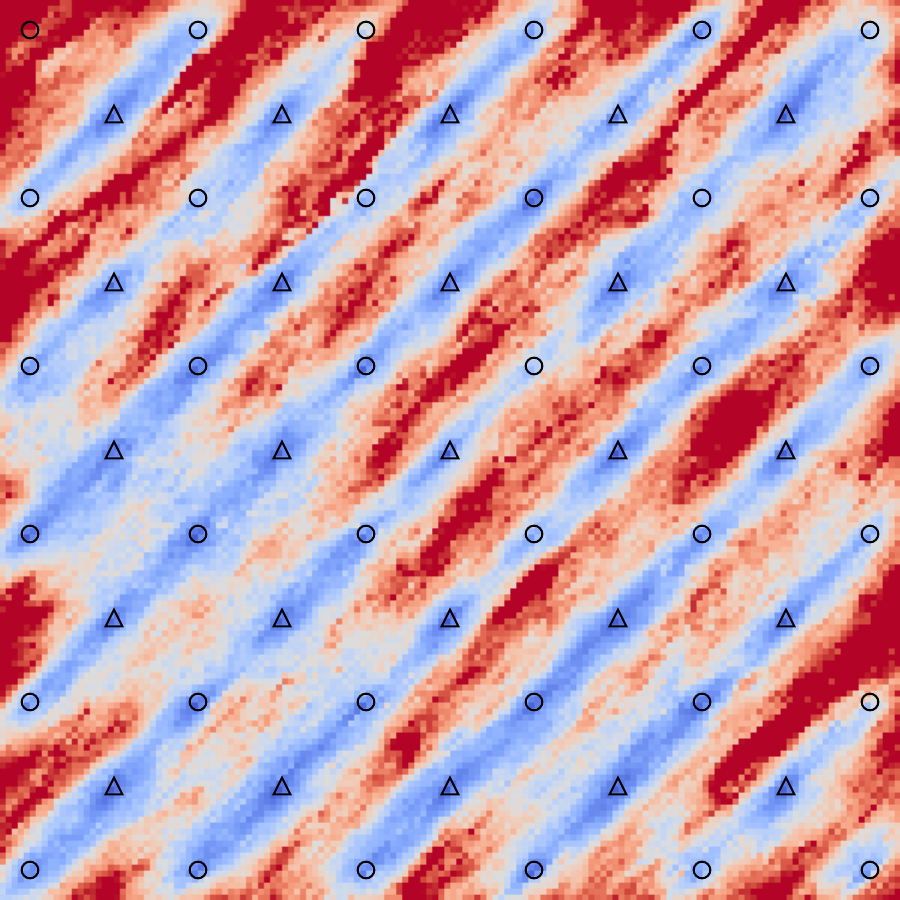}} \hspace{2mm}
    \subfloat[\scriptsize{Discrepancy ($\eta = 0.5$)}]{\includegraphics[width=0.3\linewidth]{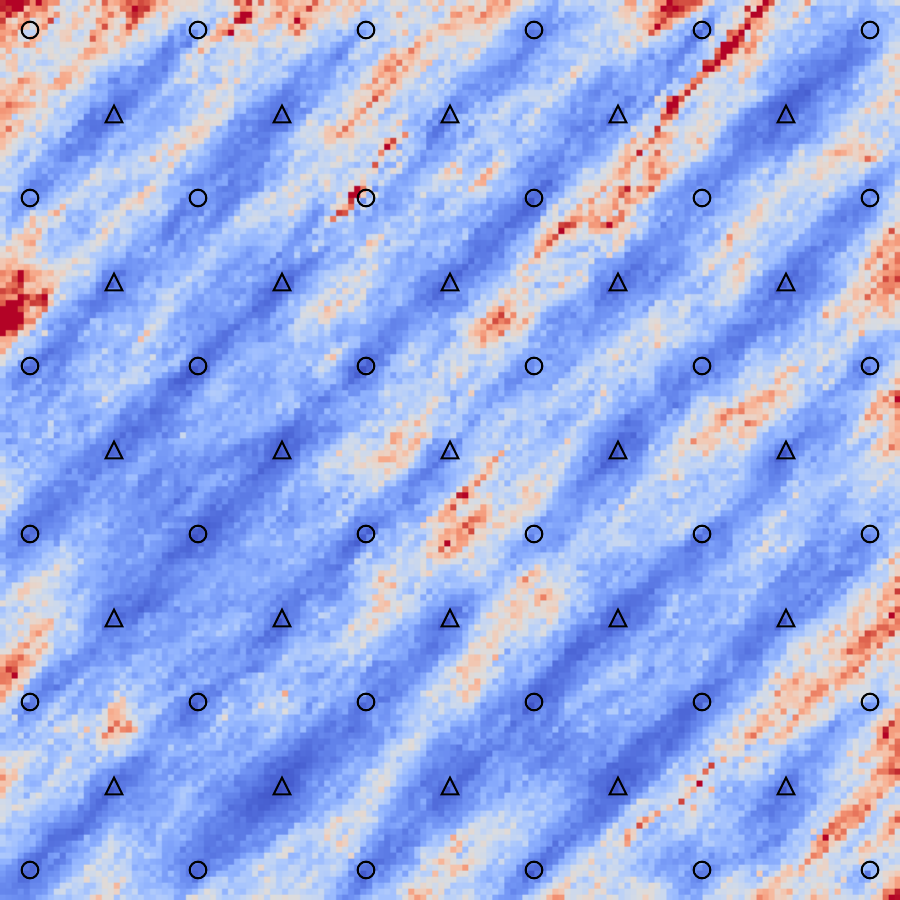}} \\
    \subfloat[\scriptsize{CGC ($\theta = \sigma$)}]{\includegraphics[width=0.3\linewidth]{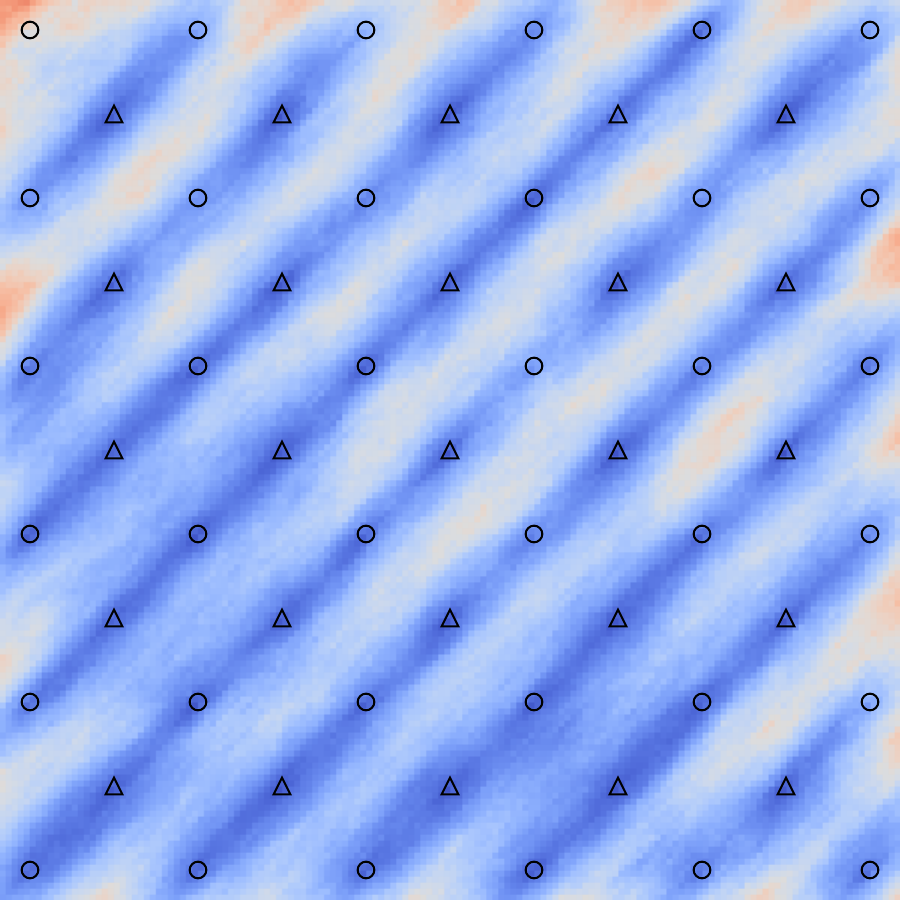}} \hspace{2mm}
    \subfloat[\scriptsize{PO}]{\includegraphics[width=0.3\linewidth]{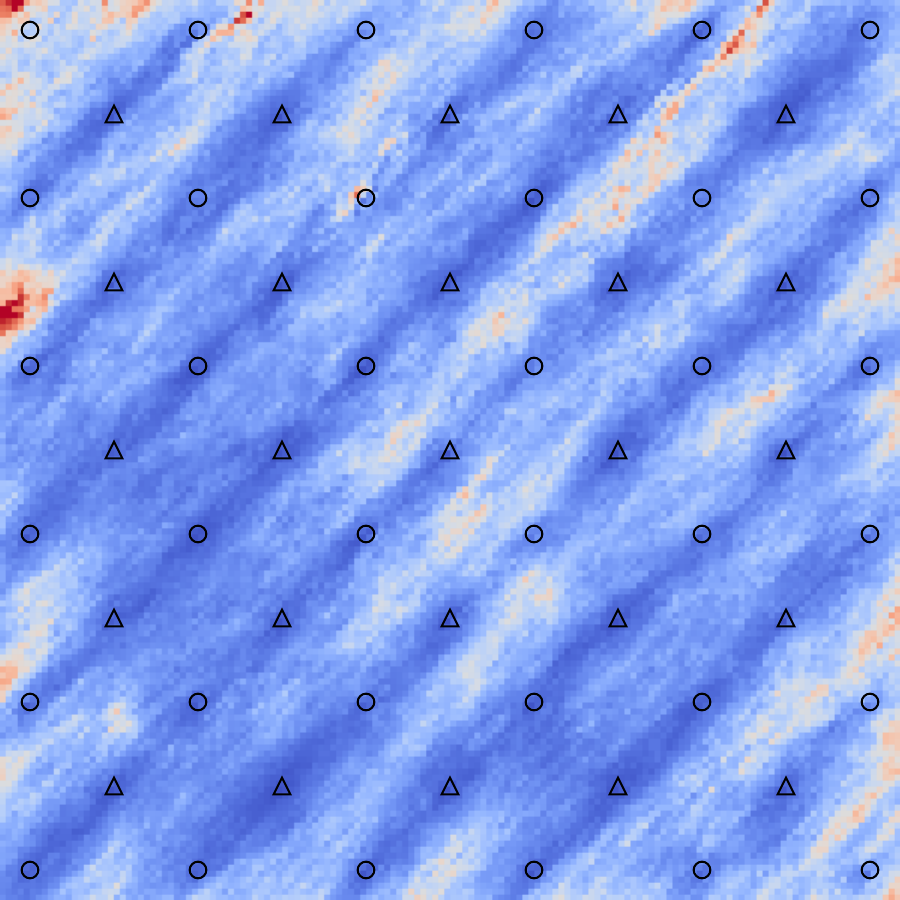}} \hspace{2mm}
    \subfloat[\scriptsize{MPO}]{\includegraphics[width=0.3\linewidth]{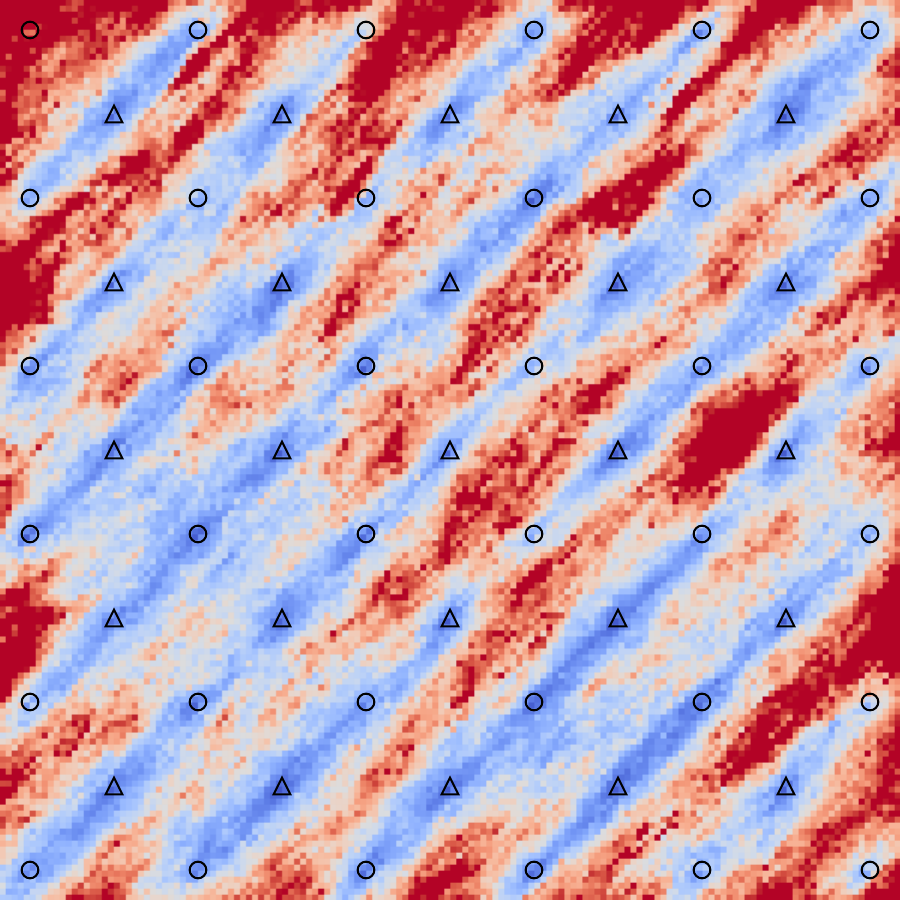}} \\
    {\includegraphics[width=0.7\linewidth]{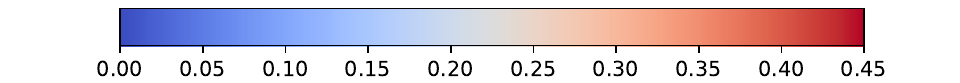}}
\caption{Normalized variance of log-permeability after data assimilation with different tapers. Test case 3.}
\label{fig:test_cases.local_updates.nv}
\end{figure}

Fig.~\ref{fig:test_cases.local_updates.nv} shows the resulting normalized variance fields for log-permeability. Most data assimilation cases underestimate the posterior variance when compared to the large-ensemble reference case. Among the cases presented, the logistic and MPO tapers produced normalized variances closer to the reference solution, indicating a better preservation of ensemble variability. It is also interesting to note that distance-based localization resulted in relevant underestimation of the posterior variance.

\begin{figure}
\centering
    \subfloat[\scriptsize{Objective function}]{\includegraphics[width=0.45\linewidth]{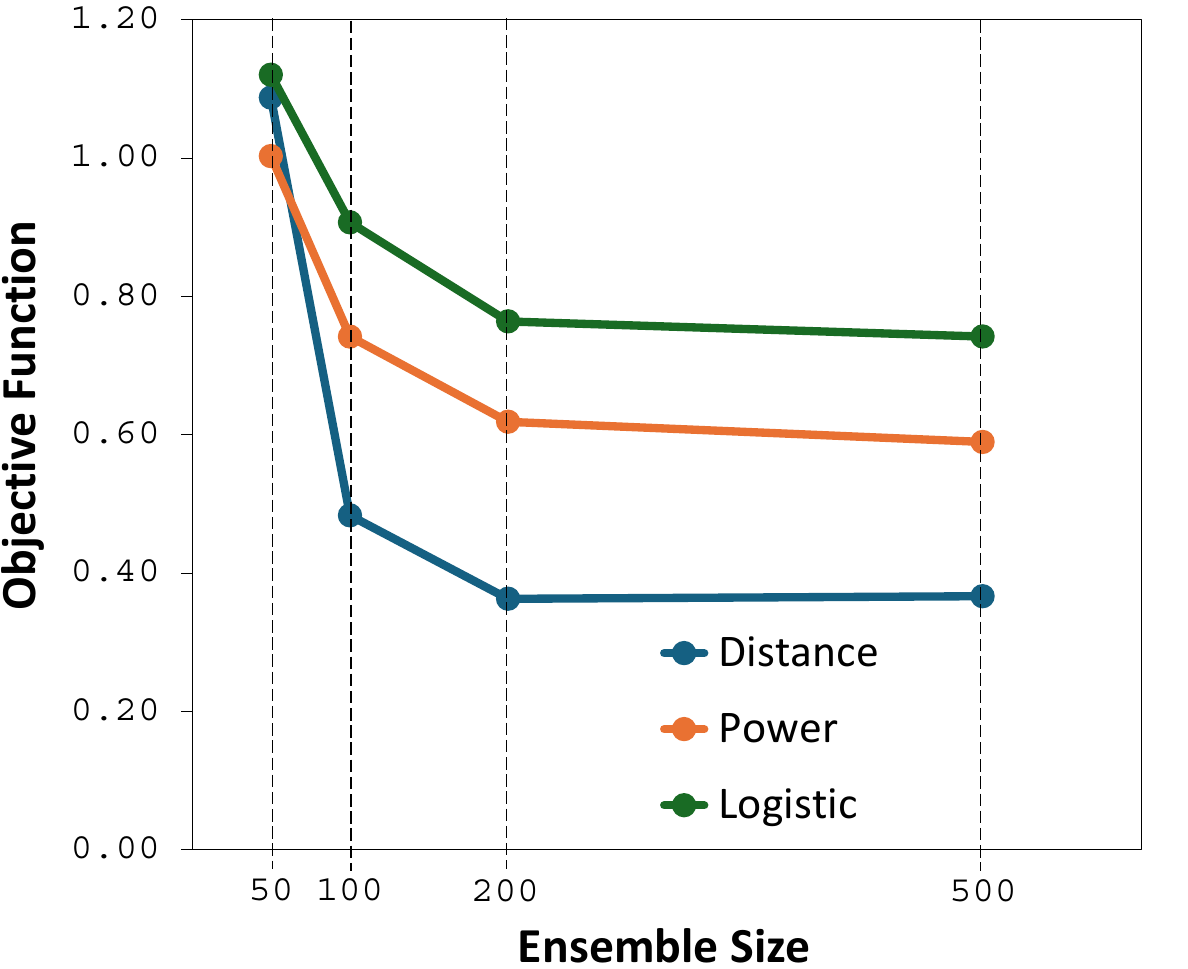}} 
    \subfloat[\scriptsize{Normalized variance}]{\includegraphics[width=0.45\linewidth]{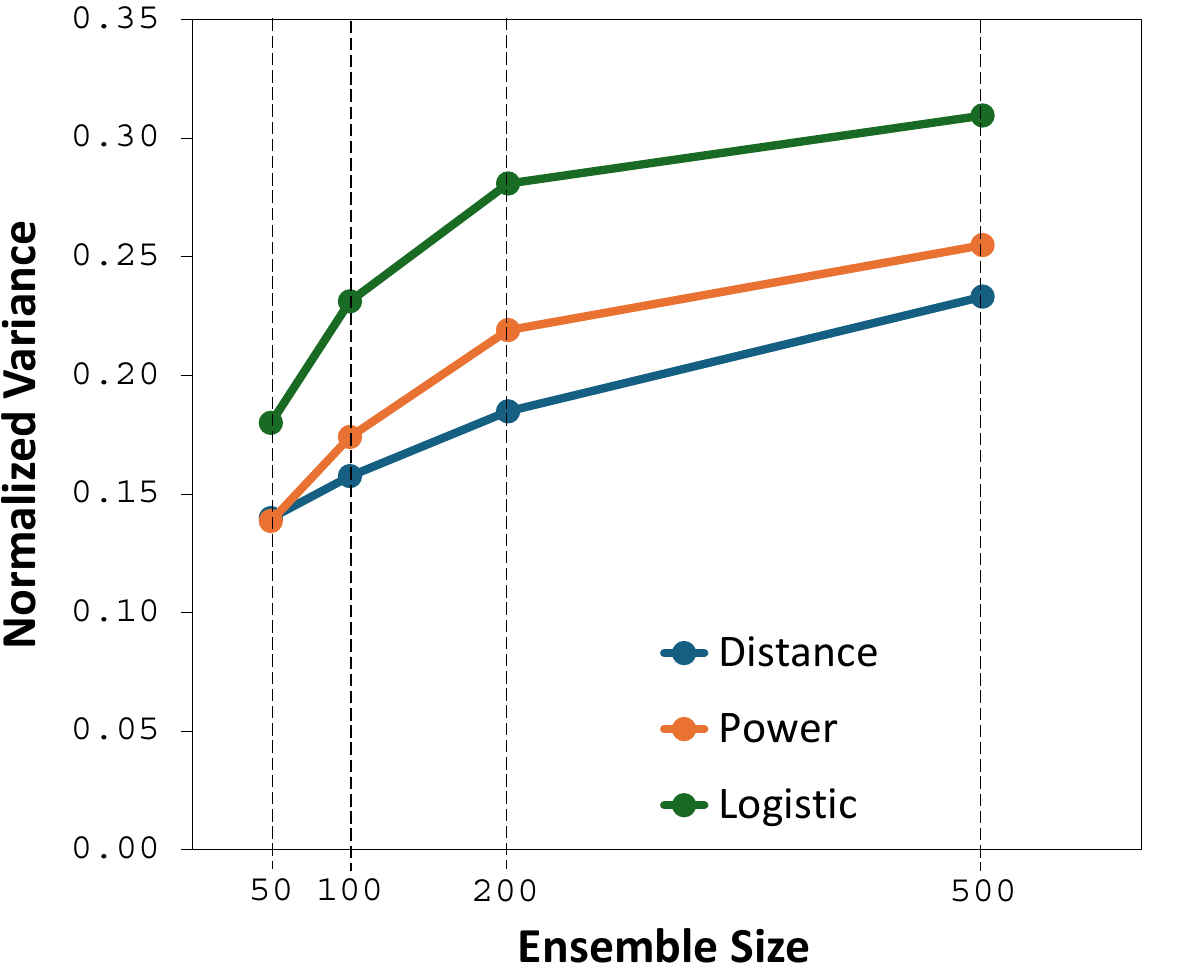}} 
\caption{Average data-mismatch objective function and normalized variance as functions of ensemble size for the different localization methods. Test case 3.}
\label{fig:test_cases.local_updates.Ne}
\end{figure}

Fig.~\ref{fig:test_cases.local_updates.Ne} shows the average objective function and normalized variance as functions of ensemble size for the cases with distance-based localization, the power-law taper, and the logistic taper. The results show that the same general trend persists across different ensemble sizes: distance-based localization resulted in $\overline{\mathcal{O}_d(\m)} \leq 0.5$, but at the cost of lower $\textrm{NV}$. The logistic taper produced the highest values of $\overline{\mathcal{O}_d(\m)}$ for all ensemble sizes, but also led to higher values of $\textrm{NV}$, indicating better preservation of ensemble variability. The results obtained with the power-law taper lie between these two cases.

\subsection{Test Case 4: Large Number of Model Parameters}
\label{sec:test_cases.large_model}

A potential limitation of correlation-based tapers is related to the dimensionality of the problem, both in terms of the number of model parameters and the number of data points. Two related issues may arise. First, as the problem dimension increases, the quality of the estimated correlations deteriorates due to sampling noise. As a consequence, we should expect the presence of more long-range spurious correlations. Second, as the model size (i.e., the number of model gridblocks) increases, the magnitude of the true correlations between model parameters and data points typically decrease, since the contribution of each individual gridblock becomes smaller. Because small correlations are inherently more difficult to estimate with finite ensembles, we expect the ability of correlation-based tapers to discriminate between meaningful and spurious correlations to be reduced.

\begin{figure}[ht!]
\centering
    \subfloat[\scriptsize{1 layer ($N_m = 45{,}000$)}]{\includegraphics[width=0.32\linewidth]{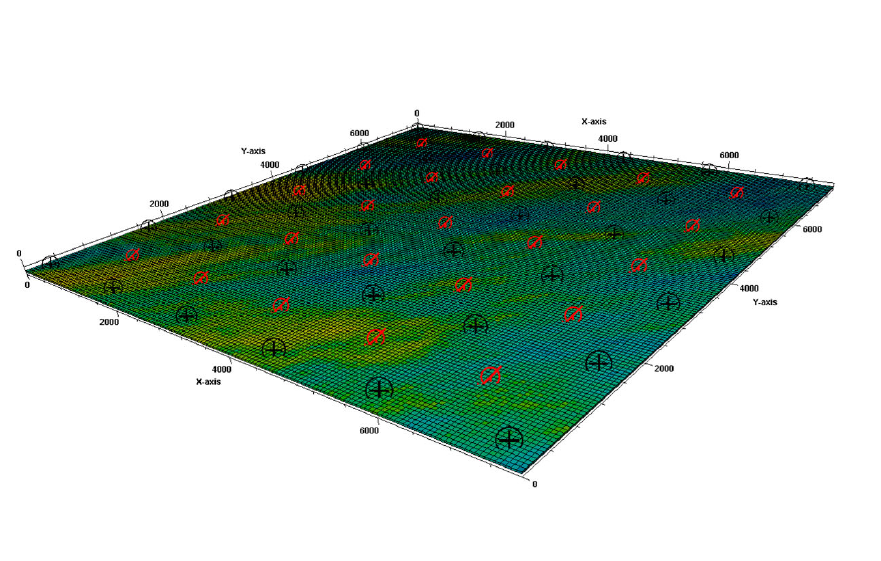}}
    \subfloat[\scriptsize{10 layers ($N_m = 675{,}000$)}]{\includegraphics[width=0.32\linewidth]{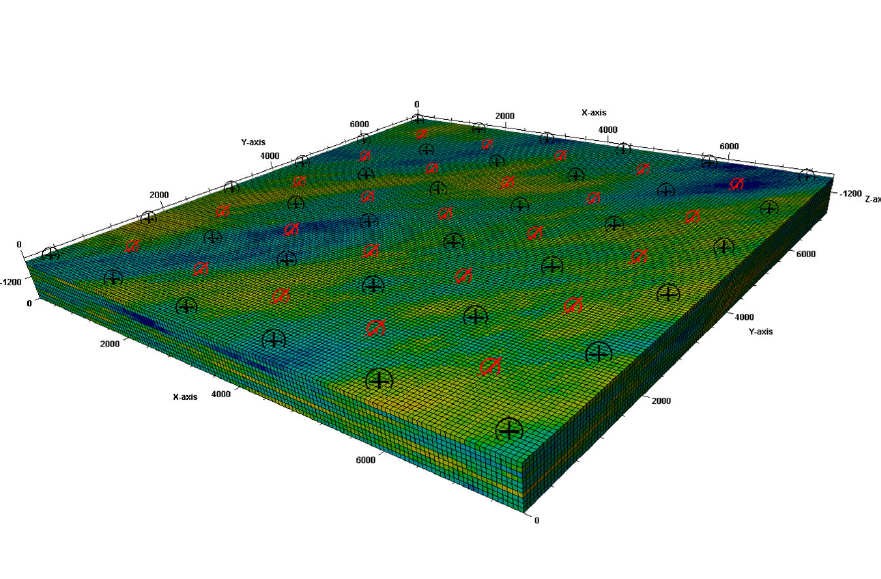}}
    \subfloat[\scriptsize{50 layers ($N_m = 3{,}375{,}000$)}]{\includegraphics[width=0.32\linewidth]{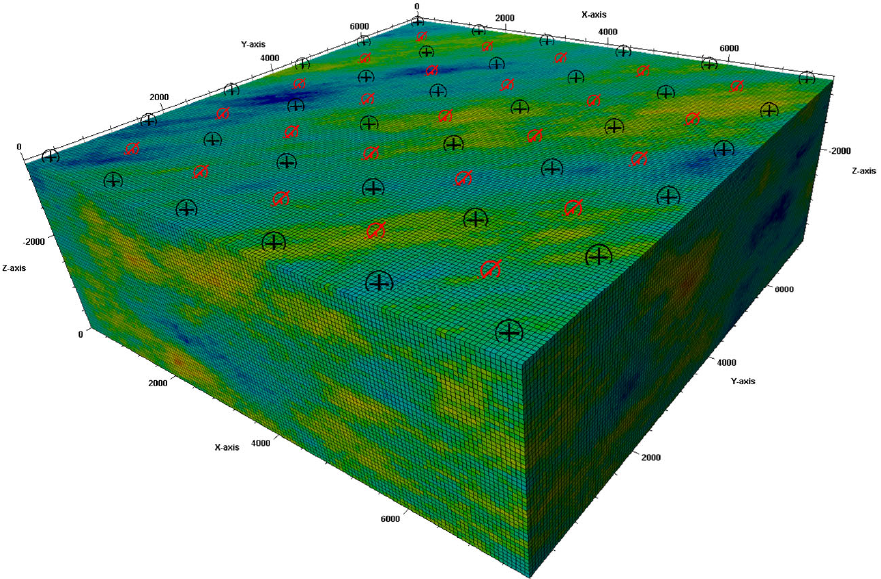}} 
\caption{Reservoir models used in the test case 4.}
\label{fig:test_cases.large_model.3D}
\end{figure}

These observations suggest that the dimensions $N_m$ and $N_d$, together with the ensemble size $N_e$, should be explicitly considered when selecting the  parameter $t_0$. To investigate this effect, we increased the size of the model in test case~3 by considering configurations with 10 and 50 layers (Fig.~\ref{fig:test_cases.large_model.3D}). In these cases, due to the presence of multiple layers, vertical permeability was also included as uncertain model parameter. The resulting problems have $N_m = 675{,}000$ and $N_m = 3{,}375{,}000$ parameters, respectively. It is worth noting that, although the three reservoir cases share similar characteristics, such as the number and location of wells and the prior geostatistical settings, they do not correspond to the same reservoir represented with different levels of discretization. Instead, they represent reservoirs with different volumes, since the number of layers was increased in each case.

For this test problem, in addition to distance-based localization, we considered the power-law and logistic tapers using two strategies for defining $t_0$. Besides adopting a fixed value of $t_0 = 2$, we also evaluated an adaptive strategy in which $t_0$ is determined from the current distribution of $t$ values. Specifically, $t_0$ was selected as the 90th percentile of the $t$ values computed for each data source, which in this example corresponds to each well in the reservoir, at each assimilation step. The rationale behind this choice is that, if the magnitude of the estimated correlations decreases as the model size increases, then $t_0$ should be adjusted accordingly rather than kept fixed. Under this adaptive strategy, each datum is assigned its own value of $t_0$ when computing the corresponding localization taper coefficients.

\begin{figure}[ht!]
\centering
    \subfloat[\scriptsize{1 layer ($N_m = 45{,}000$)}]{\includegraphics[width=0.32\linewidth]{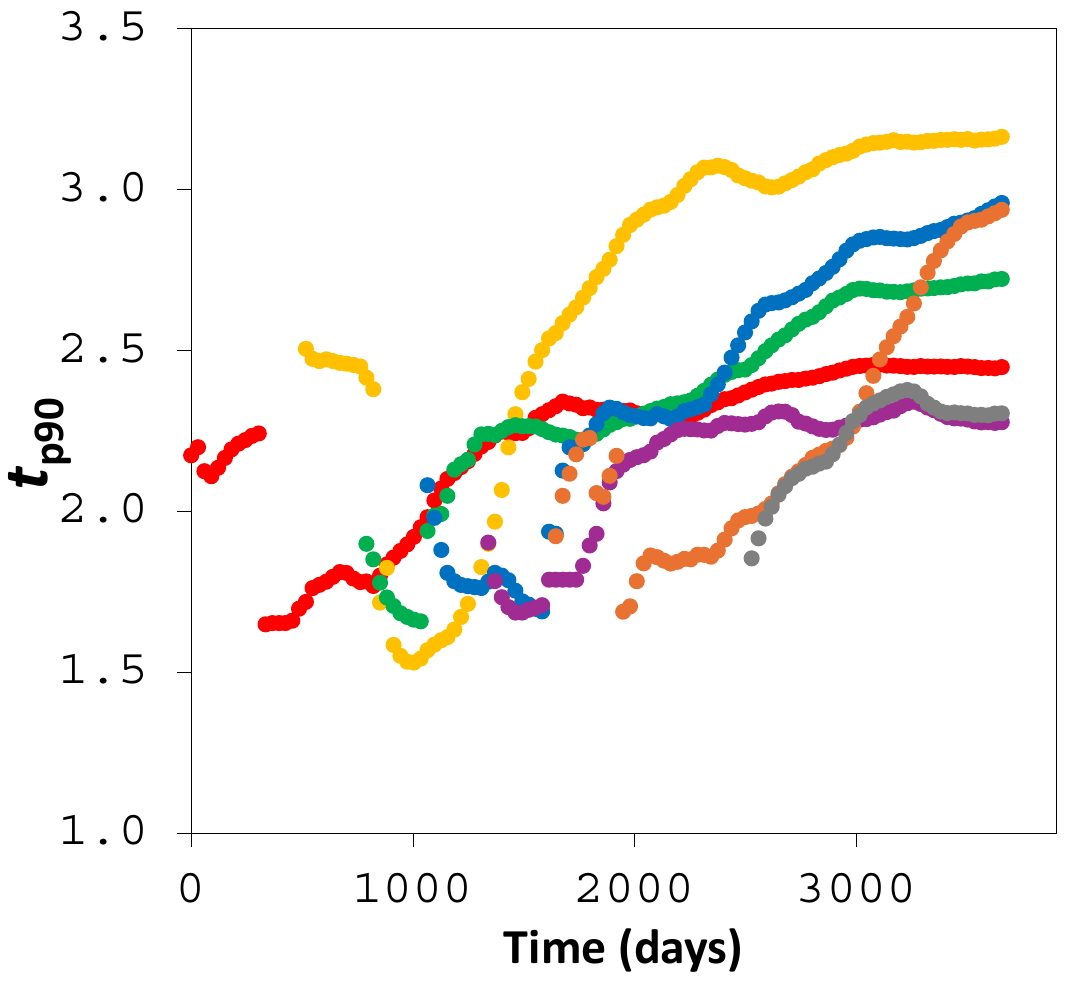}}
    \subfloat[\scriptsize{10 layers ($N_m = 675{,}000$)}]{\includegraphics[width=0.32\linewidth]{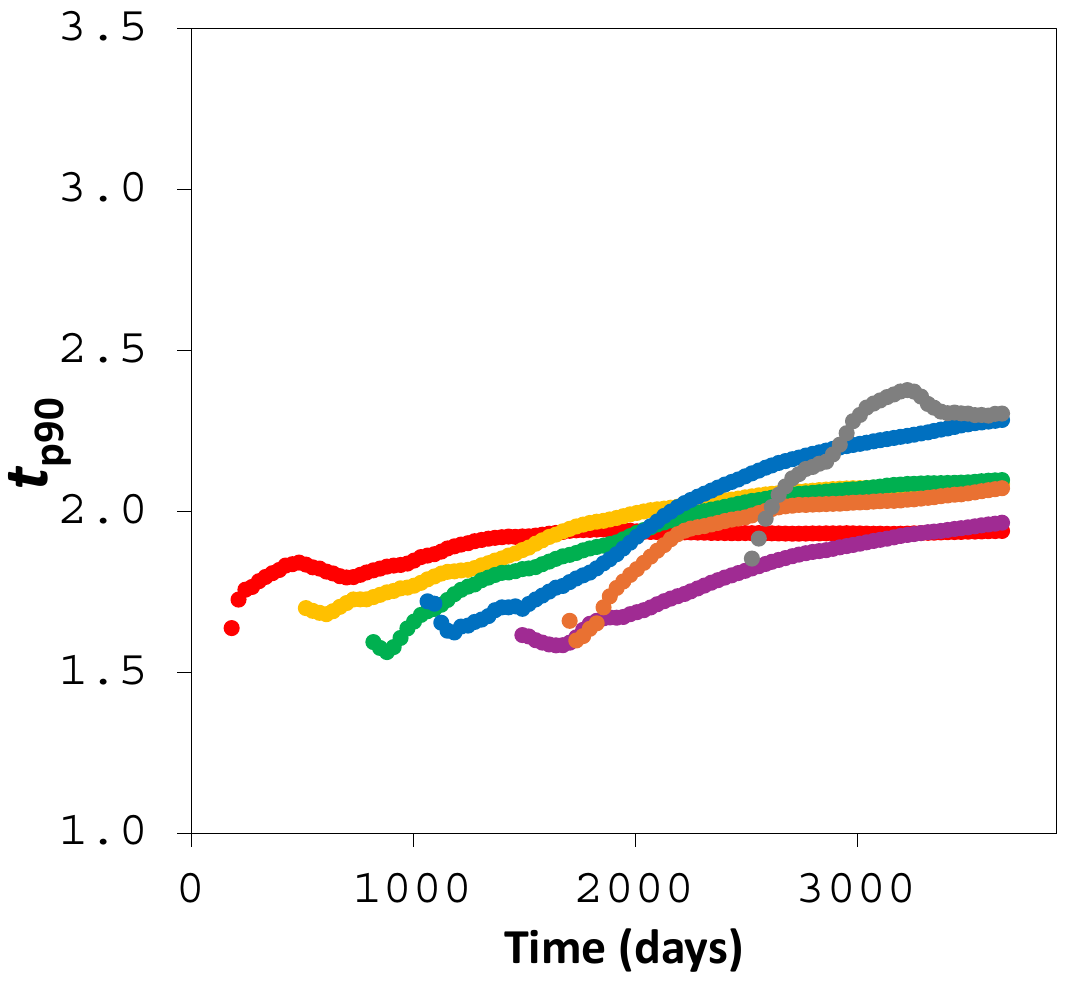}}
    \subfloat[\scriptsize{50 layers ($N_m = 3{,}375{,}000$)}]{\includegraphics[width=0.32\linewidth]{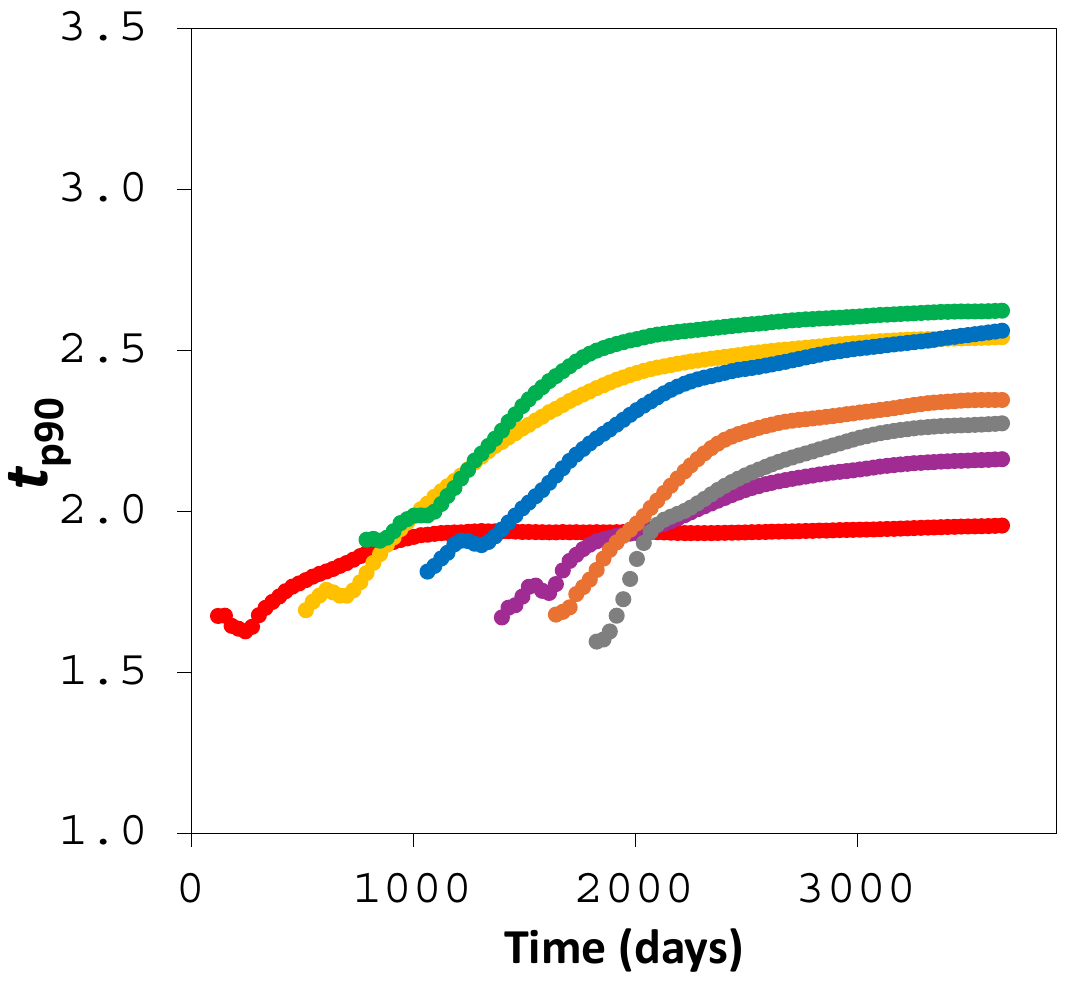}} 
\caption{90th percentile of $t$ over time for seven production wells uniformly distributed throughout the reservoir model, considering cases with 1, 10, and 50 layers. Each color corresponds to a different well. Test case 4.}
\label{fig:test_cases.large_model.t90}
\end{figure}

Fig.~\ref{fig:test_cases.large_model.t90} shows the evolution over time of the 90th percentile of the standardized correlation for water-cut data, denoted by $t_{p90}$, for seven wells uniformly distributed throughout the reservoir, considering models with 1, 10, and 50 layers. This figure shows that $t_{p90}$ values range approximately from 1.5 to 3.5. Moreover, we observe only a slight reduction in the values of $t_{p90}$ as the number of model parameters increases from $N_m = 45{,}000$ (1 layer) to $N_m = 675{,}000$ (10 layers) and $N_m = 3{,}375{,}000$ (50 layers). However, this reduction is relatively modest, suggesting that the initial choice of $t_0 = 2$ remains reasonable for all three cases. Surprisingly, compared to the case with 10 layers, the $t_{p90}$ values for the case with 50 layers increased slightly rather than decreasing further. This contradicts the intuition that the magnitude of the correlations should decrease as the model size increases.

\begin{figure}
\centering
    \subfloat[\scriptsize{1 layer}]{\includegraphics[width=0.32\linewidth]{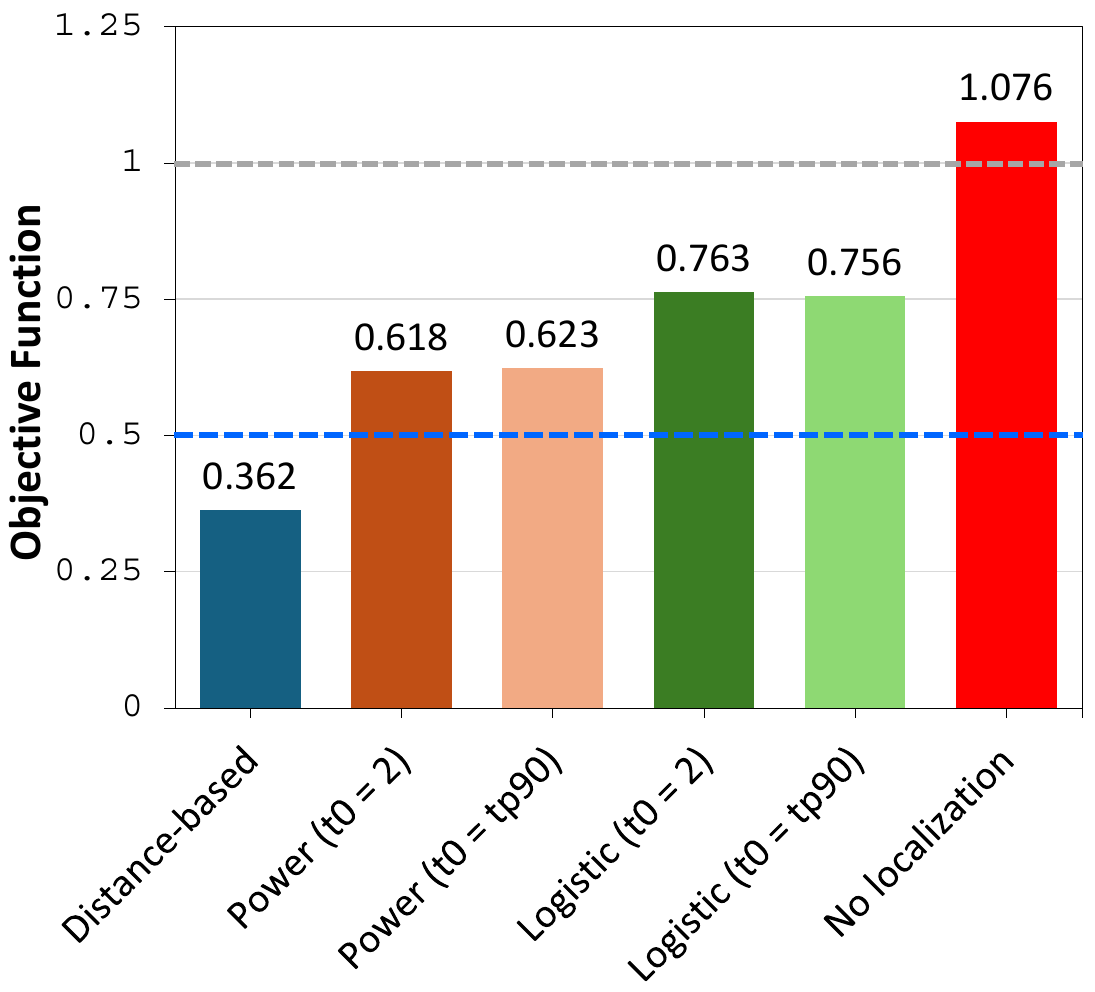}} 
    \subfloat[\scriptsize{10 layers}]{\includegraphics[width=0.32\linewidth]{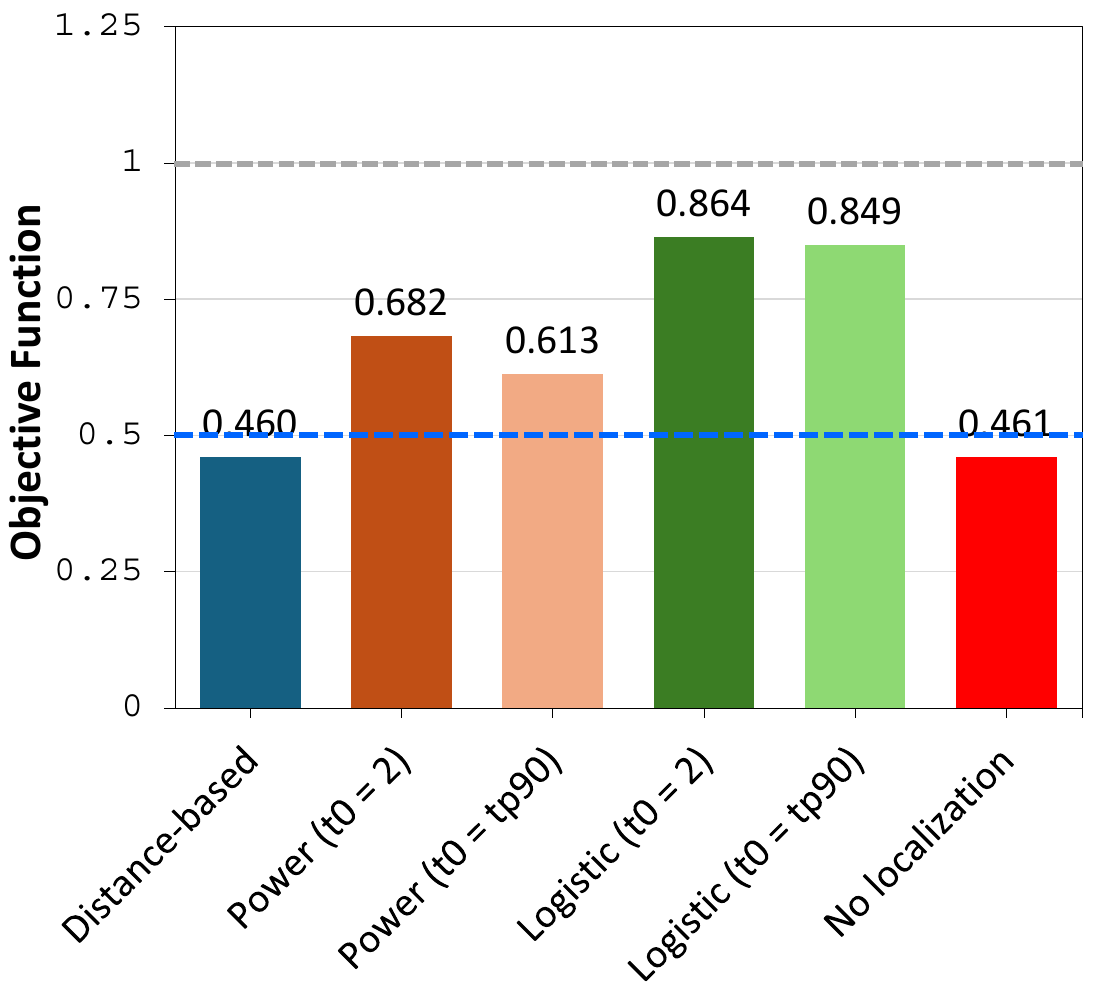}} 
    \subfloat[\scriptsize{50 layers}]{\includegraphics[width=0.32\linewidth]{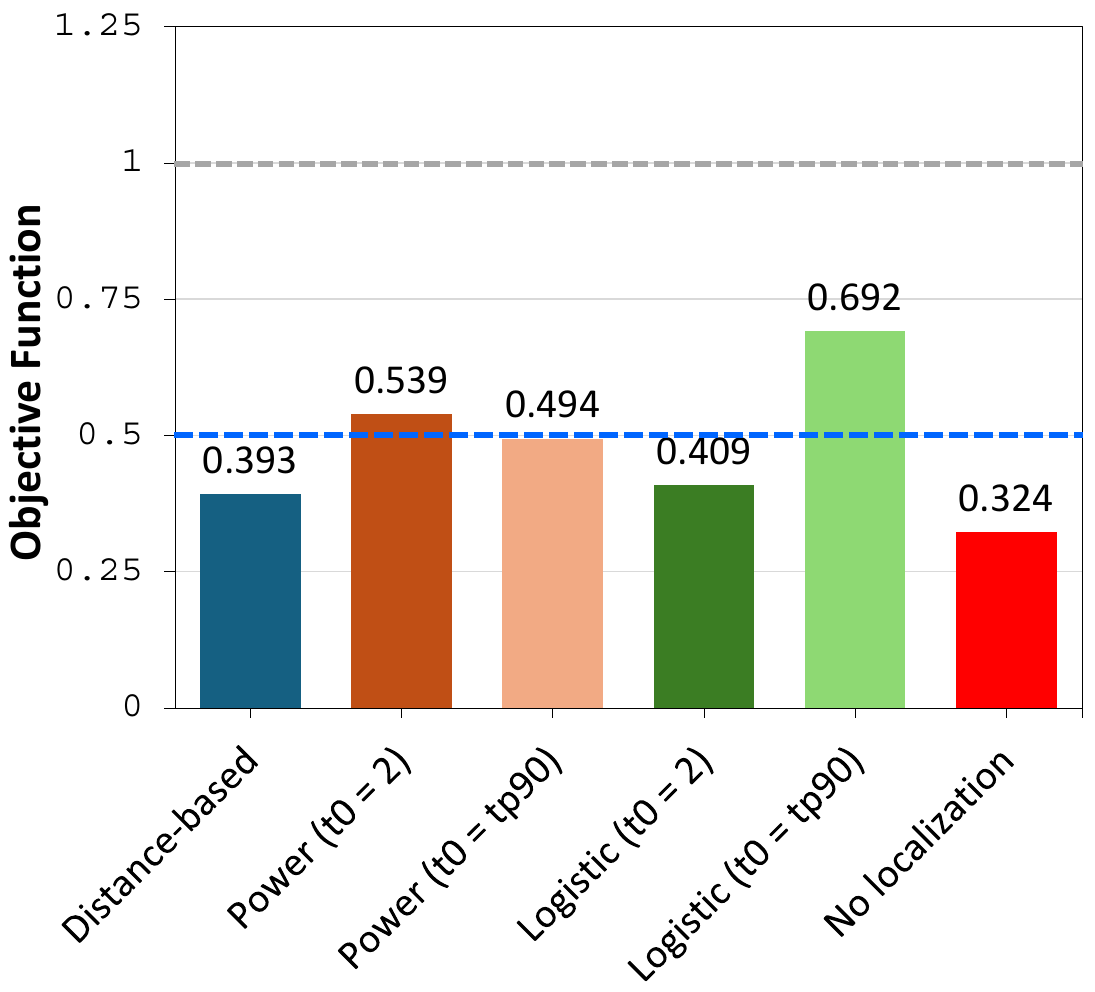}} \\
    \subfloat[\scriptsize{1 layer}]{\includegraphics[width=0.32\linewidth]{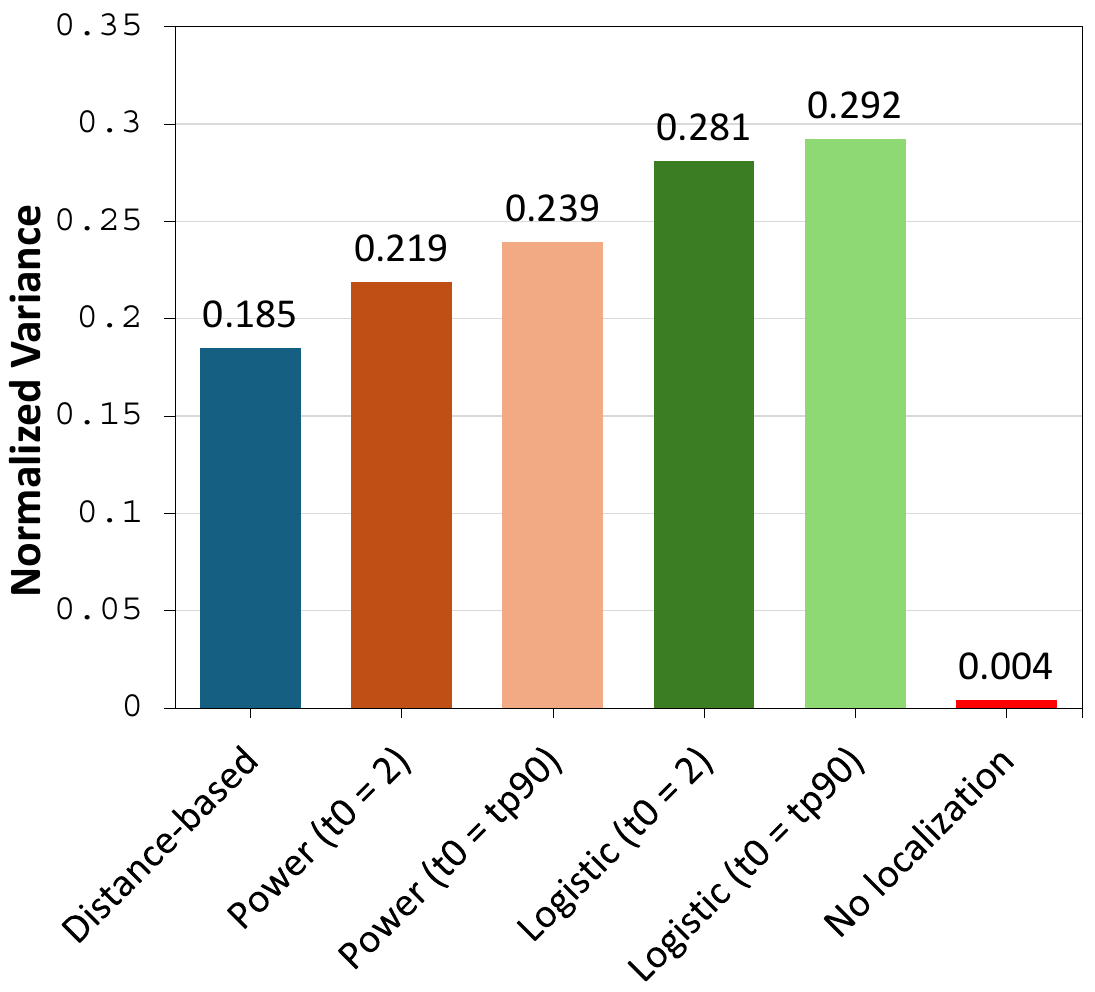}} 
    \subfloat[\scriptsize{10 layers}]{\includegraphics[width=0.32\linewidth]{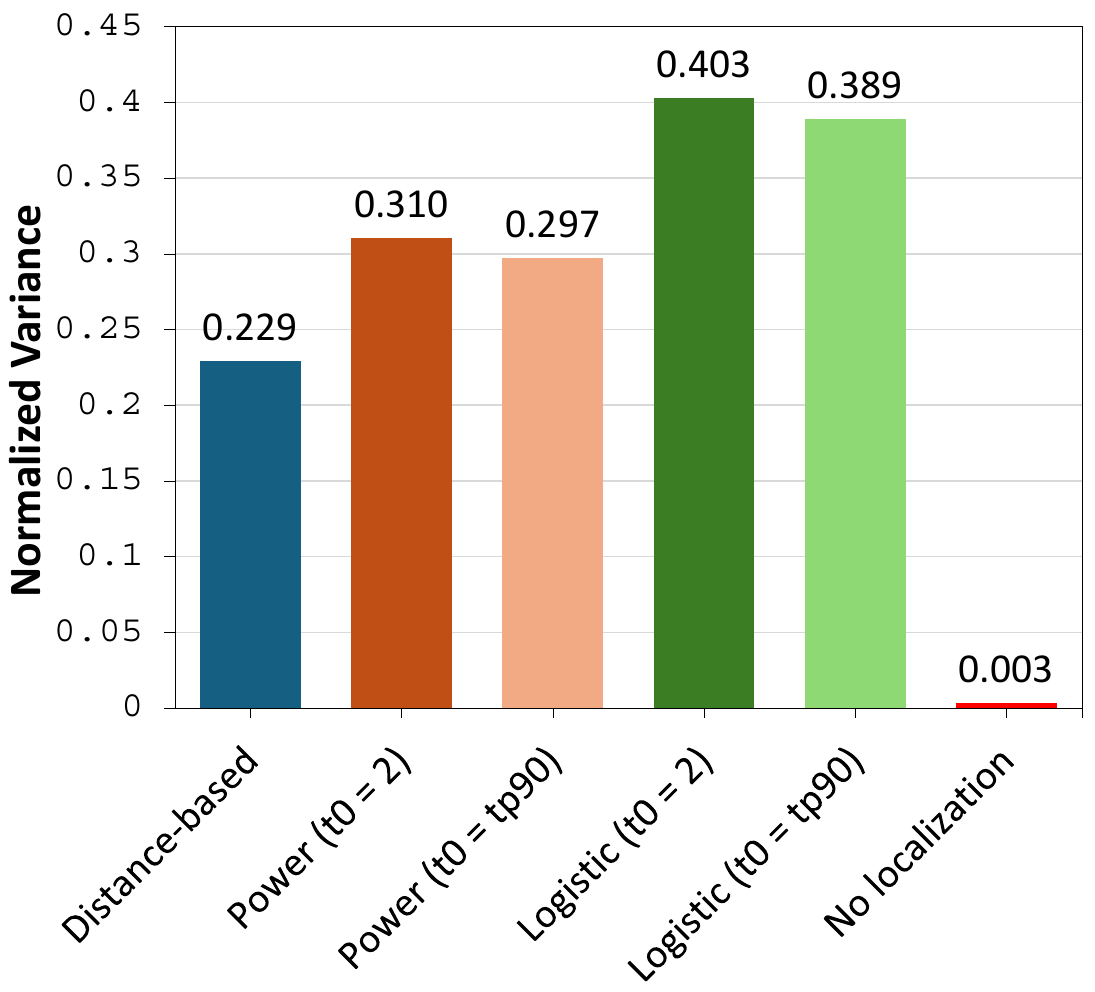}} 
    \subfloat[\scriptsize{50 layers}]{\includegraphics[width=0.32\linewidth]{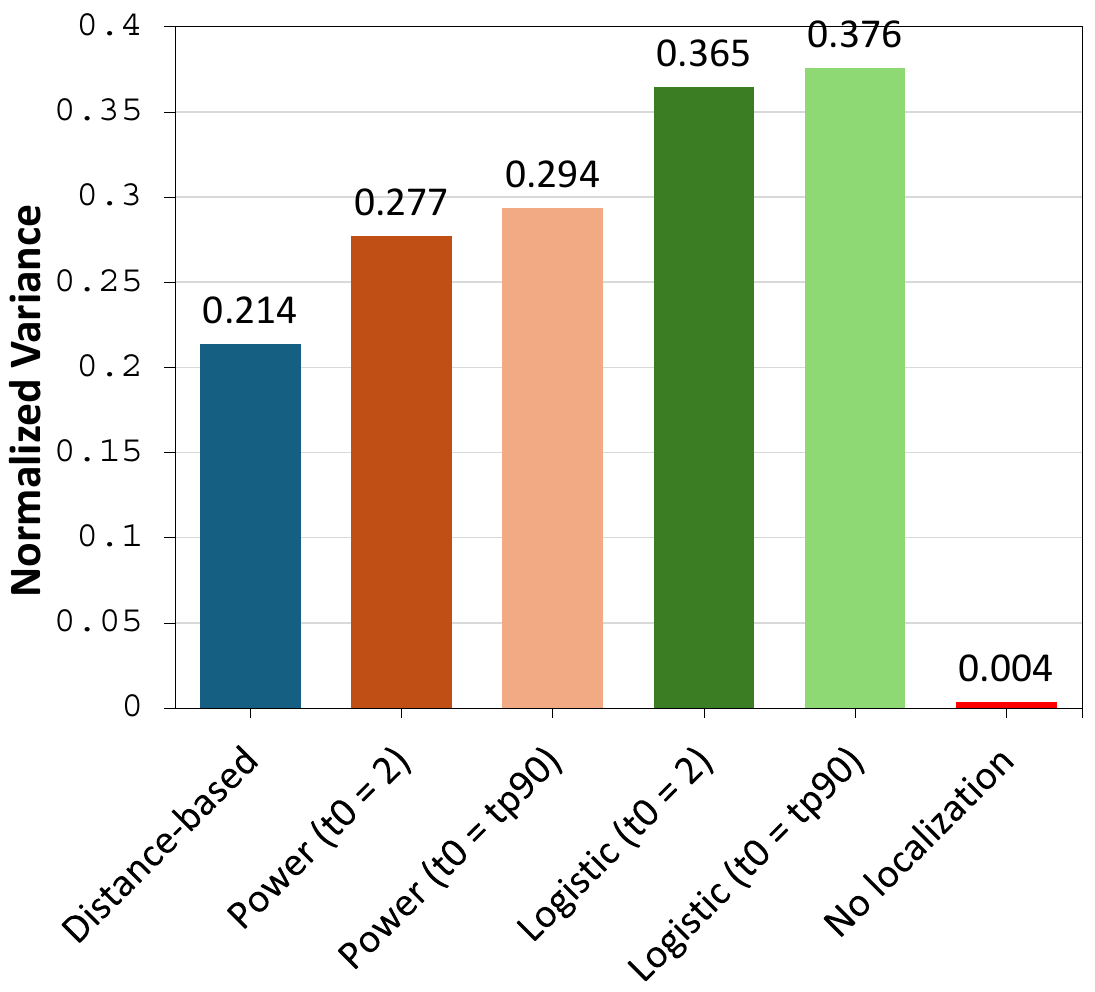}}  
\caption{Average data-mismatch objective function (first row) and normalized variance (second row) for different localization methods, considering the cases with 1, 10, and 50 layers. Test case 4.}
\label{fig:test_cases.large_model.of_nv}
\end{figure}

As a result, the two choices, $t_0 = 2$ and $t_0 = t_{p90}$, led to similar performance. Fig.~\ref{fig:test_cases.large_model.of_nv} summarizes these results. In terms of data-match quality, all localization cases resulted in $\overline{\mathcal{O}_d(\m)} \leq 1$. As observed previously, the logistic taper produced slightly higher values of $\overline{\mathcal{O}_d(\m)}$, followed by the power-law taper and distance-based localization. These results suggest that smoother tapers favor improved data matching.

Regarding normalized variance, the correlation-based tapers resulted in larger values of $\textrm{NV}$, indicating better preservation of ensemble variability than distance-based localization. Unfortunately, for the larger reservoir models, results from large ensembles were not available. Nevertheless, even in the absence of reference solutions, the larger posterior variances obtained with the correlation-based methods can be regarded as a desirable outcome, provided that they are accompanied by acceptable data-match quality.

The results presented in Fig.~\ref{fig:test_cases.large_model.of_nv} do not confirm the initial concern that the performance of correlation-based localization would degrade as $N_m$ increases, as true correlations become weaker and more difficult to estimate. To further investigate this aspect, we define the metric $N_{\textrm{eff}}$ as

\begin{equation}
    N_{\textrm{eff}} = \frac{1}{N_d} \sum_{j=1}^{N_d} \sum_{i=1}^{N_m} r_{ij},    
\end{equation}

\noindent where $r_{ij}$ is the taper coefficient associated with the $i$-th model parameter and the $j$-th data point. This metric provides an estimate of the effective number of model parameters updated, on average, by each observation. Since $r_{ij}\in[0,1]$, each taper coefficient can be interpreted as a fractional contribution of parameter $m_i$ to the update induced by data point $d_j$. Therefore, $N_{\textrm{eff}}$ measures the average effective size of the update footprint associated with each observation.

Dividing $N_{\textrm{eff}}$ by the actual number of model parameters gives

\begin{equation}
    \chi = \frac{N_{\textrm{eff}}}{N_m},
\end{equation}

\noindent where $\chi \in [0,1]$. This normalized quantity represents the average fraction of model parameters effectively updated by each observation.

Table~\ref{tab:test_cases.large_model.Neff} shows the values of $N_{\textrm{eff}}$ and $\chi$ obtained for test case 4. The results indicate that, although the effective number of updated parameters increased with model dimension, the normalized update footprint $\chi$ remained approximately constant or decreased slightly. This suggests that, in the current test case, the correlation-based tapers did not become progressively less selective as the problem dimension increased. Furthermore, the absence of strong performance degradation suggests that nominal model dimension alone may not control the robustness of correlation-based localization.

\begin{table}[h!]
\centering
\caption{Effective number of model parameters updated per observation. Test case 4.}
\label{tab:test_cases.large_model.Neff}
\begin{small}
\begin{tabular}{l c c c | c c c}
\toprule
 & \multicolumn{3}{c|}{$N_{\textrm{eff}}$} 
 & \multicolumn{3}{c}{$\chi = N_{\textrm{eff}}/N_m$} \\
\cmidrule(lr){2-4} \cmidrule(lr){5-7}
Case & 1 layer & 10 layers & 50 layers 
     & 1 layer & 10 layers & 50 layers  \\
\midrule
Distance-based 
& 5,850 & 87,705 & 438,525 
& 0.138 & 0.130 & 0.130 \\

Power-law ($\beta = 3$, $t_0 = 2$) 
& 10,526 & 108,258 & 619,994 
& 0.234 & 0.160 & 0.184 \\

Power-law ($\beta = 3$, $t_0 = t_{p90}$) 
& 8,140 & 106,772 & 528,363 
& 0.181 & 0.158 & 0.157 \\

Logistic ($\gamma = 1.5$, $t_0 = 2$) 
& 9,756 & 96,548 & 570,684 
& 0.217 & 0.143 & 0.169 \\

Logistic ($\gamma = 1.5$, $t_0 = t_{p90}$) 
& 7,218 & 94,998 & 472,044 
& 0.160 & 0.141 & 0.140 \\

No localization 
& 45,000 & 675,000 & 3,375,000 
& 1.000 & 1.000 & 1.000 \\
\bottomrule
\end{tabular}
\end{small}

\end{table}

\section{Discussion}
\label{sec:discussion}

The results presented in the previous section indicate that correlation-based tapers can provide effective localization strategies. Among the methods investigated, the power-law and logistic tapers showed the best overall performance. Both depend on two free parameters: $t_0$ and an exponent controlling the sharpness of the transition. In the numerical experiments, we adopted $\beta = 3$ for the power-law taper and $\gamma = 1.5$ for the logistic taper. With these choices, the power-law taper produced a smoother transition, whereas the logistic taper was more selective.

For $t_0$, we used $t_0 = 2$ in most test cases. This choice was motivated by the analysis of the Student-$t$ statistic, which shows that, for a significance level of 5\%, the corresponding critical value is approximately 2 for a broad range of ensemble sizes (see Appendix~\ref{sec:app.t0}). In test case 4, we also evaluated an adaptive choice based on the 90th percentile of the $t$ values, but this strategy did not lead to relevant improvements.

Overall, the logistic taper with $\gamma = 1.5$ and $t_0 = 2$ provided the best performance in terms of preserving posterior ensemble variance, although at the cost of a small degradation in data-match quality. However, similar taper profiles could likely be obtained with the power-law taper by increasing $\beta$. In this case, the practical distinction between the two tapers may become limited, and there may be no clear advantage of one over the other.

The remaining correlation-based tapers generally showed slightly inferior performance. The MSE and PO tapers do not require additional parameters, but they are too permissive in retaining weak correlations and were less robust across repeated data assimilation experiments. The MPO taper, although slightly more restrictive than the PO taper because of its embedded truncation level of $1/\sqrt{N_e}$, exhibited a similar lack of robustness. The discrepancy taper has only one parameter, but it also showed limited robustness in the repeated experiments of test case 1. Finally, the CGC taper exhibited the most distinct behavior among the correlation-based tapers: it produced a very smooth transition and, unlike the other tapers, did not converge to either zero or one as the correlation coefficient approached its limiting values.

A useful perspective for interpreting correlation-based localization is to distinguish between two related but fundamentally different objectives: \emph{significance detection} and \emph{correlation correction}. In significance detection, the goal is to determine whether an estimated ensemble correlation is statistically distinguishable from sampling noise. From this perspective, localization acts primarily as a filtering mechanism that suppresses spurious correlations caused by finite-ensemble effects. In contrast, correlation correction assumes that the estimated correlation contains useful information, but that its magnitude is distorted by sampling error. In this case, localization acts as a continuous shrinkage estimator that reduces estimation error while preserving meaningful correlation structures. The power-law and logistic tapers proposed in this work were motivated by this interpretation.

Within this framework, $t_0$ can be interpreted as a reference significance level controlling the transition between suppressing and retaining correlations, whereas the exponents $\beta$ and $\gamma$ determine how aggressively this transition occurs. Larger exponent values produce sharper transitions, approaching hard significance-based filtering, while smaller values lead to smoother shrinkage functions that progressively attenuate correlations. Thus, $t_0$ mainly controls which correlations are treated as statistically meaningful, whereas $\beta$ and $\gamma$ control the strength and smoothness of the attenuation applied during the shrinkage process.

\section{Important Remarks}
\label{sec:remarks}

\begin{itemize}
    \item \textbf{Localization based on prior statistics:} An important characteristic of the correlation-based localization methods employed in this paper is that the localization coefficients are computed exclusively from prior ensemble statistics. Although updating the taper values during the intermediate assimilation steps of ES-MDA or IES is possible in principle, our experience indicates that keeping the taper values fixed throughout the assimilation process leads to more robust results. One possible explanation is that the prior ensemble provides a more stable statistical basis for identifying the dominant model-data relationships, whereas correlations estimated during later assimilation steps may be more strongly affected by variance reduction. This behavior has also been observed in previous works \citep{luo:18a,lacerda:19a,silva:25b}. Therefore, although updating the localization coefficients remains an interesting research direction, our current recommendation is to use prior-based taper values.

    \item \textbf{Positive definiteness:} Traditional covariance localization methods are often designed to construct positive-definite correlation matrices, which ensure, through the Schur product theorem, positive-definite localized covariance matrices \citep{houtekamer:01}. In the present work, however, localization is applied directly to individual model-data correlations without explicitly constructing a global localization matrix. As a consequence, we never check or try to impose the resulting matrix to be positive-definite. This is because such matrices are prohibitively large in realistic reservoir applications involving millions of model parameters and thousands of observations. Instead, localization coefficients are computed on demand for small blocks of model-data pairs during the assimilation procedure. This strategy substantially reduces memory requirements and allows the implementation to scale to large reservoir models.

    \item \textbf{Combining correlation- and distance-based localization:} For grid property parameters, a conceivable localization strategy is to combine correlation- and distance-based localization. For example, the final taper could be obtained as the product of a distance-based and a correlation-based taper, allowing the method to simultaneously enforce the removal of long-range correlations while adaptively accounting for the estimated statistical relevance of each model-data pair. In such a strategy, one could select relatively large critical lengths for distance-based localization, so that the distance-based taper acts mainly as a safeguard against clearly nonphysical long-range updates, rather than as the primary mechanism controlling the localization pattern. The correlation-based taper would then provide a second level of adaptivity, reducing the influence of model-data pairs with weak or statistically unreliable correlations within the region allowed by the distance-based taper. This idea is used, for example, in the localization method named PANIC \citep{vishny:24a}.

    \item \textbf{Improving correlation estimates using machine-learning proxies:} A promising research direction in correlation-based localization is the use of machine-learning-based proxy models to improve the correlation estimates employed in taper construction. Since proxy models are typically much cheaper than full-physics simulations, they may allow the generation of significantly larger ensembles for estimating correlations and localization coefficients. In this context, the proxy model would not necessarily replace the forward simulator during data assimilation, but instead provide auxiliary statistical information to improve localization. This idea has been explored in recent publications \citep{lacerda:21a,silva:25b}, with encouraging results for small synthetic reservoir problems. However, the overall performance of these approaches remains limited when applied to large-scale, realistic settings.

    \item \textbf{Computational aspects:} Compared with distance-based localization, correlation-based localization is computationally more demanding. In distance-based approaches, the taper coefficients are typically prescribed for each well or data source and often remain fixed in time. In contrast, correlation-based localization requires estimating one sample correlation for each model-data pair, which may involve millions of model parameters and thousands of observations\footnote{For seismic data assimilation, $N_d$ can be very large, potentially of the same order as the number of model parameters.}. Because each correlation requires an inner product over the ensemble dimension, the total cost of computing all correlations scales as $\mathcal{O}(N_m N_d N_e)$. The subsequent computation of the standardized statistic and taper value requires only a fixed number of operations per pair, and therefore scales as $\mathcal{O}(N_m N_d)$. Note that, in practical implementations of ES-MDA, computing and storing the $N_m \times N_d$ covariance matrix $\widetilde{\C}^\ell_{\m\dsim}$ is avoided; see \citet[Chap.~8]{emerick:25bk}.

    Computing localization coefficients on demand avoids storing large localization matrices, making the approach feasible from a memory perspective. However, it does not reduce the underlying computational cost. Therefore, correlation-based localization may still require substantial CPU time in large-scale applications, especially when both the number of model parameters and observations are large.
\end{itemize}

\section{Conclusions}
\label{sec:conc}

This work investigated correlation-based localization strategies for ensemble data assimilation. The objective was to address three interrelated points:

\begin{enumerate}
    \item \textit{Propose statistically motivated correlation-based taper functions and compare them with existing formulations:} We interpreted localization as a statistical filtering or shrinkage problem, connecting it with the related concepts of significance detection and correlation correction. Based on this interpretation, we proposed three correlation-based tapers: power-law, logistic, and discrepancy-based tapers.
    
    The power-law taper was motivated as a generalization of an MSE taper, introducing additional flexibility to discriminate between meaningful and spurious correlations. The logistic taper was derived from a Bayesian formulation of the optimal tapering problem using a sparsity-promoting spike-and-slab prior. The discrepancy-based taper was motivated by applying Morozov's discrepancy principle to the localized correlation.

    \item \textit{Assess the performance of correlation-based localization in the context of non-local model parameters:} We tested the proposed tapers, together with four correlation-based tapers from the literature, in a reservoir test problem containing only scalar model parameters. Dummy parameters with standard Gaussian priors were included to provide a reference for spurious updates, since any variance reduction in these parameters should be attributed to sampling errors.
    
    All correlation-based tapers improved the results relative to the case without localization, preserving posterior variance in the actual model parameters and suppressing variance reduction in the dummy parameters. These results indicate that correlation-based localization can be useful in problems where distance-based localization cannot be directly applied.
    
    Among the correlation-based tapers, the logistic and MPO tapers produced the largest posterior variances, indicating better preservation of ensemble variability. In terms of data-match quality, all tapers showed similar average performance. However, the MSE, discrepancy-based, PO, and MPO tapers exhibited larger variability across repeated data assimilation runs, with at least one experiment resulting in a large average data-mismatch objective function.

    \item \textit{Investigate whether correlation-based localization can replace distance-based localization in subsurface applications without compromising the quality of the results:} We first tested the correlation-based tapers in a setting where distance-based localization could also be applied, but where the relevant model parameters were not necessarily located near the wells. In this case, a distance-based taper centered at the data location does not optimally represent the actual parameter-data relationships. The results showed that all correlation-based tapers improved the results compared with the case without localization, with the power-law, logistic, and MPO tapers providing the best performance in terms of posterior variance preservation.
    
    We also tested the tapers in a setting where distance-based localization is expected to perform well, since the relevant updates are predominantly local. In this case, the power-law, logistic, and MPO tapers resulted in larger posterior variances than distance-based localization, with values closer to those obtained in the reference data assimilation case using 10,000 ensemble members.
    
    The results revealed a trade-off between data-match quality and variance preservation. Smoother tapers, such as the distance-based Gaspari-Cohn taper and the power-law taper, tended to favor better data matches but produced lower posterior variances. In contrast, tapers with sharper transitions preserved ensemble variance more effectively, at the cost of some deterioration in data-match quality. These trends were consistent across different ensemble sizes and remained similar when the number of reservoir model parameters was increased by factors of 15 and 75.

\end{enumerate}

Overall, the results indicate that correlation-based localization is a viable alternative to distance-based localization. However, additional investigation is needed before concluding that it can fully replace distance-based localization in practical subsurface applications. In particular, it remains unclear how these tapers will perform in more complex data assimilation problems, especially under higher levels of nonlinearity and dimensionality. In contrast, distance-based localization is supported by substantial empirical evidence and remains a robust and computationally efficient choice in many applications.

The proposed power-law and logistic tapers also include free parameters. Our results suggest that this flexibility improves the discrimination between meaningful and spurious correlations, but it also introduces the practical difficulty of parameter tuning. In this work, the same parameter values were used across the test cases with consistently good performance, and suitable parameter ranges were recommended. Nevertheless, future work should investigate adaptive strategies for selecting these parameters automatically.

Finally, correlation-based localization can be considerably more demanding in terms of computational cost. This may become a limiting factor in large-scale applications involving large numbers of model parameters and observations, particularly in problems such as seismic data assimilation.

\section*{Acknowledgments}

The authors would like to express gratitude to Petrobras for their financial support and permission to publish the article.

\begin{small}

\end{small}

\appendix

\section{Appendix}
\label{sec:app}

\subsection{Derivation of the MSE taper}
\label{sec:app.mse}

To derive Eq.~\ref{eq:mse.rstar}, we start from the MSE

\begin{eqnarray}
  \mathcal{J}(r) & = & \Ex\left[(r\widetilde{\rho} - \rho)^2\right] \\
  \nonumber ~ & = & \Ex\left[ r^2 \widetilde{\rho}^2 - 2r\widetilde{\rho}\rho + \rho^2 \right] \\
  \nonumber ~ & = & r^2\Ex\left[\widetilde{\rho}^2\right] - 2r\Ex\left[\widetilde{\rho}\right]\rho + \rho^2.
\end{eqnarray}

\noindent Then, we compute the derivative with respect to $r$ and set to zero, which leads to

\begin{equation}
  r^\star = \frac{\rho\Ex[\widetilde{\rho}]}{\Ex[\widetilde{\rho}^2]}.
\end{equation}

\noindent Assuming $\widetilde{\rho}$ is approximately unbiased for moderate ensemble sizes, $\Ex[\widetilde{\rho}] \approx \rho$, the optimal taper can be written as

\begin{equation}
  r^\star = \frac{\rho^2}{\rho^2 + \operatorname{var}[\widetilde{\rho}]} = \frac{\rho^2}{\rho^2 + \sigma^2}.
\end{equation}

\subsection{Spike-and-Slab Distribution}
\label{sec:app.spike-and-slab_distribution}

The spike-and-slab distribution is a mixture distribution commonly used in Bayesian inference to represent sparsity \citep{mitchell:88a,ohara:09a}. It assumes that a parameter may either be exactly zero (the ``spike'') or drawn from a continuous distribution allowing nonzero values (the ``slab''). This construction is particularly useful when modeling situations where many parameters are expected to be negligible while a few may be significant.

Let $x$ denote an unknown parameter. The spike-and-slab prior can be written as a mixture distribution

\begin{equation}
  p(x) = (1-\lambda)\,\delta(x) + \lambda\,p_{\textrm{slab}}(x),
\end{equation}

\noindent where $\delta(x)$ is a Dirac delta function representing the spike at zero, $p_{\textrm{slab}}(x)$ is a continuous density representing plausible nonzero values, often assumed Gaussian, and $\lambda \in [0,1]$ is the prior probability that the parameter belongs to the slab component. 

Introducing a latent indicator variable $z \in \{0,1\}$ specifying whether $x$ belongs to the spike or slab,

\begin{align}
  x \mid z=0 &\sim \delta(x), \\
  x \mid z=1 &\sim p_{\text{slab}}(x),
\end{align}

\noindent the posterior probability given an observation $y$ with likelihood $p(y \mid x)$ of the slab component becomes

\begin{equation}
  P(z=1 \mid y) = \frac{\lambda\,p(y \mid z=1)}{(1-\lambda)\,p(y \mid z=0) + \lambda\,p(y \mid z=1)},
\end{equation}

\noindent where

\begin{align}
  p(y \mid z=1) &= \int p(y \mid x)\,p_{\text{slab}}(x)\,dx, \\
  p(y \mid z=0) &= p(y \mid x=0).
\end{align}

\noindent The posterior distribution of $x$ therefore remains a mixture,

\begin{equation}
  p(x \mid y) = P(z=0 \mid y)\,\delta(x) + P(z=1 \mid y)\,p(x \mid y,z=1).
\end{equation}

\subsection{Derivation of the Gaussian Prior Taper}
\label{sec:app.gauss_taper}

Assume

\begin{equation}
\widetilde{\rho}\mid \rho \sim \mathcal{N}(\rho,\sigma^2), \qquad \rho\sim\mathcal{N}(0,\upsilon^2).
\end{equation}

\noindent The posterior density is proportional to

\begin{equation}
  p(\rho\mid \widetilde{\rho}) \propto \exp\left(-\frac{(\widetilde{\rho}-\rho)^2}{2\sigma^2}\right) \exp\left(-\frac{\rho^2}{2\upsilon^2}\right).
\end{equation}

\noindent Expanding the exponent,

\begin{equation}
-\frac{(\widetilde{\rho}-\rho)^2}{2\sigma^2} -\frac{\rho^2}{2\upsilon^2} = -\frac{1}{2} \left[ \left(\frac{1}{\sigma^2}+\frac{1}{\upsilon^2}\right)\rho^2 - 2\frac{\widetilde{\rho}}{\sigma^2}\rho + \frac{\widetilde{\rho}^2}{\sigma^2} \right].
\end{equation}

\noindent Completing the square yields

\begin{equation}
\rho\mid\widetilde{\rho} \sim \mathcal{N}\left( \frac{\upsilon^2}{\upsilon^2+\sigma^2}\widetilde{\rho}, \frac{\upsilon^2\sigma^2}{\upsilon^2+\sigma^2} \right),
\end{equation}

\noindent so that

\begin{equation}
\Ex[\rho\mid\widetilde{\rho}] = \frac{\upsilon^2}{\upsilon^2+\sigma^2}\widetilde{\rho}.
\end{equation}

\subsection{Derivation of the Logistic Taper}
\label{sec:app.spike-slab}

Assume the observation model
\begin{equation}
  \widetilde{\rho}\mid\rho \sim \mathcal{N}(\rho,\sigma^2),
\end{equation}

\noindent and the spike-and-slab prior

\begin{equation}
  p(\rho)=(1-\lambda)\delta(\rho)+\lambda\,\phi(\rho;0,\upsilon^2),
\end{equation}

\noindent where $\phi(\cdot;\mu,\sigma^2)$ denotes the Gaussian density with mean $\mu$ and variance $\sigma^2$.

\subsubsection*{Marginal likelihoods of the spike and slab components}

If $\rho=0$, then

\begin{equation}
  \mathcal{L}_{\textrm{spike}}(\widetilde{\rho}) = p(\widetilde{\rho}\mid\rho=0) = \phi(\widetilde{\rho};0,\sigma^2).
\end{equation}

Under the slab component, the marginal likelihood is 

\begin{equation}
  \mathcal{L}_{\textrm{slab}}(\widetilde{\rho}) = \int \phi(\widetilde{\rho};\rho,\sigma^2)\, \phi(\rho;0,\upsilon^2)\,d\rho.
\end{equation}

\noindent Because the convolution of two Gaussian densities is Gaussian,

\begin{equation}
  \mathcal{L}_{\textrm{slab}}(\widetilde{\rho}) = \phi(\widetilde{\rho};0,\sigma^2+\upsilon^2).
\end{equation}

\subsubsection*{Posterior inclusion probability}

By Bayes' rule,

\begin{equation}
  f(\widetilde{\rho}) = P(\rho\neq0\mid\widetilde{\rho}) = \frac{\lambda\,\mathcal{L}_{\textrm{slab}}(\widetilde{\rho})}{(1-\lambda)\mathcal{L}_{\textrm{spike}}(\widetilde{\rho})+\lambda\,\mathcal{L}_{\textrm{slab}}(\widetilde{\rho})}.
\end{equation}

Substituting the two marginal likelihoods,
\begin{equation}
  f(\widetilde{\rho}) = \frac{ \lambda\,\phi(\widetilde{\rho};0,\sigma^2+\upsilon^2)}{(1-\lambda)\phi(\widetilde{\rho};0,\sigma^2) +\lambda\,\phi(\widetilde{\rho};0,\sigma^2+\upsilon^2) }.
\end{equation}

\noindent To simplify this expression, write the Gaussian densities explicitly:

\begin{equation}
  \phi(x;0,\sigma_x^2) = \frac{1}{\sqrt{2\pi}\sigma_x}\exp\left(-\frac{x^2}{2\sigma_x^2}\right).
\end{equation}

\noindent Then

\begin{equation}
\frac{\phi(\widetilde{\rho};0,\sigma^2)}{\phi(\widetilde{\rho};0,\sigma^2+\upsilon^2)} = \sqrt{\frac{\sigma^2+\upsilon^2}{\sigma^2}}
\exp\left(-\frac{\widetilde{\rho}^2}{2\sigma^2}+\frac{\widetilde{\rho}^2}{2(\sigma^2+\upsilon^2)} \right).
\end{equation}

\noindent Combining the terms in the exponent,

\begin{equation}
  -\frac{\widetilde{\rho}^2}{2\sigma^2} +\frac{\widetilde{\rho}^2}{2(\sigma^2+\upsilon^2)} = -\frac{\upsilon^2\widetilde{\rho}^2}{2\sigma^2(\sigma^2+\upsilon^2)}.
\end{equation}

\noindent Hence,

\begin{equation}\label{eq:app.spike-slab.gamma}
  f(\widetilde{\rho}) = \left[1 + \frac{1-\lambda}{\lambda}\sqrt{\frac{\sigma^2+\upsilon^2}{\sigma^2}}\exp\left(-\frac{\upsilon^2\widetilde{\rho}^2}{2\sigma^2(\sigma^2+\upsilon^2)}\right)\right]\inv.
\end{equation}

\subsubsection*{Posterior mean under the slab}

Conditionally on the slab component, the posterior density is proportional to

\begin{equation}
  p(\rho\mid\widetilde{\rho},\rho\neq0) \propto \exp\left(-\frac{(\widetilde{\rho}-\rho)^2}{2\sigma^2}\right) \exp\left(-\frac{\rho^2}{2\upsilon^2}\right).
\end{equation}

\noindent Expanding the exponent,

\begin{align}
  -\frac{(\widetilde{\rho}-\rho)^2}{2\sigma^2} -\frac{\rho^2}{2\upsilon^2} &=-\frac{1}{2}\left[\frac{\widetilde{\rho}^2-2\widetilde{\rho}\rho+\rho^2}{\sigma^2}+\frac{\rho^2}{\upsilon^2}\right] \\
  &=-\frac{1}{2}\left[\left(\frac{1}{\sigma^2}+\frac{1}{\upsilon^2}\right)\rho^2-2\frac{\widetilde{\rho}}{\sigma^2}\rho+\frac{\widetilde{\rho}^2}{\sigma^2}\right].
\end{align}

Completing the square shows that

\begin{equation}
  \rho\mid\widetilde{\rho},(\rho\neq0)\sim\mathcal{N}\left(\frac{\upsilon^2}{\upsilon^2+\sigma^2}\widetilde{\rho},\frac{\upsilon^2\sigma^2}{\upsilon^2+\sigma^2}\right).
\end{equation}

\noindent Therefore,

\begin{equation}\label{eq:app.spike-slab.post_mean}
  \Ex_{\textrm{slab}}[\rho\mid\widetilde{\rho}]=\frac{\upsilon^2}{\upsilon^2+\sigma^2}\widetilde{\rho}.
\end{equation}

\subsubsection*{Posterior mean under the full spike-and-slab prior}

Using the mixture representation of the posterior,

\begin{equation}
  p(\rho\mid\widetilde{\rho})=\bigl(1-f(\widetilde{\rho})\bigr)\delta(\rho)+f(\widetilde{\rho})\,p(\rho\mid\widetilde{\rho},\rho\neq0),
\end{equation}

\noindent the posterior mean is

\begin{align}
  \Ex[\rho\mid\widetilde{\rho}]&=\bigl(1-f(\widetilde{\rho})\bigr)\cdot 0+f(\widetilde{\rho})\,\Ex_{\textrm{slab}}[\rho\mid\widetilde{\rho}] \\
  &=f(\widetilde{\rho})\frac{\upsilon^2}{\upsilon^2+\sigma^2}\widetilde{\rho}.
\end{align}

\noindent Thus the taper is

\begin{equation}
  r(\widetilde{\rho}) = \frac{\Ex[\rho\mid\widetilde{\rho}]}{\widetilde{\rho}} = f(\widetilde{\rho}) \frac{\upsilon^2}{\upsilon^2+\sigma^2}.
\end{equation}

\subsubsection*{Standardized representation}

Define
\begin{equation}
  t=\frac{|\widetilde{\rho}|}{\sigma},\qquad\tau=\frac{\upsilon}{\sigma}.
\end{equation}

\noindent Then

\begin{equation}
  \frac{\upsilon^2}{\upsilon^2+\sigma^2}=\frac{\tau^2}{\tau^2+1},
\end{equation}

\noindent and

\begin{equation}
  \frac{\upsilon^2\widetilde{\rho}^2}{2\sigma^2(\sigma^2+\upsilon^2)}=\frac{\tau^2}{2(1+\tau^2)}t^2.
\end{equation}

\noindent Substituting into Eq.~\ref{eq:app.spike-slab.gamma},

\begin{equation}
  f(t)=\left[1+\frac{1-\lambda}{\lambda}\sqrt{\tau^2+1}\exp\left(-\frac{\tau^2}{2(1+\tau^2)}t^2\right)\right]\inv.
\end{equation}

\noindent Hence,

\begin{equation}
  r(t) =\frac{\tau^2}{\tau^2+1}\left[1+\frac{1-\lambda}{\lambda}\sqrt{\tau^2+1}\exp\left(-\frac{\tau^2}{2(1+\tau^2)}t^2\right)\right]\inv.
\end{equation}

\subsection{Logistic Representation}
\label{sec:app.spike-slab.logistic}

Let

\begin{align}
  r_{\max} &= \frac{\tau^2}{\tau^2+1}, \\
  b &= \frac{1-\lambda}{\lambda}\sqrt{\tau^2+1}, \\
  c &= \frac{\tau^2}{2(1+\tau^2)}.
\end{align}

\noindent Then

\begin{equation}
  r(t)=\frac{r_{\max}}{1+b\exp(-ct^2)}.
\end{equation}

\noindent Writing $b=\exp(ct_0^2)$, where

\begin{equation}
  t_0^2=\frac{1}{c}\ln b = \frac{2(1+\tau^2)}{\tau^2} \ln\left(\frac{1-\lambda}{\lambda}\sqrt{\tau^2+1}\right),
\end{equation}

\noindent we obtain

\begin{equation}
r(t) = r_{\max} \frac{1}{1+\exp\left(-c(t^2-t_0^2)\right)},
\end{equation}

\noindent which shows that the spike-and-slab taper is a scaled logistic function of the squared standardized correlation.

\subsection{Justification of Generalized Tapers via Spike-and-Slab Models}
\label{app:generalized_tapers}

In this section, we provide a probabilistic justification for the use of generalized taper functions with exponents different from two. The key idea is to interpret the taper as a posterior inclusion probability under a two-groups (spike-and-slab) model, and to relate its functional form to assumptions on the Bayes factor.

\subsubsection*{General framework}

Let $t \ge 0$ denote a statistic measuring the strength of an estimated correlation (e.g., a standardized correlation coefficient). Introduce a latent indicator variable $z \in \{0,1\}$ such that

\begin{equation}
z = \begin{cases}
  1, & \text{if the correlation is meaningful (signal)},\\
  0, & \text{if the correlation is spurious (noise)}.
\end{cases}
\end{equation}

Assume a spike-and-slab prior,

\begin{equation}
  P(z=1) = \lambda, \qquad P(z=0) = 1-\lambda,
\end{equation}

\noindent where $\lambda \in (0,1)$ is the prior inclusion probability.

We interpret the localization coefficient as the posterior inclusion probability,

\begin{equation}
  r(t) = P(z=1 \mid t).
\end{equation}

By Bayes' rule,

\begin{equation}
  P(z=1 \mid t) = \frac{p(t \mid z=1)\,\lambda}{p(t \mid z=1)\,\lambda + p(t \mid z=0)\,(1-\lambda)}.
\end{equation}

Defining the Bayes factor in favor of inclusion as

\begin{equation}
  \mathrm{BF}(t) = \frac{p(t \mid z=1)}{p(t \mid z=0)},
\end{equation}

\noindent we obtain

\begin{equation}\label{eq:posterior_inclusion}
  r(t) = \frac{\lambda\,\mathrm{BF}(t)}{(1-\lambda) + \lambda\,\mathrm{BF}(t)}.
\end{equation}

Therefore, the form of the taper function is entirely determined by how the Bayes factor depends on $t$.

\subsubsection*{Generalized MSE (power-law) taper}

Assume that the Bayes factor grows proportionally to a power of $t$, namely

\begin{equation}\label{eq:bf_power_law}
  \mathrm{BF}(t) = b\, t^\beta, 
\end{equation}

\noindent where $\beta > 0$ and $b > 0$ is a constant.

Substituting \ref{eq:bf_power_law} into \ref{eq:posterior_inclusion},

\begin{equation}
  r(t) = \frac{\lambda b t^\beta}{(1-\lambda) + \lambda b t^\beta}.
\end{equation}

Dividing numerator and denominator by $\lambda b$,

\begin{equation}
  r(t) = \frac{t^\beta}{\frac{1-\lambda}{\lambda b} + t^\beta}.
\end{equation}

Defining the parameter $t_0$ through

\begin{equation}
  t_0^\beta = \frac{1-\lambda}{\lambda b},
\end{equation}

\noindent we obtain the generalized power-law taper

\begin{equation}\label{eq:generalized_mse_taper}
  r(t) = \frac{t^\beta}{t^\beta + t_0^\beta}.
\end{equation}

This expression coincides with the standard MSE-optimal taper when $\beta = 2$. The generalization to $\beta \ne 2$ therefore corresponds to assuming that the evidence in favor of inclusion grows as a power $t^\beta$.

\subsubsection*{Generalized logistic taper}

Alternatively, assume that the log-Bayes factor grows proportionally to a power of $t$:

\begin{equation}\label{eq:log_bf_logistic}
  \ln \left(\mathrm{BF}(t) \right) = c\left(t^\gamma - t_0^\gamma\right),
\end{equation}

\noindent where $\gamma > 0$, $c > 0$, and $t_0 > 0$.

Exponentiating \ref{eq:log_bf_logistic} gives

\begin{equation}
  \mathrm{BF}(t) = \exp\!\left(c\left(t^\gamma - t_0^\gamma\right)\right).
\end{equation}

Substituting into \ref{eq:posterior_inclusion},
\begin{equation}
  r(t) = \frac{\lambda \exp\!\left(c\left(t^\gamma - t_0^\gamma\right)\right)}{(1-\lambda) + \lambda \exp\!\left(c\left(t^\gamma - t_0^\gamma\right)\right)}.
\end{equation}

Dividing numerator and denominator by $\lambda \exp\!\left(c(t^\gamma - t_0^\gamma)\right)$,

\begin{equation}
  r(t) = \frac{1}{1 + \frac{1-\lambda}{\lambda}\exp\!\left(-c\left(t^\gamma - t_0^\gamma\right)\right)}.
\end{equation}

Absorbing the prior odds $(1-\lambda)/\lambda$ into the parameter $t_0$ (or assuming equal prior odds for simplicity), we obtain the generalized logistic taper

\begin{equation}\label{eq:generalized_logistic_taper}
  r(t) = \frac{1}{1 + \exp\!\left(-c\left(t^\gamma - t_0^\gamma\right)\right)}.
\end{equation}

This shows that replacing $t^2$ by $t^\gamma$ corresponds to assuming that the log-evidence in favor of inclusion grows proportionally to $t^\gamma$.

\subsubsection*{Difference between $\gamma$ and $\beta$}
\label{app:discussion_alpha_beta}

Eqs.~\ref{eq:generalized_logistic_taper} and \ref{eq:generalized_mse_taper} provide a unified interpretation of taper functions as posterior inclusion probabilities under a spike-and-slab model. However, the roles of the exponents $\gamma$ and $\beta$ differ significantly due to the distinct ways in which they enter the taper functions.

\paragraph{Power-law taper and the role of $\beta$:}

For the power-law taper,

\begin{equation}
  r(t)=\frac{t^\beta}{t^\beta + t_0^\beta},
\end{equation}

\noindent the exponent $\beta$ acts directly on the Bayes factor, not on its logarithm. Consequently, $\beta$ controls both the behavior near $t=0$ and the rate at which the taper approaches one.

In particular:

\begin{itemize}
  \item For small $t$, one has $r(t) \sim (t/t_0)^\beta$, so the decay near zero is directly governed by $\beta$.
  \item If $\beta \le 2$, the taper tends to be too permissive, assigning non-negligible weights to weak correlations.
  \item Larger values $\beta>2$ produce stronger shrinkage of small $t$, effectively suppressing spurious correlations.
\end{itemize}

Thus, in practice, it is often desirable to select $\beta>2$ to enforce more aggressive filtering of weak correlations. From the spike-and-slab viewpoint, this corresponds to assuming that the evidence in favor of inclusion grows super-quadratically with $t$, thereby requiring stronger signals before a correlation is deemed meaningful.

\paragraph{Logistic taper and the role of $\gamma$:}

For the logistic taper,
\begin{equation}
  r(t)=\frac{1}{1+\exp\!\left(-c\left(t^\gamma - t_0^\gamma\right)\right)},
\end{equation}

\noindent the exponent $\gamma$ acts on the argument of the exponential, i.e., on the log-Bayes factor. As a result, $\gamma$ primarily controls the \emph{geometry of the transition} around the threshold $t_0$, rather than the asymptotic behavior of the taper.

In particular:

\begin{itemize}
  \item The slope of the taper at $t=t_0$ is proportional to $c\,\gamma\,t_0^{\gamma-1}$.
  \item Larger values of $\gamma$ increase this slope, leading to a sharper transition between $r(t)\approx 0$ and $r(t)\approx 1$.
  \item Conversely, values $\gamma<2$ produce a smoother and more gradual transition.
\end{itemize}

Therefore, when the goal is to avoid overly abrupt switching behavior, it is natural to consider $\gamma \in [1,2]$. In particular, $\gamma=2$ may lead to excessively sharp transitions when combined with the calibration of $c$ required to enforce small values of $r(0)$, whereas $\gamma \approx 1$ yields a more controlled and interpretable transition region.

From the spike-and-slab perspective, this corresponds to assuming that the log-evidence in favor of inclusion grows sub-quadratically with $t$, which moderates the rate at which correlations are classified as significant.

\paragraph{Interpretation of the asymmetry between $\beta$ and $\gamma$:}

The different regimes $\gamma<2$ and $\beta>2$ reflect the distinct roles played by the two taper families:

\begin{itemize}
  \item In the power-law taper, the exponent directly controls the \emph{relative weight} assigned to small versus large values of $t$. Increasing $\beta$ is necessary to sufficiently penalize weak correlations.
  \item In the logistic taper, the exponent modifies the \emph{transition region} through the log-Bayes factor. Reducing $\gamma$ smooths the transition without substantially altering the limiting values.
\end{itemize}

Therefore, although both tapers arise from the same spike-and-slab framework, their parametrizations lead to different practical requirements. The choice $\beta>2$ and $\gamma \in [0.5,2]$ reflects a balance between effective suppression of spurious correlations (power-law taper) and smoothness (logistic taper).

\subsection{Value of $t_0$ from Student-$t$ Statistic}
\label{sec:app.t0}

To provide a statistical interpretation of $t_0$, we consider the classical test of zero correlation. Under the assumption of a bivariate normal distribution and the null hypothesis $H_0:\rho=0$, the statistic

\begin{equation}\label{eq:app.t0.T}
  T = \frac{|\widetilde{\rho}| \sqrt{N_e - 2}}{\sqrt{1 - \widetilde{\rho}^2}}
\end{equation}

\noindent follows a Student-$t$ distribution with $\nu = N_e - 2$ degrees of freedom \citep{rice:07bk}.

Let $\varphi$ denote the significance level of a two-sided test, and define the critical value

\begin{equation}
  T_{\varphi} = t_{\nu,\,1-\varphi/2},
\end{equation}

\noindent where $t_{\nu,p}$ denotes the $p$-quantile of the Student-$t$ distribution with $\nu$ degrees of freedom, i.e.,

\begin{equation}
  P(T \le t_{\nu,p}) = p.
\end{equation}

\noindent The quantity $t_{\nu,p}$ does not admit a closed-form expression and must be computed numerically as the inverse of the cumulative distribution function of the Student-$t$ distribution. In practice, it is readily obtained using standard numerical routines (e.g., \texttt{t.ppf} from \texttt{scipy.stats} in Python) or from tabulated values available in statistical textbooks.

Correlations with $T < T_{\varphi}$ are not statistically significant at level $\varphi$, while correlations with $T > T_{\varphi}$ are considered significant.

In this work, we write the taper in terms of the standardized correlation coefficient, $t$. For small correlations, one can show that

\begin{equation}
  T \approx t,
\end{equation}

\noindent so that the standardized correlation $t$ can be interpreted as a Gaussian approximation of the Student-$t$ statistic. This motivates selecting $t_0$ as an approximation of the critical value $T_{\varphi}$, i.e.,

\begin{equation}
  t_0 \approx T_{\varphi}.
\end{equation}

\noindent $t_0$ induces a corresponding threshold on the magnitude of the correlation coefficient. By inverting the relation between $T$ and $\widetilde{\rho}$, the critical correlation is given by

\begin{equation}
  \rho_0 = \frac{T_{\varphi}}{\sqrt{T_{\varphi}^2 + (N_e - 2)}}.
\end{equation}

\noindent Thus, correlations with magnitude below $\rho_0$ are not statistically distinguishable from noise at significance level $\varphi$.

Table~\ref{tab:t0_values} shows typical values of $t_0 \approx T_{\varphi}$ and the corresponding correlation thresholds $\rho_0$ for different ensemble sizes and significance levels. As expected, smaller ensembles require larger thresholds, reflecting higher sampling uncertainty.

\begin{table}[h!]
\centering
\caption{Typical values of $t_0 \approx T_{\varphi}$ and corresponding critical correlations $\rho_0$ for different ensemble sizes and significance levels $\varphi$.}
\label{tab:t0_values}
\begin{tabular}{c c c c | c c c}
\hline
& \multicolumn{3}{c}{$t_0$} & \multicolumn{3}{c}{$\rho_0$} \\
\cline{2-4} \cline{5-7}
$N_e$ & $\varphi=0.10$ & $\varphi=0.05$ & $\varphi=0.01$ & $\varphi=0.10$ & $\varphi=0.05$ & $\varphi=0.01$ \\
\hline
50   & 1.677 & 2.011 & 2.682 & 0.235 & 0.279 & 0.361 \\
100  & 1.660 & 1.984 & 2.626 & 0.165 & 0.197 & 0.256 \\
200  & 1.653 & 1.972 & 2.601 & 0.117 & 0.139 & 0.182 \\
1,000 & 1.646 & 1.962 & 2.581 & 0.052 & 0.062 & 0.081 \\
\hline
\end{tabular}
\end{table}

The table highlights two important aspects. First, as $N_e$ increases, the threshold converges to the corresponding Gaussian quantile, so that $t_0$ becomes essentially independent of the ensemble size. Second, for small ensembles, larger values of $t_0$ are required, leading to stronger shrinkage of weak correlations. The values of $\rho_0$ further show that the minimum detectable correlation decreases approximately as $1/\sqrt{N_e}$.

Based on these arguments, a reasonable default choice is 

\begin{equation}
    t_0 \in[1, 3].    
\end{equation}

The selection of $t_0$ based on the Student-$t$ statistic relies on assumptions that are not strictly satisfied in ensemble data assimilation. In particular, Eq.~\ref{eq:app.t0.T} assumes independent and identically distributed samples drawn from a bivariate normal distribution. In practice, ensemble members are typically correlated due to the data assimilation updates, model errors, and finite ensemble effects. As a result, the effective number of degrees of freedom may be smaller than $N_e-2$, leading to an underestimation of the sampling variability of $\widetilde{\rho}$. Consequently, $t_0$ obtained from the nominal Student-$t$ distribution may be overly permissive. 

On the other hand, another important aspect is the impact of the problem dimensions, namely the number of model parameters $N_m$ and data points $N_d$. In high-dimensional settings, the estimation of correlations becomes increasingly challenging, as the signal-to-noise ratio of individual correlation estimates decreases and the number of estimated correlations grows rapidly. As a result, many meaningful correlations may lie below the nominal significance threshold implied by $t_0$. In such cases, adopting a large value of $t_0$ may lead to excessive tapering, removing relevant correlations and potentially deteriorating the performance of the data assimilation procedure. This highlights that the choice of $t_0$ should balance statistical significance with practical identifiability in high-dimensional problems. 

\end{document}